\documentclass[useAMS,usenatbib]{mn2e}
\usepackage{amsmath}
\usepackage{amssymb}
\usepackage{subfig}
\usepackage{color}
\usepackage{array}
\usepackage{caption}
\usepackage{graphicx}
\usepackage{setspace}
\usepackage{natbib}
\newcommand{\civ}{\ifmmode {\rm C}\,{\sc IV} \else C\,{\sc iv}\fi}
\newcommand{\ciii}{\ifmmode {\rm C}\,{\sc III} \else C\,{\sc iii}\fi}
\newcommand{\mgii}{\ifmmode {\rm Mg}\,{\sc II} \else Mg\,{\sc ii}\fi}

\makeatletter
\newcommand*{\rom}[1]{\expandafter\@slowromancap\romannumeral #1@}
\makeatother

\usepackage{fixltx2e}

\usepackage{hyperref}
\begin{document}
\title[AGN Reverberation Mapping Project]{Simulations of the OzDES AGN Reverberation Mapping Project}
\author[A. L. King et al.]
{\parbox{\textwidth}{Anthea L. King$^{1,2}$\thanks{E-mail:anthea.king@uqconnect.edu.au}, Paul Martini$^{3,4}$, Tamara M. Davis$^{1}$, K. D. Denney$^{3}$,  C. S. Kochanek$^{3,4}$, Bradley M. Peterson$^{3,4}$, Andreas Skielboe$^{2}$, Marianne Vestergaard$^{2,5}$, Eric Huff$^{3,4}$, Darach Watson$^{2}$, Manda Banerji$^{6}$, Richard McMahon$^{6}$,  Rob Sharp$^{7}$ and C. Lidman$^{8}$.}\vspace{0.4cm}\\
\parbox{\textwidth}{$^1$School of Mathematics and Physics, University of Queensland, Brisbane, QLD 4072, Australia\\
$^2$Dark Cosmology Centre, Niels Bohr Institute, University of Copenhagen, Juliane Maries Vej 30, DK-2100 Copenhagen, Denmark\\
$^3$Department of Astronomy, Ohio State University, 140 W. 18th Ave.,Columbus, OH, 43221, USA\\
$^4$Center for Cosmology and AstroParticle Physics, The Ohio State University, 191 West Woodruff Avenue, Columbus, OH 43210, USA\\
$^6$Steward Observatory, University of Arizona, 933 North Cherry Avenue, Tucson, AZ 85721\\
$^5$Institute of Astronomy, University of Cambridge, Madingley Road, Cambridge CB3 0HA, United Kingdom\\
$^7$Research School of Astronomy \& Astrophysics, Australian National University, Canberra, ACT, Australia\\
$^8$Australian Astronomical Observatory, PO Box 915, North Ryde, NSW 1670, Australia}}

\maketitle
\begin{abstract}
As part of the OzDES spectroscopic survey we are carrying out a large scale reverberation mapping study of $\sim$500
 quasars over five years in the 30 deg$^2$ area of the Dark Energy Survey (DES) supernova fields. These quasars have redshifts ranging up to 4 and have apparent AB magnitudes between $16.8<r<22.5$ mag.  The aim of the survey is to measure time lags between fluctuations in the quasar continuum and broad emission line fluxes of individual objects in order to measure black hole masses for a broad range of AGN and constrain the radius-luminosity ($R-L$) relationship. Here we investigate the expected efficiency of the OzDES reverberation mapping campaign and its possible extensions. We expect to recover lags for $\sim$35-45\% of the quasars. AGN with shorter lags and greater variability are more likely to yield a lag, and objects with lags  $\lesssim$6 months or $\sim$1 year are expected be recovered the most accurately. The baseline OzDES reverberation mapping campaign is predicted to produce an unbiased measurement of the $R-L$ relationship parameters for H$\beta$, \mgii $\lambda$2798, and \civ $\lambda$1549. However, extending the baseline survey by either increasing the spectroscopic cadence, extending the survey season, or improving the emission line flux measurement accuracy will significantly improve the $R-L$ parameter constraints for all broad emission lines. 
\end{abstract}


\section{Introduction}

There is good evidence that supermassive black holes (SMBHs) are present at the centre of all massive galaxies  \citep[e.g.][]{Kormendy1995,Richstone1998,Ferrarese2005}, and that there are tight, empirical relationships between the mass of the SMBH and properties of the host galaxy, such as stellar velocity dispersion \citep{Ferrarese2000,Gebhardt2000,McConnell2013}, light concentration \citep{Graham2001}, and bulge luminosity and stellar mass \citep{Richstone1998,Kormendy2001,Marconi2003,McConnell2013}. These relationships suggest an interplay between black hole growth and galaxy evolution; however, the true nature of this relationship is still unknown and is a major area of research in understanding galaxy evolution \citep[e.g.][]{King2003,King2005,Ferrarese2005,Murray2005,DiMatteo2005,DiMatteo2008,Park2014}. To better comprehend the origin and evolution of the SMBH-galaxy relationship and the growth of SMBHs over cosmic time, it is necessary to obtain accurate and precise measurements of black hole masses.  Direct measurements of black hole masses through stellar or gas dynamics require high  spatial resolution and are therefore limited to the local universe. 

Active galactic nuclei (AGN) provide an alternative method of black hole mass measurement. Continuum emission from the accretion disk is absorbed by gas deep within the gravitational potential of the black hole. The broad line region (BLR) gas reprocesses this radiation and emits Doppler broadened emission lines. The emission line luminosity varies in response to changes in the continuum emission in a roughly linear fashion with an associated time lag, $\tau$, which is the mean light travel time from the accretion disk to the BLR, at the responsivity weighted mean distance, $R=c\tau$. The measurement of this time lag, through detailed comparison of the emission line and continuum flux variations, is referred to as reverberation mapping \citep[RM,][]{Blandford1982,Peterson1993}.

If the BLR gas is in virial equilibrium and its motion is dominated by the gravity of the SMBH, the mass of the black hole is
\begin{equation}
M_{\rm BH} = \frac{fc\tau\Delta V^2}{G},
\label{eq:virial}
\end{equation}
where $G$ is the gravitational constant, $\tau$ is the measured reverberation time lag, $\Delta V$ is the line of sight velocity dispersion of the BLR gas estimated from the emission line width of the RMS variance spectrum, and $f$ is a dimensionless virial factor that converts the measured line-of-sight virial product into the true black hole mass. The virial factor depends on the geometry, kinematics, and orientation of the BLR, and although it is of order unity, it is expected to differ between quasars.

Reverberation mapping has yielded lags for approximately 50 AGN \citep[e.g.,][]{Peterson2004RM,Bentz2009,Denney2010,Grier2012,Bentz2013}. The lags exhibit a tight power law relationship with the continuum luminosity, $\lambda L_{\lambda}$ \citep{Kaspi2000, Bentz2009,Bentz2013}, as predicted from simple single-photon photoionization physics \citep{Davidson1972,Krolik1978}. This strong correlation is the basis for single-epoch black hole mass estimates based on a single epoch of spectroscopy \citep[e.g.][]{Laor1998,Wandel1999,McLure2002,Vestergaard2006}, and  enables black hole masses  to be estimated for a far larger sample of AGN than is possible with the full reverberation mapping method \citep[e.g.,][]{Vestergaard2004,Vestergaard2009,Shen2011,Kelly2013}.

The single-epoch mass estimation technique has been used widely in cosmology \citep[e.g.,][]{Willott2010,Mortlock2011,Vestergaard2008,Vestergaard2009,Schulze2010,Trump2011,Shen2012,Kelly2013}. As a consequence, it is important to determine the $R-L$ relationship accurately and precisely, and any dependencies,  because both random and systematic uncertainties in the $R-L$ relationship are necessarily transferred into the single-epoch masses and all subsequent studies.  The origin of the scatter around the $R-L$ relationship has been investigated by several authors \citep{Bentz2009,Bentz2013,Watson2011,KilerciEser2014}, and recent evidence suggests an additional dependence on the $R-L$ relationship with the Eddington ratio \citep{Du2015}. 

Despite `dark energy' appearing to be the dominant energy component of the universe \citep[e.g.][]{Conley2011,Blake2011cos,Blake2011distz,Padmanabhan2012,Anderson2012,Hinshaw2012,Planck2013CosmoParams}, its nature is unknown and gaining additional understanding of its properties remains a high priority.
It may be possible to reverse the $R-L$ relationship and use the time lag to infer the intrinsic luminosity of an AGN, and therefore its luminosity distance \citep{Watson2011}. The resulting distance measurements can be used to independently probe the acceleration of the universe and dark energy. The high luminosity and prevalence of AGN would make them valuable probes of the expansion history of the universe over a greater redshift range than other methods, and thus could be valuable for investigating the time evolution of dark energy \citep{Czerny2013,King2014}.

Previous reverberation mapping campaigns have generally only observed a small number of AGN in a single campaign ($\sim$10), using small telescopes,  and the majority of campaigns took place over short time-scales (under a year) \citep[e.g.,][]{Clavel1991,Robinson1994,Wanders1997,Collier1998,Peterson1998,Peterson1999_NGC5548,Kaspi2000,Peterson2002,Kaspi2007,Bentz2009LAMP,Denney2010,Barth2011,Rafter2011,Grier2012,Rafter2013,Du2014}. As a consequence, they have focused on the brightest and most variable objects, leading to a bias towards local, low-luminosity AGN.  While there have also been long, multi-year campaigns \citep[up to 8 years][]{Peterson1999,Kaspi2000,Peterson2002,Kaspi2007}, they only monitored a small number of quasars so the bias towards the brightest and most variable objects remains.  Both higher redshift and higher luminosity quasars have longer lags due to time dilation and the $R-L$ relation, and higher luminosity quasars also have lower variability amplitudes \citep{VandenBerk2004, Macleod2010}. To account for these effects we require larger telescopes and longer observation campaigns. This Chapter investigates a large scale reverberation mapping campaign being run as part of the ongoing Dark Energy Survey (DES), in conjunction with the OzDES spectroscopic survey. This 5 year campaign covers a large range in magnitude and redshift, allowing reverberation mapping studies of a much broader AGN sample over a large redshift range. 

DES is an optical survey aimed at understanding the expansion of the universe using four complementary methods: Type Ia supernovae (SNe), baryon acoustic oscillations, weak lensing, and galaxy cluster counts.  DES officially began in the 2nd half of 2013, and it plans to image 5000 square degrees with 5 filters ($g,r,i,z,Y$) over 5 years.  The supernova component of the survey will consist of repeated observations of 30 square degrees of sky in the $g$, $r$, $i$, $z$ filters, divided into two deep and eight shallow SNe fields, to detect and monitor supernova and other transients. 

OzDES is the leading spectroscopic counterpart to DES. It will repeatedly monitor the DES SNe fields using the 2dF multi-object fibre spectrograph \citep[AAOmega,][]{Saunders2004} at the Anglo-Australian Telescope (AAT). Its main science goal is to measure the redshifts of  Type Ia SNe host galaxies. In addition, a  number of fibres in each field will be dedicated to monitoring a selected group of quasars to perform reverberation mapping. OzDES, in conjunction with DES, will monitor $\sim500$ quasars for the full five years of the survey over the redshift range $0<z\lesssim4$. This is approximately a ten-fold increase in number and redshift range over the existing  reverberation mapping samples, and comparable in number to the ongoing Sloan Digital Sky Survey (SDSS) reverberation mapping project \citep{Shen2014}.

While the OzDES reverberation mapping programs will monitor hundreds of quasars, the likelihood that this program will successfully recover reverberation lags depends on the frequency and accuracy of the light curve measurements, the length of the survey, and the intrinsic variability of the monitored AGN sample. OzDES is expected to target each field approximately 25 times over the five year period. This number of observation epochs is significantly smaller than traditional reverberation mapping campaigns \citep[e.g.][]{Peterson2002,Bentz2009,Denney2009,Barth2011}, which have found that emission-line lag recovery generally requires 30-50 well-spaced epochs of observations, and favourable continuum flux variations.
 Our current study aims to investigate the expected efficiency of the OzDES reverberation mapping campaign by generating realistic AGN light curves and attempting to recover the input lags. We will then use our findings to determine how to optimally select our target AGN sample, and make predictions about the scientific results for our sample. We also investigate ways to improve the program design and execution, through increased cadence, changes to the survey length, and improved measurement accuracy, as a means to maximise the scientific output.

The outline of the Chapter is as follows, a technical summary of the DES and OzDES surveys is given in Section \ref{sec:ozdes}, followed by a description of the survey simulation in Section \ref{sec:methods}. The predictions of the efficiency of the survey are presented in Section \ref{sec:results}, along with the expected improvements for several possible survey extensions. We examine the predicted scientific results in Section \ref{sec:science}. Finally, the results are summarised in Section \ref{sec:discussion}.
\section{DES/OzDES Survey} \label{sec:ozdes}
\subsection{Fields}
The DES SNe fields were chosen to have extensive past observation histories and to overlap with the Visible and Infrared Survey Telescope for Astronomy (VISTA). The ten chosen fields were  the Elias fields (E1,  E2), SDSS Stripe 82 (S1, S2), the Chandra Deep Fields (C1, C2, C3) and the XMM Large Structure Survey fields (X1, X2, X3).  The coordinates of the fields are given in Table \ref{tab:fields}.

\begin{table}
\caption{DES SNe fields}
\label{tab:fields}
\begin{center}
\begin{tabular}{llll}
\hline
Target name &	RA (h m s) &	Decl. ($^{\circ}$ '	'') &	Type\\
\hline
E1 &	00:31:29.9 &	$-$43:00:34.6 &	shallow\\
E2 &	00:38:00.0 &	$-$43:59:52.8 &	shallow\\
S1 &	02:51:16.8  &	 +00:00:00.0 &	shallow\\
S2 &	02:44:46.7 &	$-$00:59:18.2 &	shallow\\
C1 &	03:37:05.8  &	 $-$27:06:41.8 &	shallow\\
C2 &	03:37:05.8  &	 $-$29:05:18.2 &	shallow\\
C3 &	03:30:35.6 &	 $-$28:06:00.0 &	deep\\
X1 &	02:17:54.2  &	 $-$04:55:46.2 &	shallow\\
X2 &	02:22:39.5  &	 $-$06:24:43.6 &	shallow\\
X3 &	02:25:48.0 &	 $-$04:36:00.0 &	deep\\
\hline
\end{tabular}
\\ Note: RA and Decl. are given for J2000.
\end{center}
\end{table}%

\subsection{Target selection}\label{sec:targetselection}
\begin{figure*}
\centering 
\subfloat{\label{fig:2airmass}\includegraphics[width=0.45\textwidth]{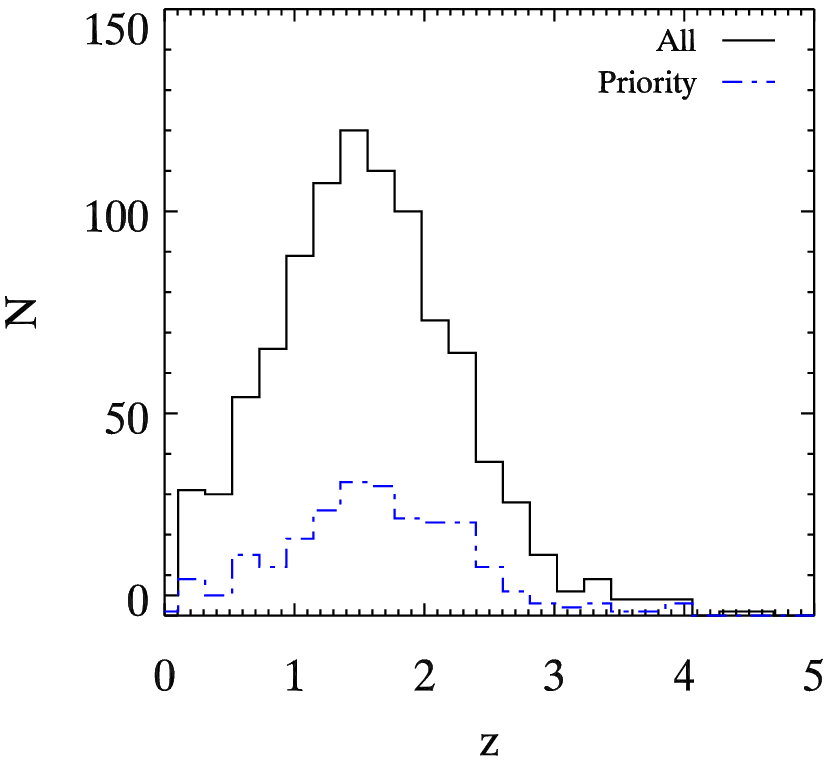}}
\subfloat{\label{fig:3airmass}\includegraphics[width=0.45\textwidth]{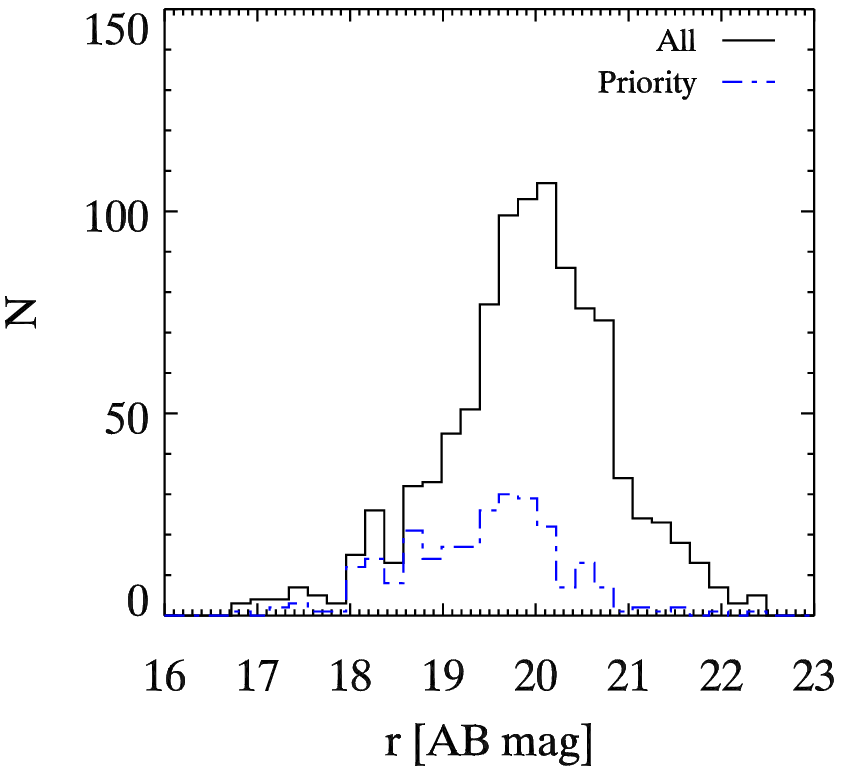}}
\caption{The redshift (left) and magnitude (right) distributions of the current OzDES quasar sample. The priority sample consists of the highest quality candidates based on visual inspection of the spectra. }
\label{fig:ozdesdist}
\end{figure*}

The quasar candidates were initially chosen from: 1) known quasars in the DES SNe fields with $m_{\rm r,psf}<21.2$ mag; 2) point sources with $m_{\rm r,psf}<21$ mag selected through the KX method\footnote{This method selects quasars based on excess flux in the K band relative to stars, which is due to the power law nature of the quasar spectral energy distribution.} \citep{Warren2000} using data from DES and the VISTA Hemisphere Survey \citep[VHS,][]{McMahon2013,Sutherland2014}, and; 3) point sources with $m_{\rm r,psf}<21$ mag selected through photo-z template fitting using DES, VHS and Wide-field Infrared Survey Explorer \citep[WISE,][]{Wright2010} photometry. The details of this selection process are described in \citet{Banerji2015}. We obtained spectra of 3331 quasar candidates in 2013, and after visual inspection, the sample was reduced to the current 989 objects. Fig. \ref{fig:ozdesdist} shows the redshift and $r$-band magnitude distribution of this sample. We have ranked the sources based on the quality of the spectra and the number of emission lines present in the spectra. The highest priority objects are shown in blue. When we decrease this sample to 50 per field, we will incorporate the results of the simulations presented here into our target selection criteria and also select for the most variable quasars (for which lag recovery is most probable) based on the first two years of data.

\subsection{Photometry}
\begin{figure}
\centering 
\includegraphics[width=0.45\textwidth]{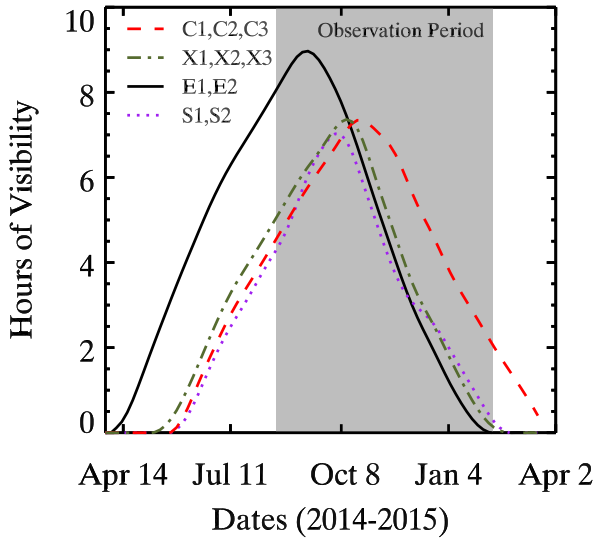}
\caption{The number of hours the DES supernova fields are visible throughout the year with an airmass of $<2$. The values are calculated for the period April 2014 - April 2015 using \textit{JSkyCalc}\protect\footnotemark. The shaded observation period roughly represents the time when photometric and spectroscopic data will be taken with the current DES and OzDES program design. Section \ref{sec:ext} investigates the improvements afforded by extending this observation period to fully encompass the time when the fields are visible.}
\label{fig:visibility}
\end{figure}

DES uses the DECam instrument \citep{Flaugher2010} on the Blanco 4-meter telescope at the Cerro Tololo Inter-American Observatory in Chile. Images of the ten SNe fields are planned to be taken approximately every 5$-$7 days 
between September 1 and February 15 every year between 2013 and 2017.  Photometry from November 2012  -- February 2013  were also taken as part of the DES Science Verification period.  For a given field, images are taken in all four filters  in the same night, when possible\footnote{All four filters are observed in one night more than 80\% of the time}.
Otherwise, images in the remaining filters are taken during the next available night. The approximate observation period of DES is shown in Fig. \ref{fig:visibility}, along with the visibility of the 10 DES SNe fields. During every night of DES imaging, the SNe fields that have not been observed in the last five nights are given highest priority, with special preference for the deep fields, C3 and X3. To date, the median gap between consecutive observations is 6.5 days and the maximum gap ranges between 12 to 21 days. For these simulations we assume that photometric observations are taken every 7 days. The nominal exposure times and corresponding limiting magnitudes for both the deep and shallow fields are given in Table \ref{tab:exptimes}.

\begin{table}
\caption{DES exposure times and limiting magnitudes of the SN fields.}
\label{tab:exptimes}
\begin{center}
\begin{tabular}{@{}ccccc@{}}
\hline
\multicolumn{5}{c}{~~~~~~~~~ Shallow Field~~~~~~~~~~~~~~~  Deep Field}\\
 Filter &	Exposure &	Limiting  &	Exposure &	Limiting\\
  &	Time (s) &	 Mag (AB) &	Time (s) &	Mag (AB)\\
\hline
g &	175 &	24.9 &	600 &	25.6\\
r &	150 &	24.3 &	1200 &	25.4\\
i &	200  &	 23.9 &	1800 &	25.1\\
z &	400 &	23.8 &	3630 &	24.8\\
\hline
\end{tabular}
\end{center}
\end{table}%

\subsection{Spectroscopy}

   The OzDES spectroscopic observations are being taken with the AAOmega spectrograph fed by  the Two Degree Field (2dF) multi-object system on the AAT. The 2.1 degree diameter field of view of 2dF is almost identical to that of DECam, making it the ideal instrument for the spectroscopic followup of DES targets \citep{Yuan2015}, and OzDES will run over a similar period of time to DES. The 2dF multi-object system is a robotic fibre positioner that allows simultaneous observations of up to 392 targets anywhere within  it field of view. The projected fibre diameter of the instrument is approximately two arcsec. The 2dF fibres feed AAOmega, a 
double beam spectrograph with a wavelength coverage of 3750\AA-8900\AA\  and a resolution of $R\sim1500$.
   
    During each run, OzDES integrations will be 2 hours long in each DES field.  Over the five years of the survey, a total of 100 nights have been allocated for OzDES with a graduated allocation plan.  A larger number of nights have been allocated each year as the survey progresses: 12 nights in 2013, 16 in 2014, 20 in 2015, 24 in 2016, and 28 in 2017.   It takes approximately four nights to observe all 10 DES SN fields (including a 33\% allowance for bad weather). We expect that each field will be visited approximately 25 times over the five year period, and the cadence of the measurements will be approximately monthly within each year. The 2014 observations are now complete and each field has five or more epochs of measurements. To date, we have been able to devote fibres to 100 AGN per field to help optimise our final sample selection. This fibre allocation is scheduled to be reduced to a final 50 per field in 2015.

\begin{figure}
\centering 
\includegraphics[width=0.48\textwidth]{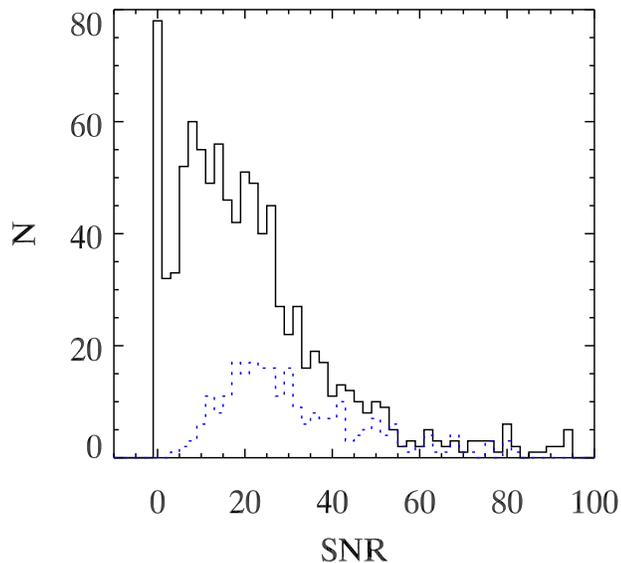}
\caption{The distribution of emission line signal-to-noise ratios (SNR) in the current OzDES quasar sample for the full and priority sample (defined in section \ref{sec:targetselection}). The emission line SNR values were estimated for H$\beta$, \mgii,  and \civ, in each object for all observation epochs, using the reduced un-flux calibrated OzDES spectra. However, the SNR assigned to each object corresponds to the median SNR value of the emission line with the highest median SNR value. }
\label{fig:snrdist}
\end{figure}

To give insight into the quality of the current OzDES spectra, the distribution of preliminary line flux signal-to-noise ratio (SNR) measurements is shown in Figure \ref{fig:snrdist}. The median SNR value for the total sample is SNR$_{\rm all}\sim11$ and for the priority sample is SNR$_{\rm priority}\sim21$. More details on the calculation of these preliminary SNR values and examples of reduced un-flux calibrated OzDES spectra across the redshift and magnitude range of the survey are given in Chapter  \ref{chap:targetselection}.
OzDES is also monitoring 10-15 F stars per field. These observations are important for accurate absolute spectrophotometric calibration of the spectra, which is expected to be good to $\lesssim$10\% based on the results from the GAMA project \citep{Hopkins2013}.

\footnotetext{\url{http://www.dartmouth.edu/~physics/labs/skycalc/flyer.html}}

\section{Simulation Setup}\label{sec:methods}
 Our ability to accurately recover lags is highly dependent on the presence of prominent features in the continuum and emission-line light curves. Accurate detection and characterisation of light curve features depends on (1) the frequency of measurements, (2) the length of the survey, (3) the accuracy of the measurements, and (4) the intrinsic variability of the object.  
In order to predict how well OzDES will be able to recover lags, we analysed mock catalogues of quasars with realistic continuum and emission line light curves. Below we describe our methods for simulating and subsequently recovering lags from such light curves.

\subsection{Mock Catalogue}

We began by constructing mock catalogs of 520 AGN distributed uniformly in 40 redshift bins and 13 magnitude bins over the range $0 < z < 4.0$ and  $18.0 < r < 20.5$.  This range roughly corresponds to the capabilities of the 2dF spectrograph within the framework of the OzDES program design.  The small number of AGN brighter than 18th magnitude will be automatically targeted and are not considered for this investigation. 

\subsection{Monochromatic continuum luminosity estimation}\label{sec:lag}
\begin{figure}
\begin{center}
\includegraphics[width=0.45\textwidth]{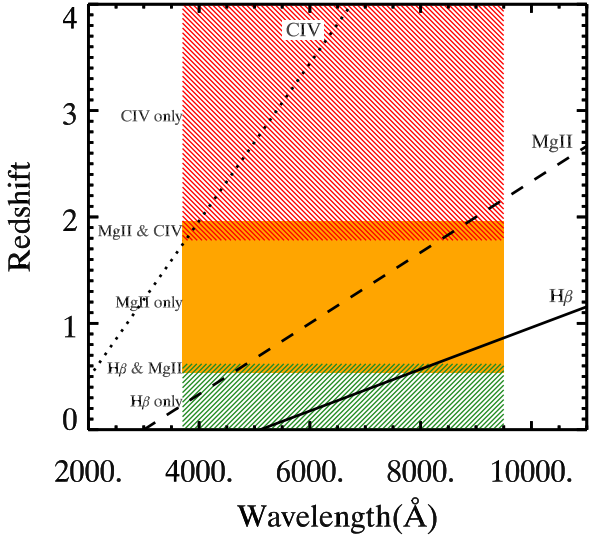}
\caption{The five redshift regimes in which we are sensitive to different broad lines (or pairs of lines) given the properties of the 2dF instrument. The black curves show the observed frame wavelength of each broad emission line considered in our simulations as a function of redshift.}
\label{fig:waveredshift}
\end{center}
\end{figure}

  The observed $R-L$ relationship for each broad emission line is constructed empirically using the monochromatic luminosity of a nearby continuum region as a proxy for the ionising luminosity. 
The most common emission-line--continuum region pairs used for constructing $R-L$ relationships, are H$\beta$ with L$_{5100{\text\AA}}$ \citep[e.g.][]{Peterson2004,Vestergaard2006,Bentz2009,Bentz2013}, \mgii\ with L$_{3000{\text\AA}}$ \citep{Vestergaard2009,Trakhtenbrot2012}, and \civ\ with L$_{1350{\text\AA}}$ \citep{Vestergaard2002,Peterson2004RM,Vestergaard2006, Kaspi2007,Vestergaard2009,Park2013,Trevese2014}, and we used these pairs in our analysis. As we are restricted to the observed optical spectrum, the redshift range is broken into five different sections according to the emission lines we are able to observe: $z<0.54$ H$\beta$ only, $0.54<z<0.62$ H$\beta$ and \mgii, $0.62<z<1.78$ \mgii\ only, $1.78<z<1.96$ \mgii\ and \civ, and $z>1.9$ \civ\ only (see Fig. \ref{fig:waveredshift} for illustration). Accurate measurement of the surrounding continuum is required to measure the line flux correctly. Therefore, these redshift ranges allow a generous amount of continuum on either side of the broad line, and avoid bluer wavelengths where the spectrograph's throughput is low.  However, our choice of redshift ranges are somewhat conservative with respect to how far into the blue and/or red each emission line or continuum region can be well-measured.  

The monochromatic fluxes at 5100\AA, 3000\AA, and 1350\AA\ for each AGN are estimated using  the known DECam $r$-band magnitudes and redshifts, and a K-correction based on the filter response curves of the DECam $r$-band filter and the Sloan Digital Sky Survey (SDSS) quasar template \citep{VandenBerk2001}. The SDSS quasar template spans the rest-frame wavelength range 800\AA-8555\AA\ and is based on a sample of quasars that covers a similar redshift and magnitude range to that being observed by OzDES, so it should be a representative template for this work. The quasar template is scaled to the input magnitude of each target under the assumption that the bolometric luminosity scales as $L_{bol} = 9.0\lambda L_{\lambda}$(5100\AA) \citep{Kaspi2000}. This bolometric correction is the approximate midpoint between the results of \citet{Richards2006} and \citet{Krawczyk2013} who found values of $10.3\pm2.1$ and $7.79 \pm 1.69$, respectively. 
 The luminosity is then estimated assuming a $\Lambda$CDM cosmology with $H_0$ = 70 km s$^{-1}$ Mpc$^{-1}$, $\Omega_M = 0.3$, and $\Omega_{\Lambda} = 0.7$. 
 
 This process does not take into account host galaxy contamination, Galactic extinction, or intrinsic variations in the AGN spectral energy distribution (SED). Nonetheless, it provides a sufficient luminosity estimate for our present purposes.  When host galaxy contamination is included in the simulations, no significant change in the results was observed because the loss in sensitivity due to contamination from the host galaxy is generally balanced by the increased variability expected for fainter objects (see section \ref{sec:lightcurve}). For more discussion on this point see Section \ref{sec:discussion}.

\subsection{Lag estimation}

We calculate the time lags associated with each source from the monochromatic continuum luminosity and previously published $R-L$ relationships for H$\beta$, \mgii\ and \civ.  We adopt the H$\beta$ $R-L$ relationship  
\begin{align}
\log_{10}{(R_{H\beta} ~{\rm lt~days}}) =& -21.2\pm{2.2}+0.517\pm0.033\\
&\times~\log_{10}{\left(\lambda L_{\lambda}(5100\text\AA)~{\rm erg~s}^{-1}\right)},\nonumber
\end{align}
presented by \citet{Bentz2009}, which is derived from RM measurements of 35 AGN spanning four orders of magnitude in luminosity. 
Switching to the \citet{Bentz2013} relation would not appreciably change our results.
The \civ\ $R-L$ relationship is not as well determined as the H$\beta$ $R-L$ relationship because \civ\ lags have only been measured for a few objects \citep{Koratkar1991,Peterson2004,Kaspi2007}. Most of these objects have similar luminosities, and there is only a single object at each of the low and high luminosity ends. The \civ\ $R-L$ relationship we use,
\begin{align}
\log_{10} (R_{CIV} ~{\rm lt~days}) =& -23.3\pm2.6+~(0.55\pm0.04)\\
&\times~\log_{10}{\left(\lambda L_{\lambda}(1350{\text\AA})~{\rm erg~s}^{-1}\right)},\nonumber
\end{align}
is taken from \citet{Kaspi2007} and is based on only seven objects. \mgii\ is not yet well-studied with reverberation mapping, and there are only a few studies in which the time lag has been measured \citep[][though the latter two papers only present marginal detections]{Metzroth2006,Reichert1994,Dietrich1995}. However, because the \mgii\ lag measurement was found to be consistent with the H$\beta$ lag in two of the studies \citep[][using \citet{Metzroth2006}]{Reichert1994,Stirpe1994,Bentz2006NGC4151}, there is a strong correlation between the width of \mgii\ and H$\beta$ emission lines\footnote{However, this correlation is not a tight one-to-one relationship between the \mgii\ and H$\beta$ width \citep{Trakhtenbrot2012}, as is commonly reported in the literature.} and H$\beta$ and \mgii\ have similar ionisation parameters, it is generally assumed that the two lines originate at the same radius from the ionising source. \citet{Trakhtenbrot2012} used this assumption to estimate a \mgii\ $R-L$ relationship of
\begin{align}
\log_{10} (R_{MgII} ~{\rm lt~days}) =& -25.72\pm0.62+(0.615\pm0.014)\\&\times~\log_{10}{\left(\lambda L_{\lambda}(3000{\text\AA})~{\rm erg~s}^{-1}\right)}.\nonumber
\end{align}
This estimate was created using an empirical correlation between $\lambda L_{\lambda}(5100\text\AA)$ and $\lambda L_{\lambda}(3000\text\AA)$, and an existing H$\beta$ $R-L$ relationship. 
 As a consequence, it is less certain than an $R-L$ relationship derived from direct reverberation mapping measurements, but it is our only option in the absence of such measurements for \mgii.

\begin{figure*}
\centering 
\includegraphics[width=0.8\textwidth]{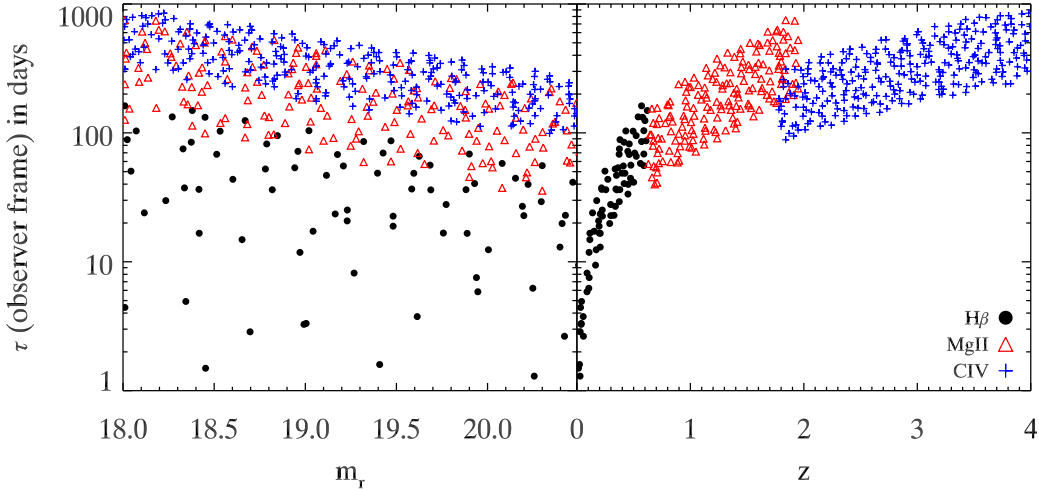}
\caption{The distribution of lags in a mock AGN sample, as a function of $r$-band magnitude (left) and redshift, $z$ (right).  The different colours and symbols correspond to the different emission lines (black circles: H$\beta$, red triangles: \mgii, blue crosses: \civ). }
\label{fig:lagdist}
\end{figure*}

The resulting distributions of the observed time lags (including time dilation) with redshift and magnitude  are shown in Fig. \ref{fig:lagdist}. The \civ\ lags tend to be a factor of $\sim2$ smaller than the \mgii\ lags for the same objects. This is consistent with the findings of \citet{Kaspi2007}. 

\subsection{Light Curve estimation}\label{sec:lightcurve}
 We model the AGN continuum light curves as a damped random walk (DRW) characterised by a time scale, $\tau_D$\footnote{Note that \citet{Macleod2010} use $\tau$ without the subscript for this parameter, but we use $\tau_{\rm D}$ following \citet{Zu2011} so as not to confuse this parameter with the reverberation lag, which we denote as $\tau$.}, and the asymptotic amplitude of the structure function over long time scales, $S_F(\infty)$. \citet{Zu2013} show this model is a good representation of quasar variability on the time scales of the OzDES survey.
Our ability to recover the time lag for each AGN is highly dependent on these variability parameters. If $S_F(\infty)$ is large and $\tau_D$ is short, it is more likely that significant variations in the light curve will be seen during the observation period. 

 \citet{Macleod2010} found that the values of $\tau_D$ and $S_F(\infty)$ scales with the luminosity of the AGN, the  observed wavelength, and the mass of the black hole, following the power law,
\begin{align}
\log_{10}(\alpha) =& A_{\alpha} + B_{\alpha}\log_{10}(\lambda/4000.)+C_{\alpha}(M_i+23)\nonumber\\ &+D_{\alpha}\log_{10}(M_{\rm BH}),
\label{eq:macleod1}
\end{align}
where $\alpha$ refers to either $\tau_D$ (in days) or $S_F(\infty)$ (in mag), $\lambda$ (in \AA)  is the continuum wavelength of interest (1350\AA, 3000\AA\ or 5100\AA), the coefficients are
\[
\begin{tabular}{l@{~}l@{~}l@{~}l}
$A_{\tau_D}= 2.4$,& 
$B_{\tau_D} = 0.17$,&
$C_{\tau_D} = 0.03$,& 
$D_{\tau_D} = 0.21$,\\
$A_{S_F} = -0.51$,& 
$B_{S_F} = -0.48$,&
$C_{S_F} = 0.13$,&
$D_{S_F} = 0.18,$
\end{tabular}
\]
$M_i$ is the absolute magnitude of the quasar and is calculated using the known $r$-band magnitude and the K-corrections prescribed in \citet{Richards2006} for SDSS, and $M_{\rm BH}$ is the mass of the black hole.  The black hole masses were randomly assigned using the probability distribution described by \citet{Macleod2010},
\begin{equation}
P(\log_{10}M_{\rm BH} |M_i) = \frac{1}{\sqrt{2\pi}\sigma}\exp\left[-\frac{(\log_{10} M_{\rm BH} - \log_{10} \overline{M}_{BH})^2}{2\sigma_{M_{\rm BH}}^2}\right],
\end{equation}
where $ \log_{10}\overline{M}_{BH}=2.0-0.27M_i$ and the spread in the distribution, $\sigma_{M_{\rm BH}} = 0.58+0.011M_i$. The mass of the black hole is in Solar units. The DES  and SDSS \textit{gri} filters are quite similar, so the use of the \citet{Richards2006} K-corrections are justified \citep{Honscheid2008}. Additionally, \citet{Macleod2010} assumed the same cosmological model as we use here, so the absolute magnitude scales are identical.

\subsubsection{Continuum light curve}
The continuum light curve, in magnitude, is defined as a combination of a mean $\mu$ and variable term $\Delta C(t_i)$, such that $C(t) =\mu + \Delta C(t)$. The value of $\mu$ is defined as the monochromatic continuum flux at given wavelength (5100\AA, 3000\AA, and 1350\AA; Section \ref{sec:lag}) converted to a magnitude. For a DRW, the variable component is constructed by initialising the light curve at $t_0$ as $\Delta C(t_0) = \sigma G(1)$ where $G(1)$ is a Gaussian random deviate of unit dispersion, and $S_F(\infty) = \sqrt{2}\sigma$. Subsequent points are created using the recursion 
formula \citep[see][]{Kelly2009,Macleod2010,Kozlowski2010},
\begin{align}
            \Delta C(t_{i+1}) =& \Delta C (t_i) \exp\left(-|t_{i+1}-t_i|/\tau_D\right) \\
                 +& \sigma \left[ 1 - \exp\left(-2|t_{i+1}-t_i|/\tau_D\right)\right]^{1/2} G(1).\nonumber
\end{align}

\subsubsection{Emission line light curve}
The emission line light curve is the response to the continuum light curve. If we only consider the temporal response of the overall emission line, the resulting light curve is given by
\begin{equation}
\Delta L(t) = \int \Psi(\tau)\Delta C(t-\tau) d\tau,
\label{eq:convolve}
\end{equation}
where $\Delta L(t)$ is the emission line light curve flux relative to its mean value, $\Delta C(t)$ is the variable component of the continuum light curve defined above, and $\Psi(\tau)$ is the transfer function, which describes the emission line response to a delta function outburst in the continuum \citep{Blandford1982}.  The transfer function is related to the overall structure and kinematics of the broad line region and its true form remains an active area of research \citep{Pancoast2011,Zu2011,Pancoast2012,Grier2013}. For simplicity we have chosen a top hat transfer function, mirroring JAVELIN \citep{Zu2011}, given by
\begin{equation}
\Psi(t)= \left\{ 
  \begin{array}{l l}
    \frac{A}{2w} & \quad \text{for $\tau-w<t<\tau+w$}\\
    0 & \quad \text{otherwise}
  \end{array} \right.
  \end{equation}
where $A$ is a scaling term and $w$ is the half-width of the top hat function. We  set the scaling term, $A=1$, and the half-width to $w=0.1\tau$. This half-width value was motivated by previous reverberation mapping campaigns \citep{Grier2013}. The choice of amplitude makes a strong assumption about the line's responsivity to continuum variations. Photoionisation research suggest that different lines will have different associated responsivities \citep{Goad1993,Korista2000, Korista2004}, and there is observational evidence that the responsivity for individual lines is not consistent between monitoring programs \citep[e.g. \mgii\ responsivity;][]{Woo2008,Cackett2015}. Additionally, the value of $\Delta C(t)$ implemented in this analysis is only a proxy of the true ionising flux and as a consequence, the transfer function quoted will be different to the inherent transfer function of the system.
 We investigate the effects of our choice of transfer function in  Section \ref{sec:discussion}.

 \subsubsection{Measurement uncertainty}\label{sec:uncertain}
 We created mock light curves with daily cadence, starting sufficiently before the start of the observing campaign to allow time for the emission line response (described by the convolution in Equation \ref{eq:convolve}) to lie in the observed time period. We then down-sample these light curves following the OzDES program design and any extensions described below.
 
 
A Gaussian error is added to each mock light curve measurement based on the expected flux uncertainties. The expected flux measurement uncertainty for OzDES RM project is 0.01 mag for photometry and  0.1 mag for spectroscopy. We do not include magnitude dependent (photon counting) errors on the line flux measurements as our uncertainties are generally expected to be dominated by our overall absolute flux calibration.  The absolute flux calibration of the spectroscopy is performed using the following procedure:
 1) the observed F stars are matched with a F star with equivalent g-r colours (measured from DES photometry) from an existing stellar catalogue; 2) the catalogue spectrum is then warped to exactly replicate the colours of the observed F star; 3) next, the observed spectrum is divided by the warped catalogue spectrum; 4) the resulting function is smoothed and represents the sensitivity curve of the observed spectrum. This process is performed separately for the red and blue arm; 5) and repeated for all the observed F-stars; 6) the median sensitivity is then computed for each season. The sensitivity is dependent on the observed wavelength and radial position of the fibre, and the median scatter in the sensitivity curve is $\sim$5\%; 7) At that time, all AGN spectrum are corrected according to the measured sensitivity curve and synthetic $g$, $r$, and $i$ magnitudes are calculated; 8) Finally, the spectrum is compared to the nearest photometric $g$, $r$, and $i$ magnitudes (almost all within four days)  and scaled under the assumption that there will be insignificant changes in flux on this time scale. The photometric uncertainty in this case is on order of a few percent. Current OzDES data shows that residuals of these fits are roughly Gaussian in form, therefore it is reasonable to assume Gaussian uncertainties in our light curve measurements. For realism, we further randomly shifted 10\% of the data points without adjusting their error bars to introduce an element of the non-Gaussianities present in real data. The shift was chosen from a Gaussian distribution with a mean value of zero and a standard deviation twice the size of the measurement uncertainty. An example of a simulated light curve is shown in Fig. \ref{fig:examples}.

The accuracy of the emission line flux measurement also sensitive to the method of determining the emission line flux.  The method for measuring the emission line flux in the OzDES data will differ depending on the quality of the spectrum. If the signal-to-noise ratio in the individual spectrum is sufficient, the spectrum will be modelled to separate the emission lines of interest from the continuum, absorption line and contaminating emission line signal.  Otherwise, the emission line flux is simply calculated as the integrated flux above an approximate linear continuum fitted between two pseudo-continuum regions near the line. \mgii\ $\lambda$2798 flux measurements are especially affected by Fe {\sc ii} contamination \citep{Vestergaard2001} and therefore care must be taken when calculating the flux for this line.
 Additionally, AGN with strong absorption (e.g. broad absorption lines; BALs) will be discarded from the final sample. 
 \subsection{Possible Survey Extensions}
We also consider various extensions to the baseline survey to determine if modifications to the program design can increase the scientific return. These extensions include (i) decreasing the effective seasonal gap; (ii) increasing the cadence; (iii) improving the data quality; and (iv) extending the total survey duration.  While ultimately uncontrollable, we also consider the effect of a reduction in the number of observed epochs due to extreme weather losses. 

\begin{description}
\item{\textbf{Seasonal Gap [Full season/Year]:}}
We consider two extended spectroscopic observation windows to investigate the effect of the seasonal gap on lag recovery. The longest possible observation window in which any one field can continuously be observed is approximately between May 1 and Feb 14 for fields E1 and E2 (see Fig. \ref{fig:visibility}). This will be our first possible extension and will be referred to as `full season'. As the first two years of observations are completed or underway, we can only apply this extension to the following three years.
We also consider a case in which observations are taken over the full year, with no seasonal gaps. This is our second possible extension and will be referred as `year'. Due to the restricted observability of the DES SN/OzDES fields resulting from their position in the sky, this observational setup is not possible for OzDES, but illustrates the potential for future surveys targeting polar fields. In both cases, it is assumed that photometric measurements are taken weekly and spectroscopic measurements are taken monthly. 
\item{\textbf{Cadence [Weekly]:}}
We also consider the case where  spectroscopic measurements are taken weekly over the baseline observation window of the DES photometry as defined in Section \ref{sec:ozdes}. This will test how the spectroscopic measurement frequency affects the recoverability of the lag. Previous reverberation mapping campaigns have found that high quality and high cadence in the continuum light curve is required for accurate lag recovery as it is the driver of the line light curve. However, it is also important that the  line flux measurements are taken with sufficient cadence to map the line response accurately. This possible extension will be referred to as `weekly'. Again, we can only apply this extension to the last three years of the survey. We should note that this extension increases the number of spectroscopic measurements significantly, which also plays a major role in the recovery of lags \citep{Horne2004, Shen2014}. 
\item{\textbf{Data quality [Goal]:}}
We investigate the effects of data quality by changing the spectroscopic measurement uncertainty. We use the baseline sampling rate of OzDES and reduce the spectroscopic measurement uncertainties from 0.1 mag to 0.03 mag, which corresponds to the optimistic goal for the calibration of the OzDES spectral data. The photometric measurement uncertainty is kept constant at 0.01 mag. This extension will be referred to as `goal'. We also consider a spectroscopic measurement uncertainty of 0.03 mag combined with the `year' extension, referred to as `year+goal'.
\item{\textbf{Survey Length [Long]:}}
Finally we investigate an extension of the survey by 2 years with sampling and cadence equivalent to the last year of the planned DES/OzDES survey  (i.e. 7 spectroscopic measurements per additional year). This possible extension will be referred to as `long' and it should allow the recovery of longer lags and more accurate recovery of lags for a broader AGN population. 
\item{\textbf{Weather:}}
The expected 25 epochs, used in the simulations, already takes into account the expected weather loss for the AAT. However, to consider the effects of extreme weather we simulated losing an additional 3-5 spectroscopic epochs over the 5 year period.
\end{description}

\subsection{Recovering time lag}\label{sec:recover}
Traditionally, reverberation lags have been recovered using simple linear interpolation and cross correlation techniques \citep{Gaskell1986,White1994,Edelson1988}, but recently other approaches have been implemented that take into account our existing knowledge of AGN behaviour as a means to optimize, and in some cases, to improve the likelihood of accurate lag recovery. We adopt one such approach, using the program JAVELIN, an updated version of SPEAR \citep[see][for details]{Zu2011}. Instead of linearly interpolating between data points, JAVELIN uses a damped random walk to model the AGN continuum light curve, and attempts to fit the emission line light curve by convolving the continuum light curve with a top hat transfer function. 

JAVELIN  uses the amoeba minimisation method \citep{Press1992} to recover a model of the continuum light curve (including DRW parameters) and transfer function that best fits  the continuum and the line data. JAVELIN has been  generally found to be consistent with traditional cross-correlation (CCF) methods  \citep{Zu2011,Grier2012,Peterson2014} and has the advantage of fitting multiple emission line light curves at once. 
The biggest limitation of JAVELIN is that its error estimates assume well-characterised Gaussian noise, so caution must be used when interpreting the parameter uncertainties if these assumptions are violated. 

\subsubsection{Implementation}
We allow JAVELIN to explore the lag range between zero days and three times the input lag, with a maximum allowed lag of 1931 days, corresponding to the separation between the first photometric and last spectroscopic measurement planned for OzDES. Lags longer than 1931 days cannot be constrained by the data; however, we expect very few, if any lags this large in the considered magnitude and redshift range (see Figure \ref{fig:lagdist}).
 The window of $0<t<3{\tau}$ was found to be sufficient to fully enclose the recovered likelihood distribution of the lags, and there is no evidence for an artificial cutoff in the likelihood distribution (see Fig. \ref{fig:examples}). 

We analysed the five different redshift ranges separately. For the redshift ranges where two emission lines are present, the two lags were fit both individually and simultaneously. For the single line case we used 150 MCMC chains with 150 iterations per chain for both the burn in period\footnote{The burn in period refers to an initial portion of a Markov chain sample that is discarded to minimise the effect of initial values on the posterior inference.} and to sample the posterior probability distribution. This is equivalent to 22500 total burn in iterations. For the case where both emission lines light curves are fitted simultaneously, we used 200 MCMC chains with 200 iterations per chain. The larger number of iterations accounts for the additional complexity in the double line fit.  For the baseline OzDES setup we analysed ten realisations of the mock catalogue of 520 AGN. For the investigation of different survey extensions, we only analysed two realisations of the mock catalogue per extension.

\subsubsection{Output}
For each continuum and emission-line light curve, JAVELIN produces an estimate of the posterior distribution for all the fitted parameters, including the lag. 
 In a substantial number of cases, a clear peak is present and easily identified in the lag posterior distribution, corresponding to the best fit lag value. In the remaining cases, multiple peaks of comparable size were present in the distribution or in some rare cases no distinct peak was detected at all. 

\begin{figure*}
\centering 
\subfloat{\label{fig:y1}\includegraphics[width=0.3\textwidth]{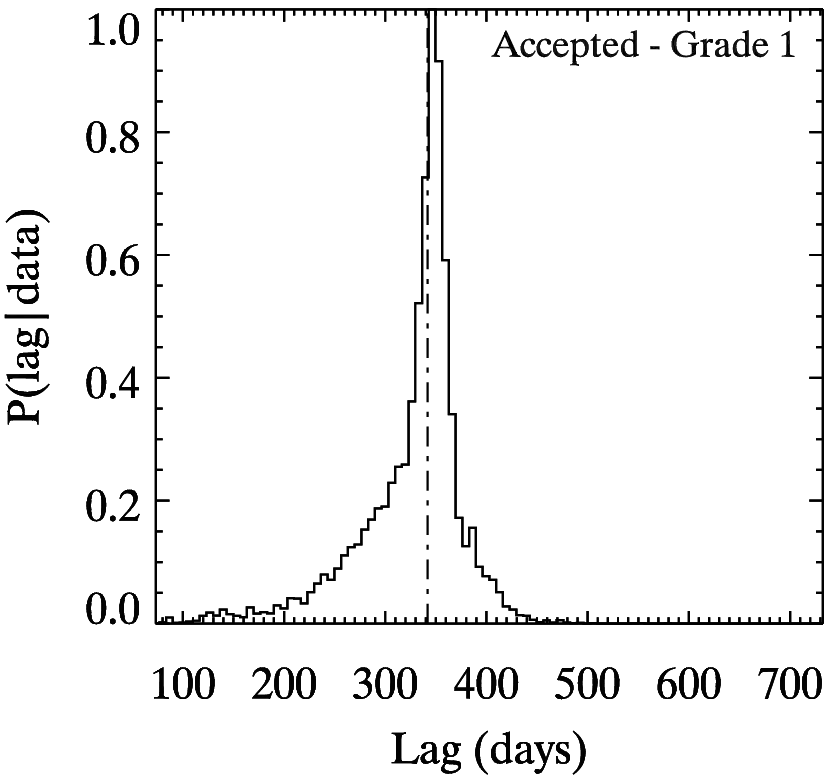}}
\subfloat{\label{fig:y1lc}\includegraphics[width=0.54545456\textwidth]{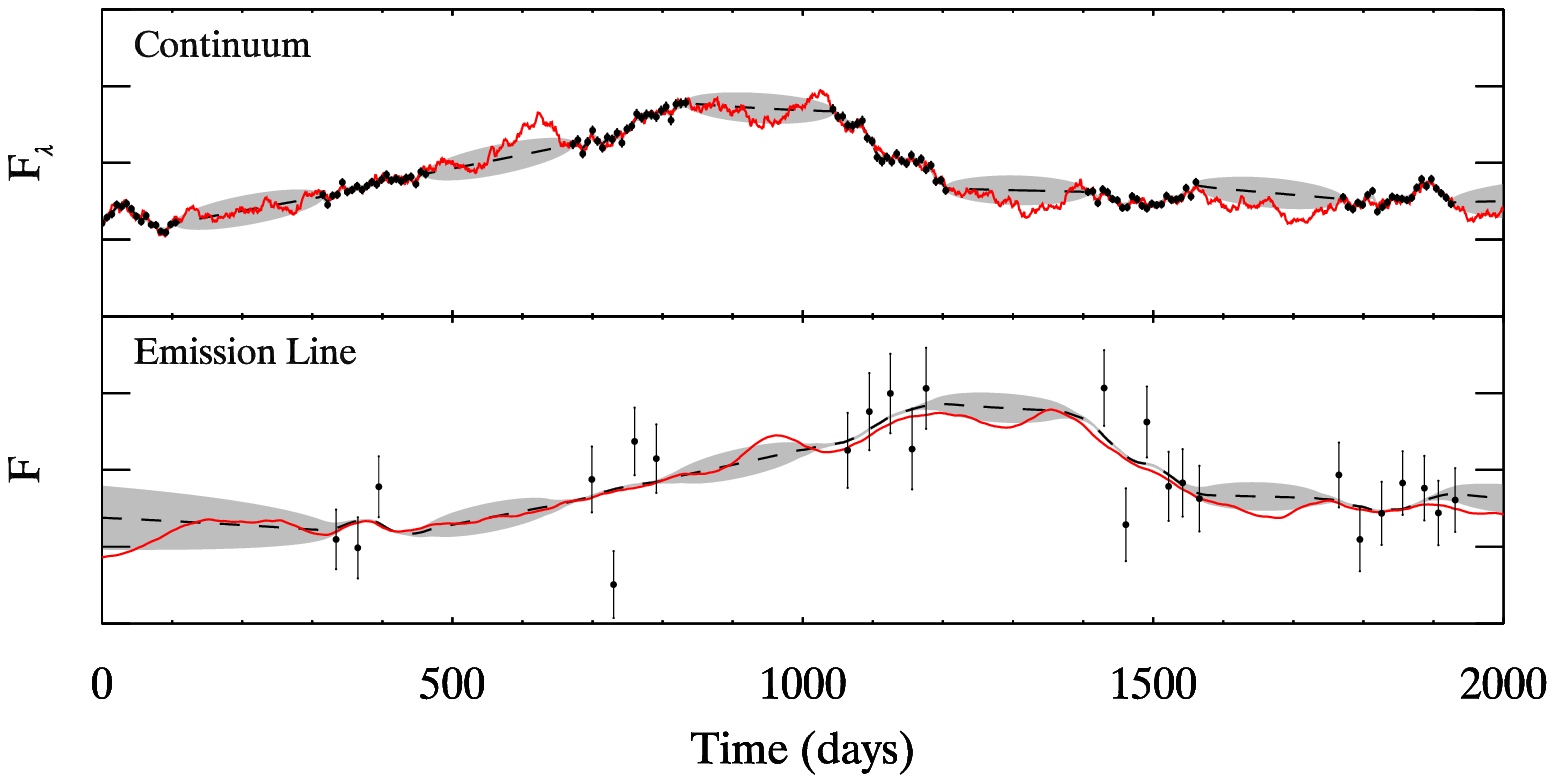}}\\
\subfloat{\label{fig:y2}\includegraphics[width=0.3\textwidth]{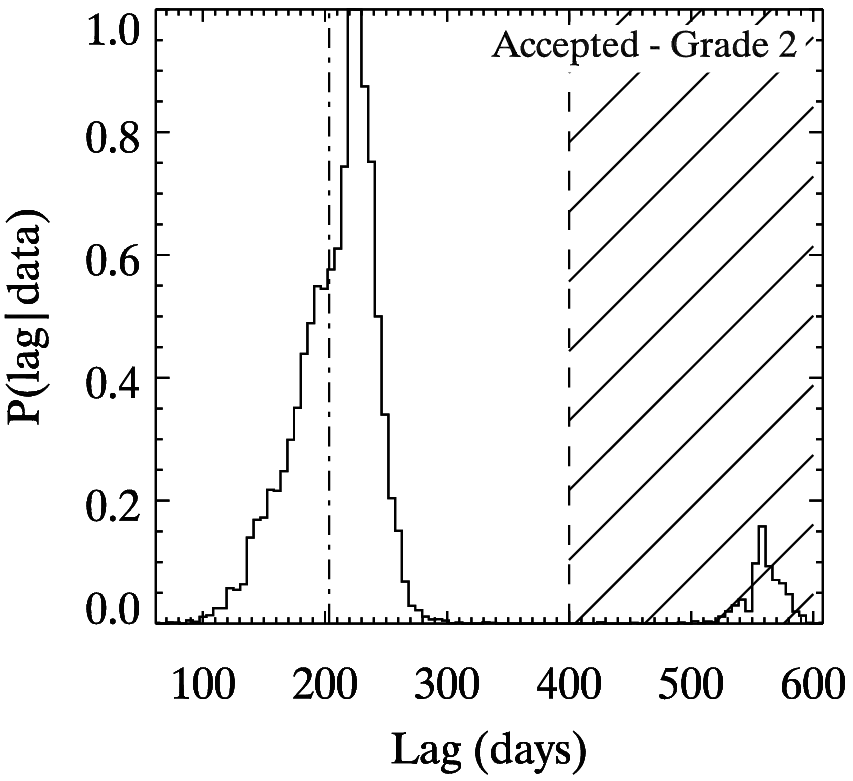}}
\subfloat{\label{fig:y2lc}\includegraphics[width=0.54545456\textwidth]{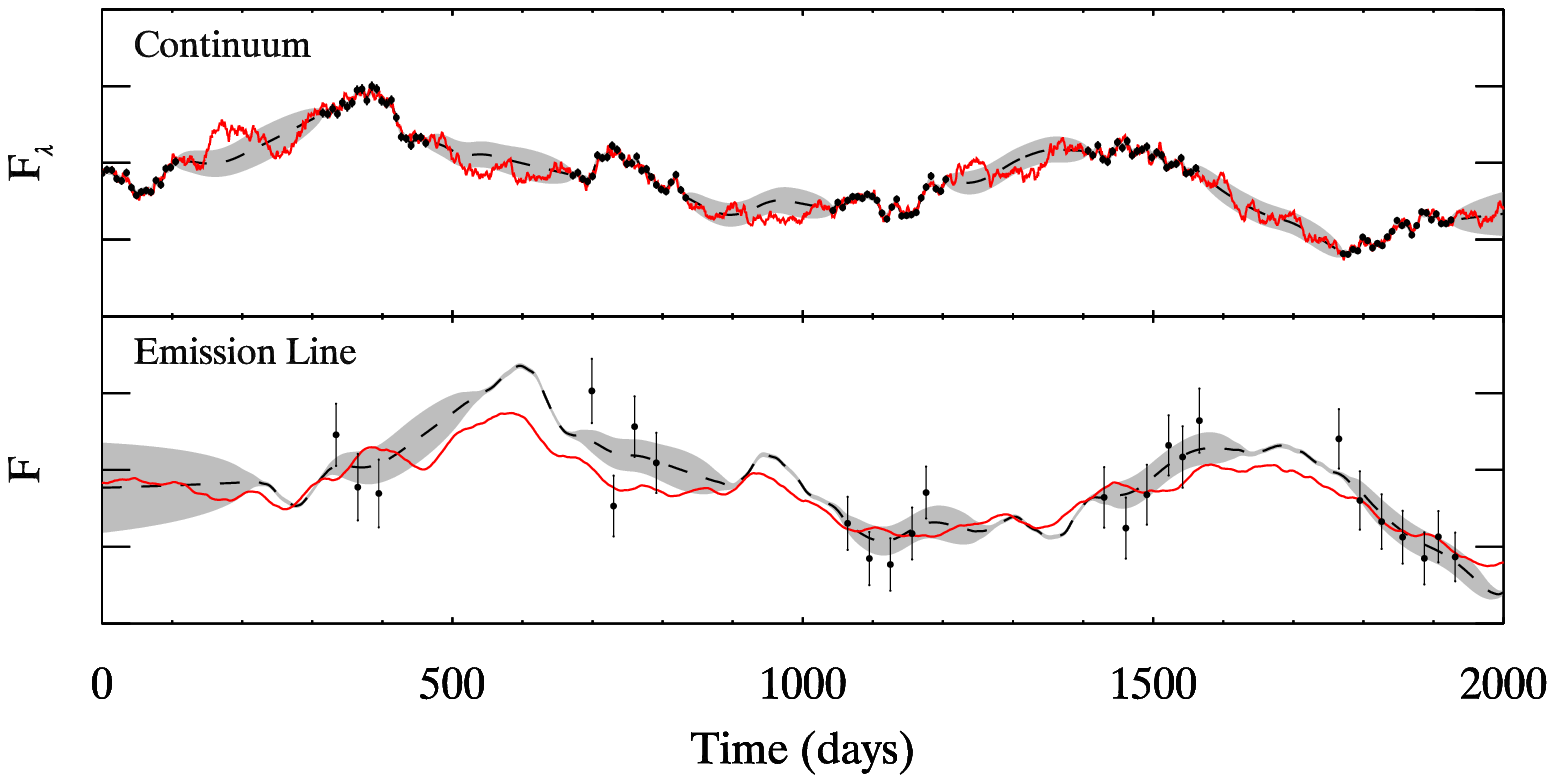}}\\
\subfloat{\label{fig:u}\includegraphics[width=0.3\textwidth]{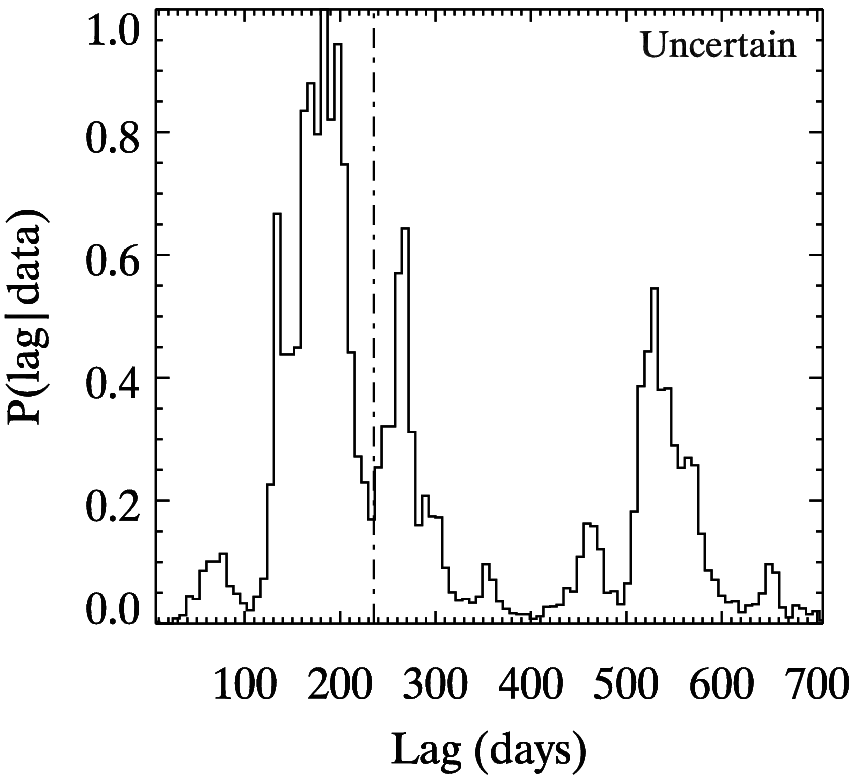}}
\subfloat{\label{fig:ulc}\includegraphics[width=0.54545456\textwidth]{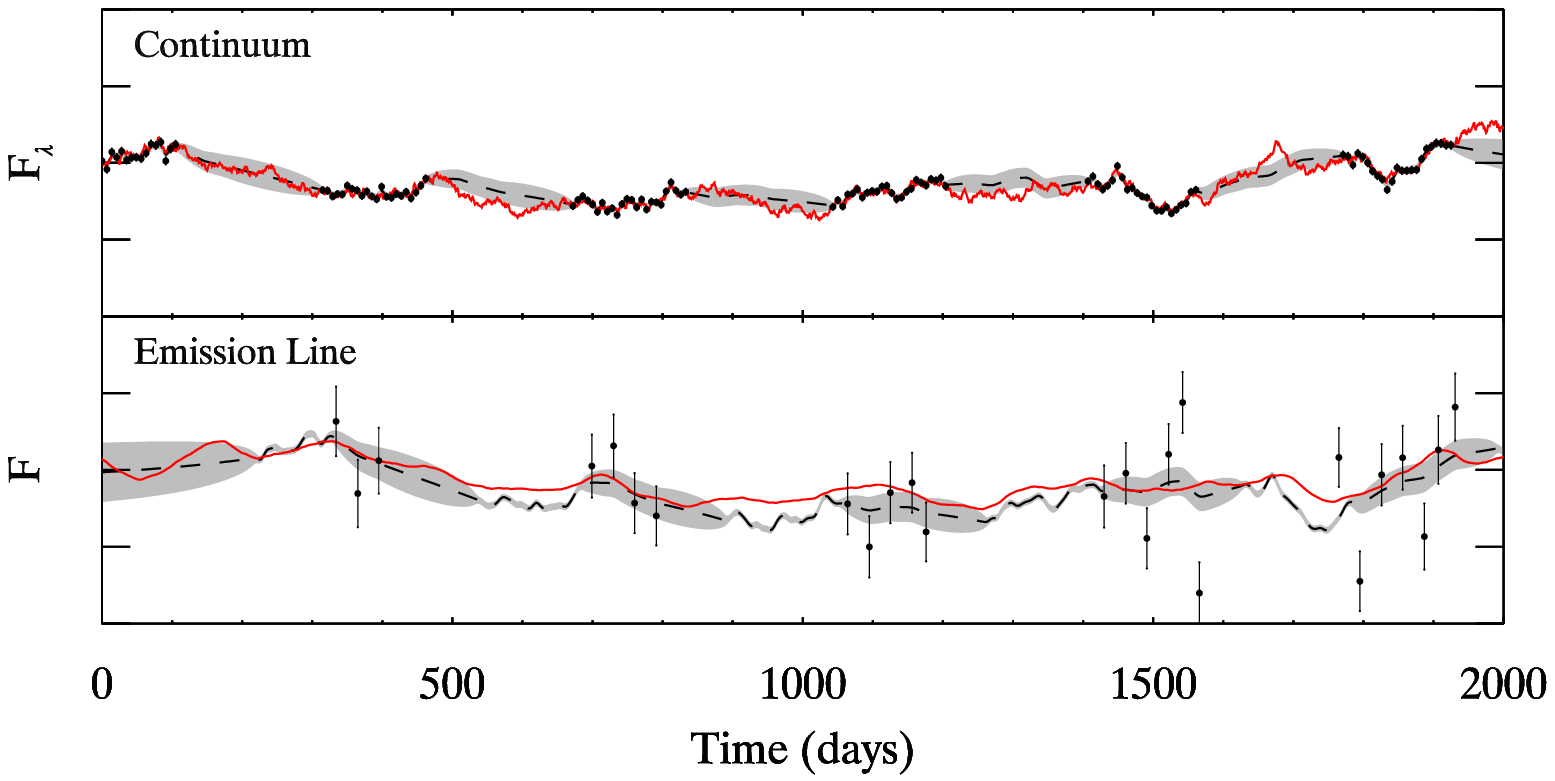}}\\
\subfloat{\label{fig:n}\includegraphics[width=0.3\textwidth]{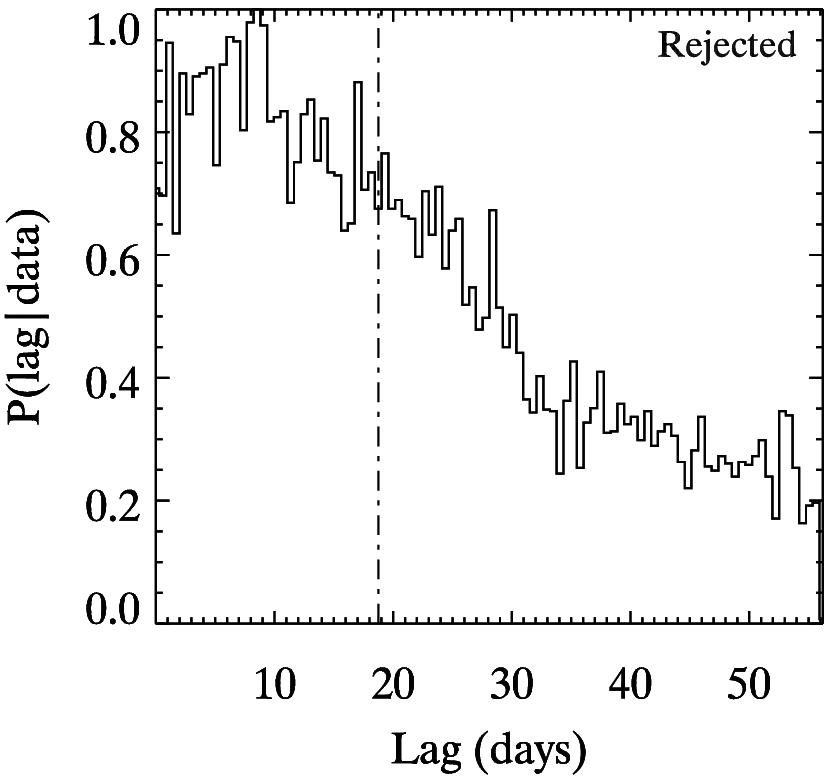}}
\subfloat{\label{fig:nlc}\includegraphics[width=0.54545456\textwidth]{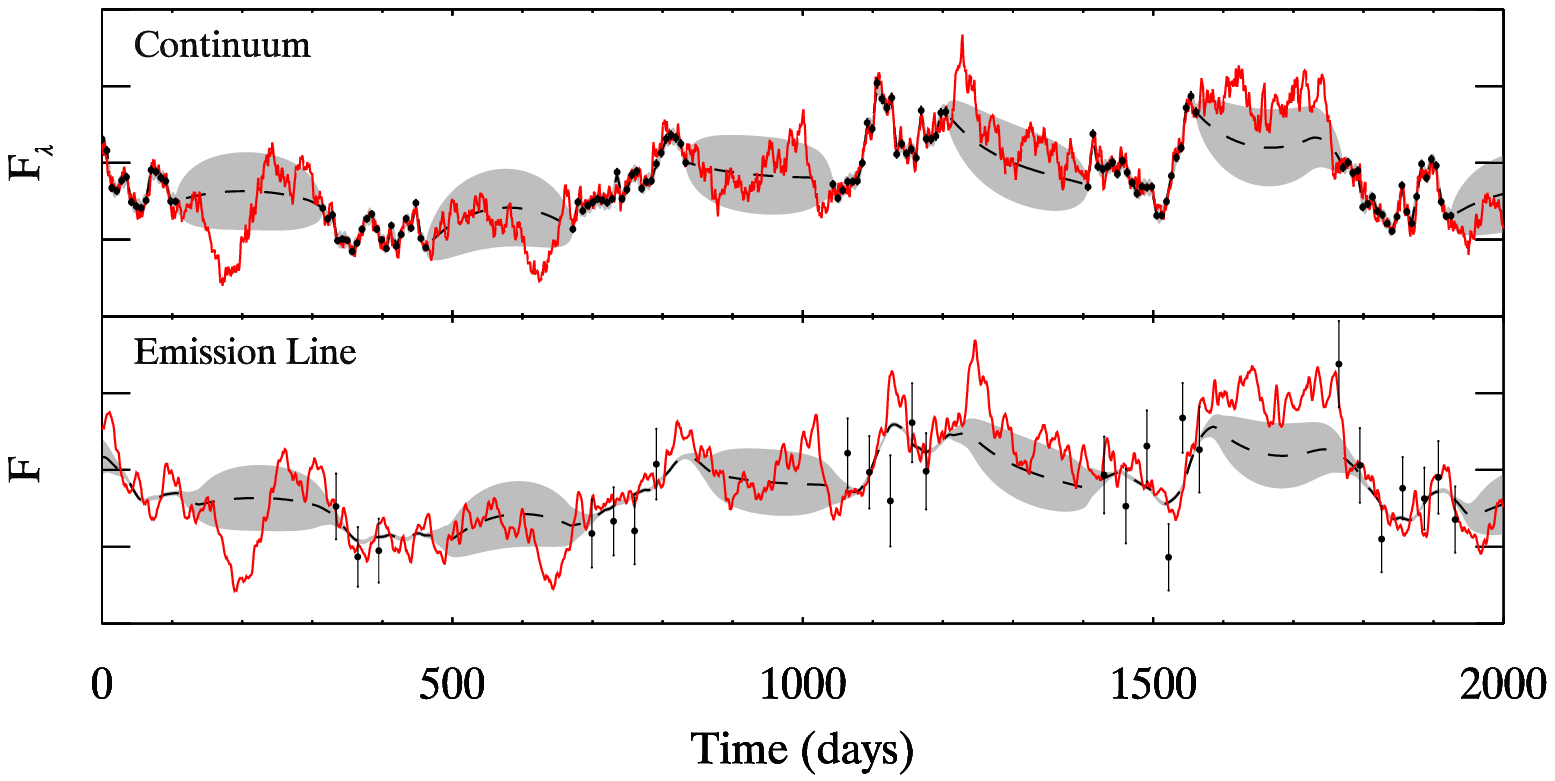}}
\caption{Examples of typical lag posterior distributions (left) for the accepted - grade 1 (top), accepted - grade 2 (middle top), uncertain (middle bottom) and rejected (bottom) classifications. The dot-dashed vertical line  represents the true lag value. The hatched region in the accepted - grade 2 case demonstrates the likelihood region excluded from the final lag estimate. The right panels show the input continuum and emission line light curves (circles) and true underlying light curve (red) compared to the weighted average of light curves that fit the data well from JAVELIN (black line), and the corresponding dispersion of these light curves (grey shaded region).}
\label{fig:examples}
\end{figure*}

We devised a simple method to classify each posterior, trained on a manually classified sample of posteriors. In the most basic terms, we smooth the posterior distribution, and identify the dominant peak and any secondary peaks. We classify the quality of the lag based  on the existence and relative size of the secondary peaks, which corresponds to the relative probability of the lag being associated with each peak.  We classified the output probability distributions into four categories: Accepted - grade 1, Accepted - grade 2, uncertain, and rejected. Accepted - grade 1 means that no secondary peaks were present in the posterior distribution. Accepted - grade 2 means that a secondary feature is present but the ratio between the probability of the primary peak and relative probability of the secondary peak is smaller than 15\%. The secondary peak may be a bump associated with the main peak. The relative probability of the secondary peak is defined as the difference between the maximum probability value of the secondary peak and either zero or where the secondary connects to the main peak, whichever is the smallest value.  This definition is chosen to allow small deviations in the probability distribution of the main peak. 
Uncertain posteriors had multiple peaks of similar magnitude, while rejected posteriors did not have any clear maxima. The classifications were made without prior knowledge of the true lag. Some examples of typical accepted, uncertain, and rejected lag posterior distributions are shown in Fig. \ref{fig:examples}. Once classification has been performed, we redefine the lag window boundaries for the Accepted - grade 2 cases to exclude the secondary peak where applicable and calculate the credible regions around the primary peak only. This algorithm performed reasonably well compared to human classification, with a 5\% misidentification rate.  In practice, the real data and lag fits will be subjected to manual inspection which will reduce the number of misidentified lags. The recovered lag used for the rest of this Chapter is defined as the median lag value from the restricted JAVELIN Monte Carlo chain for accepted lags only, and the lag uncertainty is given by the 16th and 84th percentile range about the primary peak. 

\subsubsection{Performance metrics}
To quantify the performance of the OzDES reverberation mapping campaign and its possible extensions, we define three performance metrics: 1) the \textit{recovered fraction}, describing how many lags we recover, 2) $\sigma_{\Delta}$, quantifying how accurately the recovered lags are measured, and 3) the number of \textit{misidentified lags} in the recovered fraction.
\begin{description}
\item{\textbf{Recovered Fraction:}}
Using the classification described in the previous section for accepted, uncertain and rejected posteriors, we define the recovered fraction as the fraction of Accepted-grade 1 and Accepted-grade 2 lags. We determine the natural spread in the recovered fraction by performing bootstrap resampling and calculating the recovered fraction for 1000 iterations of the sample. The stated recovered fraction is given as the median recovered fraction from the resampling and the uncertainty is given by the 68th percentile values.
\item{\textbf{Accuracy ($\sigma_{\Delta}$):}}
We next quantify the accuracy of the recovered lags. The accuracy is defined by the logarithmic ratio between the median recovered lag and the true lag,  
\begin{equation}
\Delta = \log\left(\frac{{\tau}^{\rm rec}}{{\tau}^{\rm real}}\right).
\end{equation}
Fig. \ref{fig:realvsrec2line} shows that the distribution of $\Delta$ values is approximately Gaussian, for both the single and two-emission-line simulations, and the skewness of the distributions is minimal. The underlying width of the $\sigma_{\Delta}$ distribution characterises the accuracy with which we can expect to recover lags. As each recovered lag has its own uncertainty, we cannot simply calculate the unweighted standard deviation. Instead, we find the weighted standard deviation by finding  the value of $\sigma_{\Delta}$  that minimises the likelihood,
\begin{equation}
\mathcal{L}(\Delta,e_t|\sigma_{\Delta}) = \prod_{i=1}^N \frac{1}{\sqrt{2\pi(\sigma_{\Delta}^2 + e_i^2)}}\exp\left[\frac{-\Delta_i^2}{2(\sigma_{\Delta}^2 + e_i^2)}\right]
\end{equation}
where $e_i = ({\tau}^{\rm rec}_{i_{84\%}}-{\tau}^{\rm rec }_{i_{16\%}})/2$, is the average measurement uncertainty associated with each lag measurement. To find the uncertainty on $\sigma_{\Delta}$ we again use bootstrap resampling.

\item{\textbf{Misidentified lags:}}
A lag is designated as misidentified if it is 3$\sigma_{\tau}$ away from the true lag, where $\sigma_{\tau}$ is the estimated uncertainty in the lag. This is equivalent to $e_i$ for symmetric uncertainties around the true lag. However, as the uncertainties are asymmetric, we consider the error estimates towards the true lag value. Note that while we count the number of misidentified lags, we do not remove them from our sample when calculating $\sigma_{\Delta}$ or any further analysis. This is because we will not know the true lag length when we analyse the real data, and we expect misidentified lags will contaminate the true sample at some level.
\end{description}

\subsubsection{Two line fitting - Simultaneous vs. Individual Fitting}

\begin{figure*}
\centering 
\includegraphics[width=1.0\textwidth]{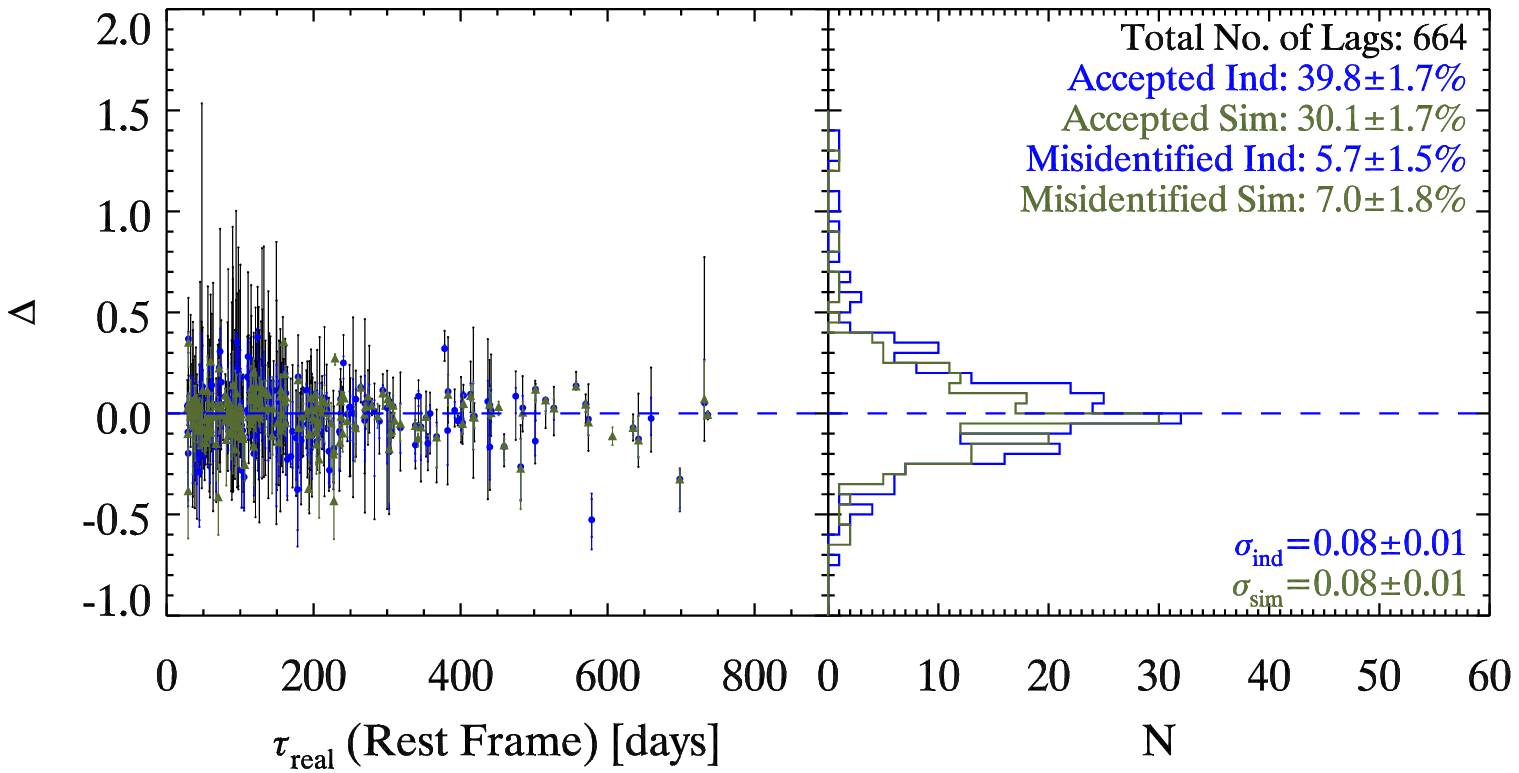}
\caption{The relative offset of the recovered lag, $\Delta$, for the individually fitted (blue circles) and simultaneously fitted (olive triangles) samples (left). The error bars represent the inner 68 percentile errors.  The distribution of $\Delta$ is shown in the right panel. }
\label{fig:realvsrec2line}
\end{figure*}

As previously mentioned, when two emission lines are present in the spectrum we fit the corresponding lags both individually and simultaneously.
Simultaneous fitting has the potential to constrain the continuum light curve more tightly, which could enable a better recovery of the two lags. However, we find the accuracy of the recovered lags and the fraction of misidentified lags to be  consistent between the two methods, although the simultaneous fits had a lower acceptance rate than the individual fits (Fig. \ref{fig:realvsrec2line}). In general, the lag posterior distribution for the simultaneous fitting case showed a peak at the same location as the individual fits, but other secondary peaks were sometimes present, making the lag classification more uncertain. It is likely that this occurs when the lag signal is weak, leading a spurious peak in one lag to amplify an otherwise weak peak in the other lag posterior, leading to multiple peaks. The reverse also occurred, although it was less common, and in this case, the information from the simultaneous fitting nullified the spurious signal. We use the results from the individual line fits for the rest of the analysis.

\section{Results}\label{sec:results}
\subsection{Basic Setup}
\subsubsection{Recovery Fraction}
\begin{figure*}
\centering 
\includegraphics[width=1.\textwidth]{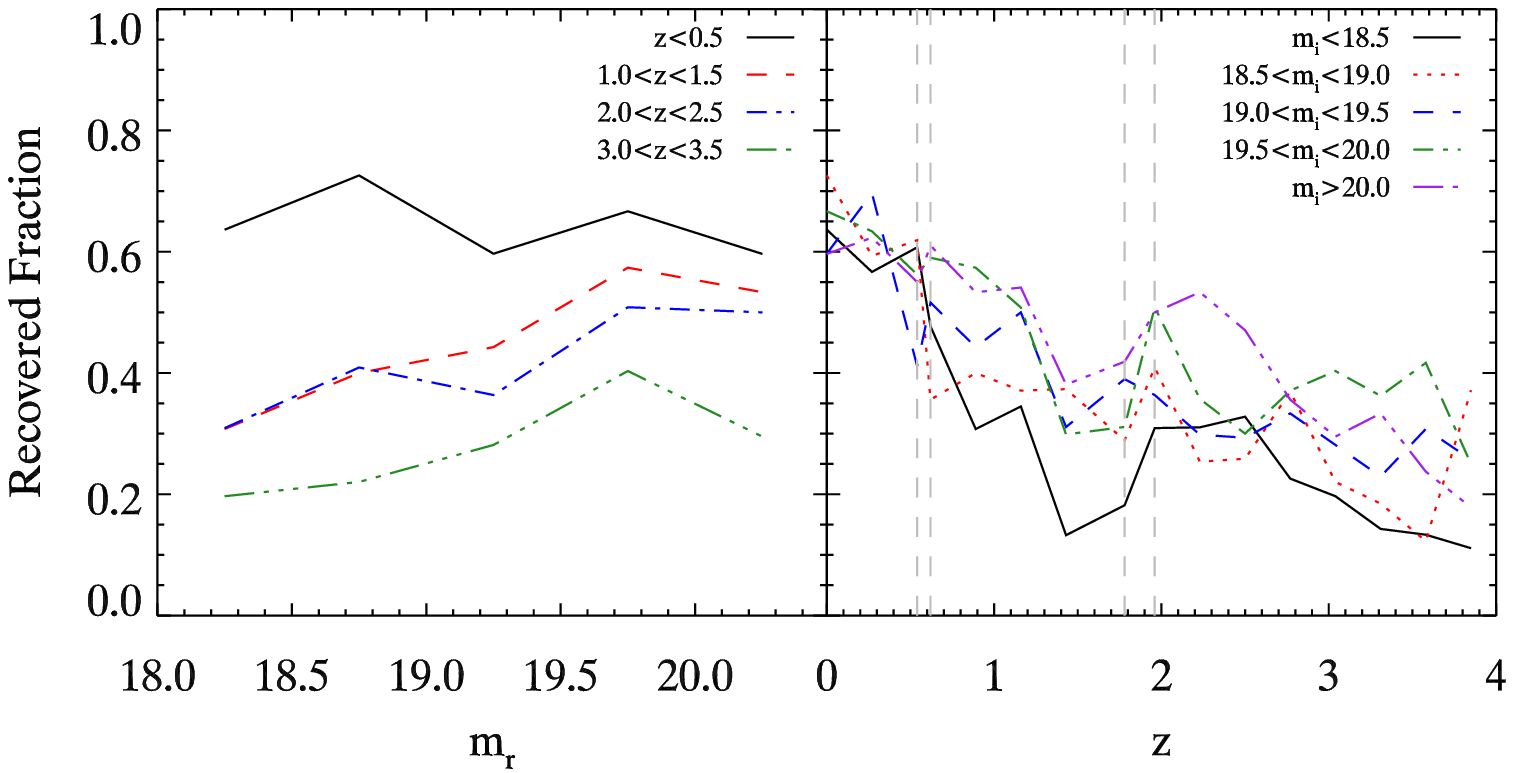}
\caption{The recovered fraction in each redshift-magnitude bin for the 10 mock catalogues, as a function of magnitude (left) and redshift (right). There is a clear trend both in redshift and magnitude to favour lower redshift, less luminous objects. The dashed vertical lines correspond to the bounds of the different redshift ranges for the different emission lines.}
\label{fig:yaccept}
\end{figure*}

\begin{figure}
\centering 
\includegraphics[width=0.45\textwidth]{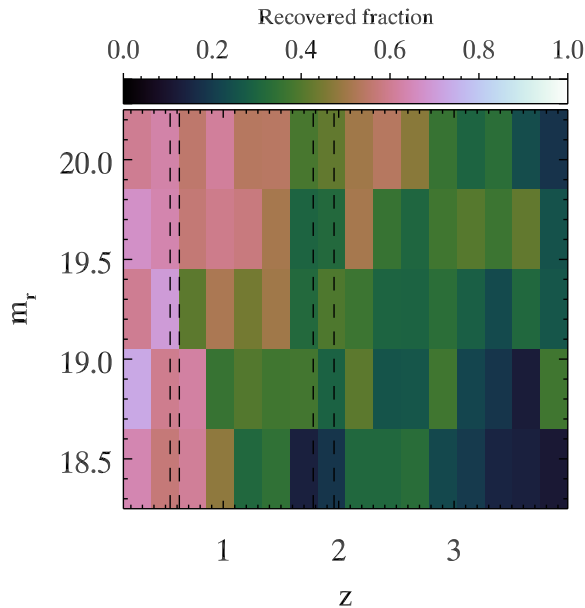}
\caption{The 2D distribution of the recovered lag fraction in redshift-magnitude bins. This distribution is similar in shape to the distribution of lags (Fig. \ref{fig:lagdist}), indicating that the acceptance rate is highly dependent on the intrinsic observed frame lag.}
\label{fig:acceptcontour}
\end{figure}

We first investigate the recovered fraction in bins of redshift and magnitude. This will help to optimise our target selection choices.  The resulting fractions of accepted lags are shown in Figures \ref{fig:yaccept} and \ref{fig:acceptcontour}. There is a clear trend in favour of low redshift, faint objects. 

\begin{figure}
\centering 
\includegraphics[width=0.45\textwidth]{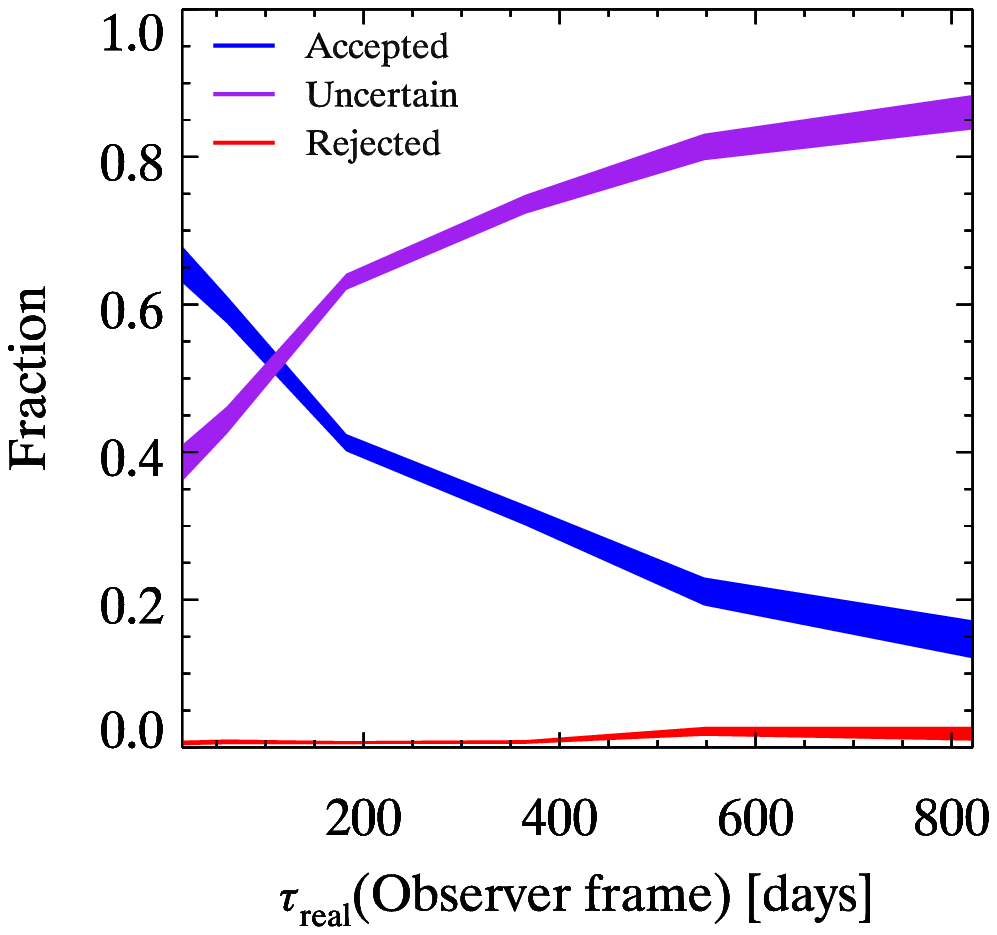}
\caption{The fraction of accepted (blue), uncertain (purple) and rejected (red) lags as a function of the observed lag. }
\label{fig:lagvsaccept}
\end{figure}

Since the observer-frame lag also depends on magnitude (through the R-L relationship) and on redshift (due to time dilation), we also investigated how the recovered fraction correlates with the input observed frame lag. Fig. \ref{fig:lagvsaccept} shows the resulting fraction of each lag quality classification (accepted, uncertain, and rejected) as a function of input observed frame lag. There is a steady decline in the recovered fraction with lag length, and a reciprocal increase in the fraction of uncertain and rejected lags. The likelihood of successfully recovering the lag increases when the overall survey length is appreciably longer than the lag length \citep{Horne2004}, because more light curve features can be traced by both the continuum and emission line light curves. However, program cadence is also important for lag recovery, where sampling the light curves at the ``Nyquist frequency'' is required to accurately resolve the lag.  Consequently, objects with an expected observer frame lag shorter than the minimum emission line light curve sampling rate ($\sim30$ days) are less likely to have a reliable lag estimate. 

\begin{figure}
\centering 
\includegraphics[width=0.45\textwidth]{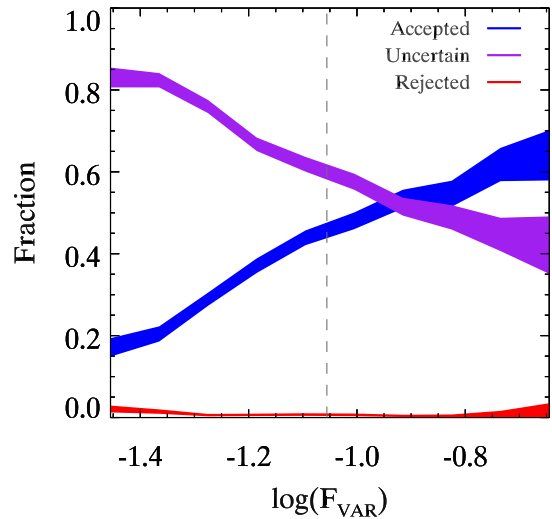}
\caption{The effect of the rms variability amplitude of the continuum, $F_{RMS}$, on the fraction of accepted (blue), uncertain (purple) and rejected (red) lag likelihoods. In general more variable objects are more likely to have a recoverable lag.}
\label{fig:sfaccept}
\end{figure}

The observed dependence of the recovered fraction on lag length will also be affected by the magnitude dependence of the damped random walk parameters, specifically $S_F(\infty)$. The $S_F(\infty)$ parameter describes the long term amplitude of the variable continuum light curve component. If $S_F(\infty)$ is small, as is true for more luminous objects, it is less likely that the observed light curve will vary by a significant degree compared to the noise, and produce a lag measurement. More accurately, the recovered fraction will be affected by the mean fractional variation of the continuum light curve, $F_{VAR}$, defined as the rms variability amplitude of the continuum magnitude \citep[see][]{RodriguezPascual1997}, rather than the true $S_F(\infty)$ value. By definition, $F_{VAR}$ is closely related to $S_F(\infty)$ as $S_F(\infty)=\sqrt{2}\sigma$, where $\sigma$ is the long term standard deviation in the continuum magnitude and $F_{VAR}$ is the fractional standard deviation in continuum flux over a finite observation period.  The approximate transformation, $10^{0.4\sigma}\sim 1+ F_{VAR}$, can be used when the  observation period is sufficiently long. This approximation tends to underestimate the input $S_F(\infty)$ value by roughly 30\% over the five year observation baseline. AGN with $F_{RMS}$ values lower than the expected measurement uncertainty of OzDES are more likely to have an uncertain lag classification. This is illustrated in Fig. \ref{fig:sfaccept}.

In practice the lag length and light curve variability are related through the magnitude dependence of the parameters (R-L relationship and Equation \ref{eq:macleod1}), so the recovered fraction can generally be characterised simply by the absolute magnitude and intrinsically brighter objects are less likely to have a lag measurement. 

\subsubsection{Accuracy of recovered lags}

\begin{figure}
\centering 
\includegraphics[width=0.45\textwidth]{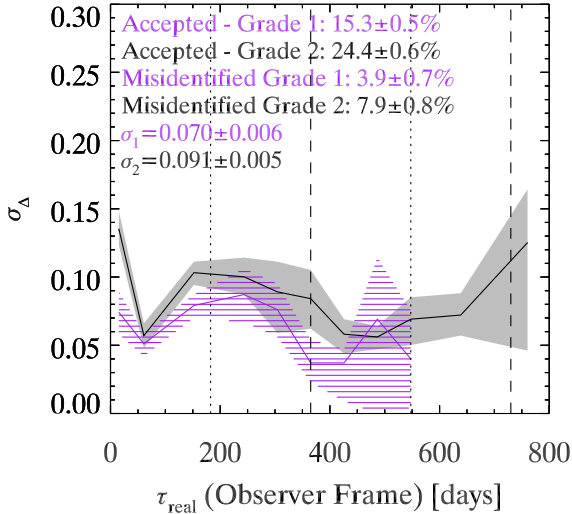}
\caption{The scatter around the true lag, $\sigma_{\Delta}$, as a function of true observed frame lag for the accepted grade 1 sample (purple hashed) and accepted grade 2 sample (black solid). The dotted lines correspond to half year intervals, and the dashed lines correspond to full year intervals. There are no accepted grade 1 lags recovered for $\tau_{real}>1.5$ years.}
\label{fig:avereloffset}
\end{figure}

Next we look at the accuracy of the recovered lags. 
Fig. \ref{fig:avereloffset} shows how the accuracy changes with observed frame time lag.
We divided the sample into accepted-grade 1 and accepted-grade 2. The accepted-grade 1 lags were more accurate in general, with a $\sigma_{\Delta}=0.070\pm0.006$, compared to $\sigma_{\Delta}=0.083\pm0.004$ for the grade 2 sample. However, the number of grade 1 lags is less than half the number of grade 2 cases, so excluding them would have a significant impact on our final sample size. 
For both samples we see a slight increase in $\sigma_{\Delta}$ for $\tau<30$, and $\tau\sim0.5$ years. Lags shorter than 30 days are not accurately recovered due to the limited temporal sampling of the survey. The increase at half a year is likely a consequence of the seasonal gaps in the survey. Lags can only be accuractely recovered when there is some overlap between the observed continuum and emission line light curve features (e.g. associated rise and drop in light curve). Lags close to $365\times n/2$ days, where $n$ is an odd number,  will have very few observed continuum and emission line light curve measurements that can trace out associated light curve features, leading to larger uncertainties and lower accuracy.  Conversely, for lags close to a year, associated light curve features are well traced by both the observed continuum and emission line light curve, leading to a well defined and accurate lag estimate. This effect strongly suggests extending our observing season, as we investigate in the next section.

\subsection{Extensions}\label{sec:ext}
\begin{figure*}
\centering 
\includegraphics[width=0.8\textwidth]{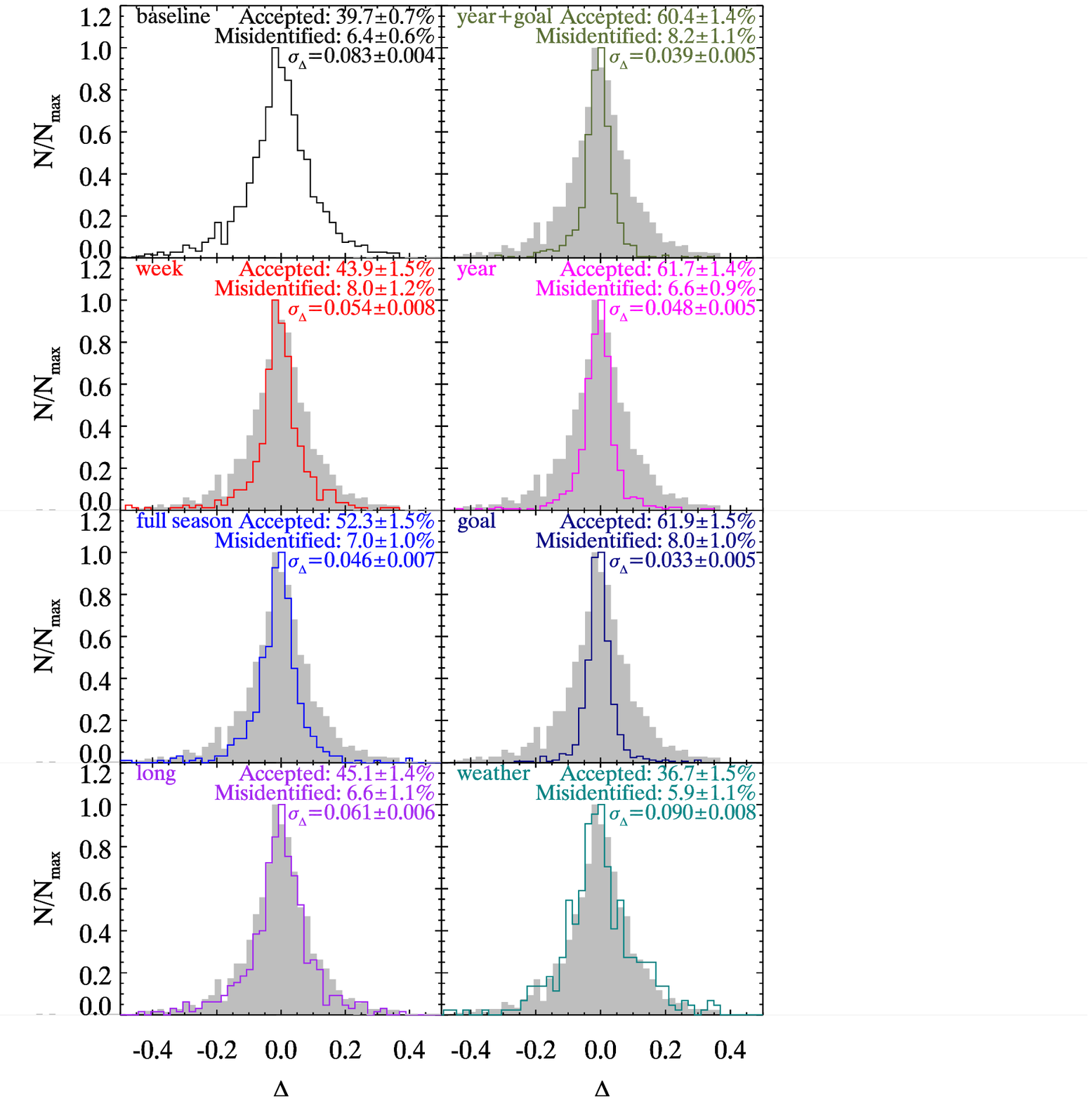}
\caption{The predicted scatter in the recovered lags and recovered fraction for (from top to bottom) the long (2 extra years of observations), weekly (weekly cadence of spectroscopic measurements), full season (9 month observation season for the last 3 years of survey), year (hypothetical 5 year survey), goal (0.03 mag uncertainty in spectroscopic measurement), year+goal and weather (3-5 epoch of spectroscopic data were lost over the 5 year period) extensions compared to the baseline OzDES results (grey underlay, and top left). }
\label{fig:offset_ext}
\end{figure*}

\begin{figure}
\centering 
\includegraphics[width=.45\textwidth]{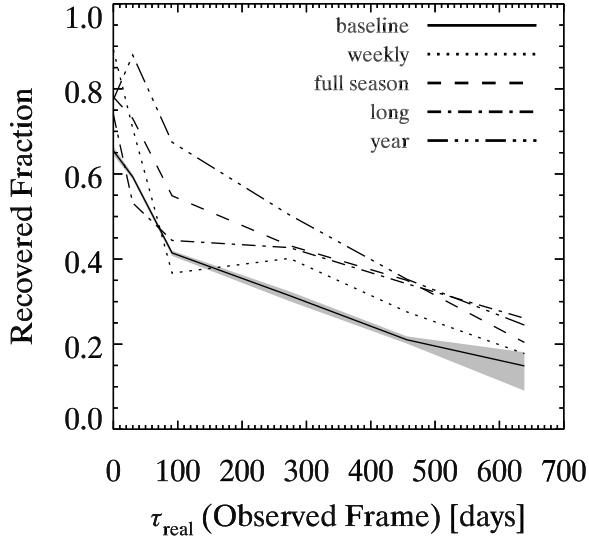}
\caption{The recovered fraction of the weekly [dotted; weekly cadence of spectroscopic measurements], full season [dashed; 9 month observation season for the last 3 years of survey], long [dot dashed; 2 extra years of observations], and year [dot dot dot dashed; 5 year survey with no seasonal gaps and monthly spectroscopic cadence] survey extensions as a function of the observed time lag compared to baseline survey [solid with grey shading; corresponding to the median value and inner 68 percentile values from bootstrap resampling].}
\label{fig:acc_ext}
\end{figure}

\begin{figure}
\centering 
\includegraphics[width=.44\textwidth]{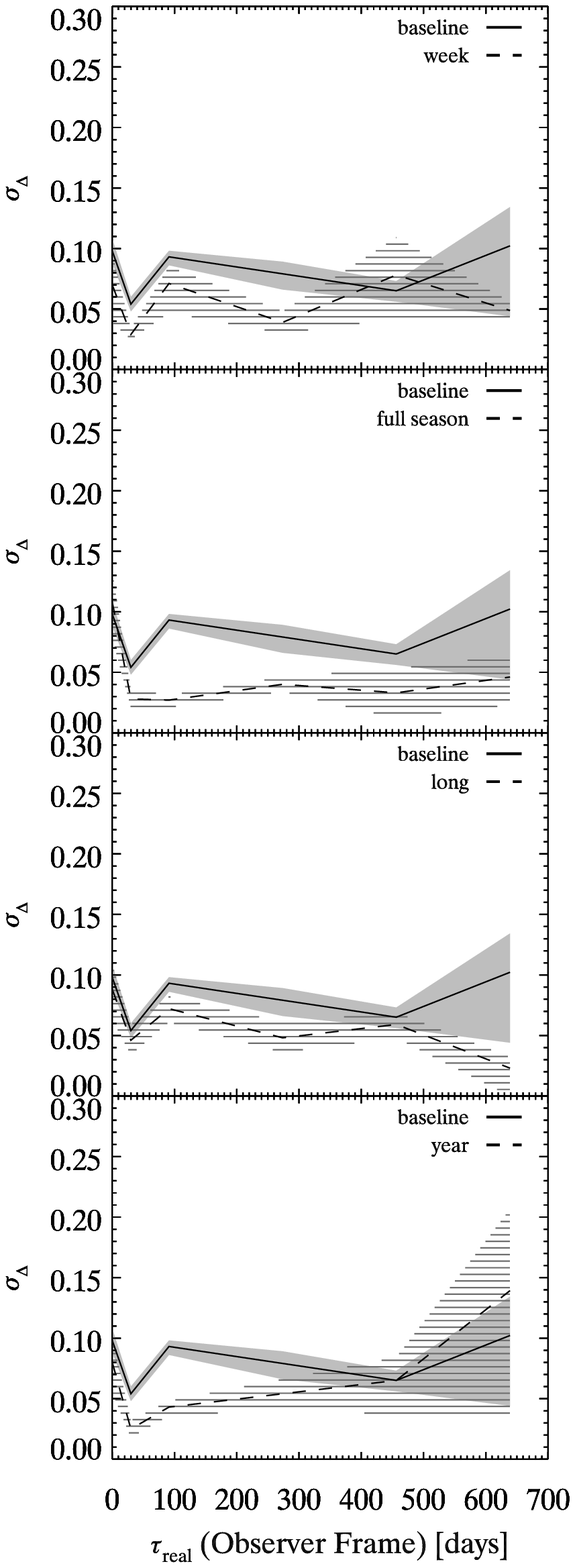}
\caption{The accuracy $\sigma_{\Delta}$ of, from top to bottom, the weekly (weekly cadence of spectroscopic measurements), full season (9 month observation season for the last 3 years of survey), long (2 extra years of observations), and year (5 year survey with no seasonal gaps and monthly spectroscopic cadence) survey extensions as a function of the observed time lag.}
\label{fig:wabs_ext}
\end{figure}

We next consider how the possible survey extensions affect the lag recovery relative to the baseline survey. The main results from these various extensions are summarised graphically in Figures \ref{fig:offset_ext}-\ref{fig:wabs_ext}, where Fig. \ref{fig:offset_ext} shows the distribution of $\Delta$, Fig. \ref{fig:acc_ext} shows the recovered fraction, and Fig. \ref{fig:wabs_ext} shows $\sigma_{\Delta}$ as a function of the real lag for the different extensions.

\begin{description}
\item{\textbf{Seasonal Gap:}}
Implementing the full season extension is expected to increase the recovered fraction to $\sim$50\%, which corresponds to a 20\% relative increase over the baseline survey. The value of $\sigma_{\Delta}$  also decreases significantly, especially for lags at $365\times n/2$ days. This is expected, as the smaller seasonal gaps reduce both the need for interpolation and the range of lags for which common features cannot be traced by both the measured continuum and emission line light curve. The year extension shows still greater improvements. However, the full season and year extensions only show significant improvement in accuracy for short lags, and no appreciable improvement was seen for long lags. In fact, the year long extension accuracy was diminished compared to the baseline survey for long lags, as the total duration of the baseline survey is extended beyond five years due the inclusion of the DES science verification data. Without extending the survey length, long lags remain hard to recover because the continuum and emission line light curves still only have a few common features.
\item{\textbf{Cadence:}}
The recovered fraction and accuracy of the recovered lags improved significantly with the weekly extension.   The finer sampling enables superior recovery of shorter lags, as can be seen in Fig. \ref{fig:acc_ext}.  It is expected that some of this improvement is due to the increased number of epochs. Although, the number of epochs is greater in this scenario than the full season extension, and the overall improvement in the lag recovery is not as large. This suggests that if additional telescope time is awarded in the next three years of observations, it should be used to extend the observation season rather than to have finer sampling.
\item{\textbf{Data quality:}}
Reducing the measurement uncertainties relative to the baseline significantly increases the recovered fraction and accuracy of the recovered lag. However, improving the measurement uncertainty in the year scenario (`year+goal'), appears to have a negligible effect. In fact, it appears as though a reduction of the emission line light curve measurement uncertainty to 0.03 mags has a gain equivalent to carrying out the year survey.  Therefore, if we can reduce the uncertainties on the light curve measurements, additional epochs will be of less consequence. Alternatively, once we reach the limit of reducing the uncertainties, we can improve the scientific outcome with more epochs.
\item{\textbf{Survey Length:}}
The long extension shows an increased recovered fraction and accuracy primarily for the objects with longer lags, as expected. We also see a moderate increase in accuracy for all lags. However, a decrease in recovered fraction and accuracy at half a year is still present due to the seasonal gaps.
\item{\textbf{Weather:}}
Finally, if 3-5 additional epochs of spectroscopic data were lost over the five years, only a slight drop in recovered fraction and accuracy is expected.
\end{description}

\section{Prospects}\label{sec:science}
\begin{figure}
\centering 
\includegraphics[width=0.45\textwidth]{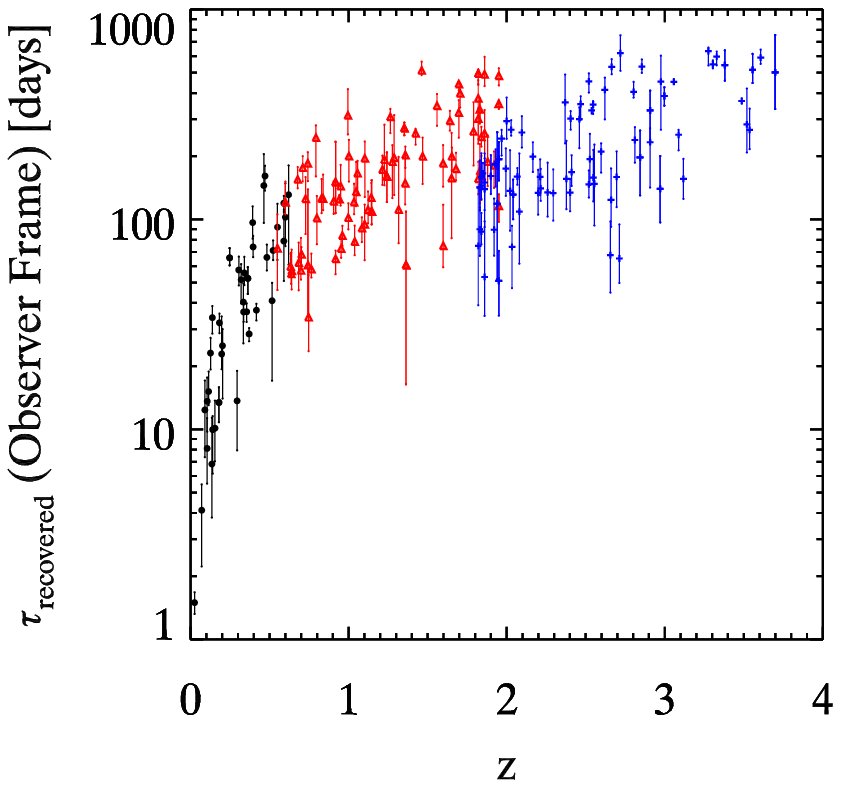}
\caption{Forecast of lag detections for the OzDES reverberation mapping campaign.  The different colours and symbols correspond to the different emission lines (black circles-H$\beta$, red triangles-\mgii, blue crosses-\civ).}
\label{fig:simulatedlagvsz}
\end{figure}

Maximising the return is not simply a question of maximising the number of recovered lags. Next we use the results of the simulations to optimise target selection for measuring black hole masses and constraining the $R-L$ relationship. Additionally, we fold in the redshift and magnitude distribution of the OzDES target quasars into our predicted results. Fig. \ref{fig:simulatedlagvsz} shows an example of the recovered lags we expect for the baseline OzDES survey if the final 500 AGN were chosen randomly from the currently observed 989 candidate quasars. The acceptance rate is $\sim 35\%$ and the sample covers a luminosity range of $10^{39}<\lambda L_{\lambda}(5100\text\AA)<10^{46}$ with mean $\mu\left\{\log\left[\lambda L_{\lambda}(5100\text\AA)\right]\right\}=45.0$, and extends to redshift $\sim4$.

\subsection{Black hole mass measurements}

The uncertainty in a RM black hole mass estimate is,
\begin{equation}
\frac{\sigma_{M_{\rm BH}}}{M_{\rm BH}} = \left[\left(\frac{\sigma_{f}}{f}\right)^2+\left(\frac{2\sigma_{\Delta V}}{\Delta V}\right)^2+\left(\frac{\sigma_{R}}{R}\right)^2\right]^{1/2},
\end{equation}
assuming the errors on the virial factor, $f$, the line width, $\Delta V$, and BLR radius, $R$, are independent.
Generally, the formal errors reported for $M_{\rm BH}$ only include the uncertainties from $R$ and $\Delta V$ and ignore the uncertainty in $f$  \citep[e.g.][]{Bentz2008,Grier2012} even though it is generally the largest source of uncertainty in the mass determination. 
The typical uncertainty in the reverberation masses due to uncertainty in $f$ is $\sim0.43$ dex \citep{Woo2010}.
 
The velocity dispersion, $\Delta V$,  is measured from the emission line width in either the rms variance spectrum (henceforth referred to simply as the rms spectrum) or the mean spectrum. The rms spectrum should be used whenever possible as it is the best representation of the responding (variable) component of the emission, and will therefore provide the most accurate mass estimate. However, sometimes the rms spectrum is not of high enough quality or the rms profile is unusual in shape, and the interpretation of width is not straight forward (e.g. NGC4151; \citet{Bentz2006NGC4151} and Mrk 817; \citet{Denney2010}). In this case the mean spectrum should be used instead. 

The quality and uncertainty in the width measurement are dependent on the line width characterisation used and the SNR of the spectra \citep{Denney2009, Fine2010, Jensen2012}. The common ways to characterise the velocity dispersion from the line width are the full width at half maximum (FWHM), the second moment of the line ($\sigma_{\rm line}$; otherwise known as the line dispersion\footnote{The `line dispersion' is distinct from  `velocity dispersion' and is defined as, $\sigma^2_{\rm line} = \langle\lambda\rangle^2-\langle\lambda^2\rangle = \frac{\int \lambda^2F(\lambda)d\lambda}{\int F(\lambda)d\lambda}- \left(\frac{\int\lambda F(\lambda)d\lambda}{\int F(\lambda)d\lambda}\right)^2$.} ), and the inter-percentile velocity \citep[IPV;][]{Whittle1985}.  We expect the fractional uncertainty in $\Delta{V}$, measured from the mean spectrum to be $\lesssim$5\% ($\lesssim0.04$ dex in $M_{\rm BH}$). However, the rms spectrum signal has an extra dependence on the level of variability, which will impact the S/N of the line profile and therefore the accuracy and precision of the line width measurement. Compared to the current reverberation mapping sample, of primarily low luminosity objects ($\mu\left\{\log\left[\lambda L_{\lambda}(5100\text\AA)\right]\right\}=43.5$), the level of variability in the OzDES sample is likely to be smaller. Therefore, we expect the line width uncertainties measured from the rms spectra of these quasars to be higher, with fractional uncertainties of order of 20-30\% (0.15-0.20 dex in $M_{\rm BH}$; K. D. Denney, private correspondence). Additionally, in some objects we expect the emission line flux errors per pixel to be on the same order as the variability signal, and consequently, the rms spectrum signal will be too low to resolve unambiguously.

It should be noted, that care should be taken when measuring the mean spectrum to ensure that any non-variable components, like narrow emission lines, are modelled and removed. This is particularly important for \civ. It has been found that using line dispersion is more robust than FWHM in this case, as the profile changes with width \citep{Rafiee2011,Denney2012}. Additionally, when using the mean spectrum a different f-factor is required as the mean spectrum width tends to differ from the rms spectrum width systematically. 

When absorption or contamination is present in the emission line profile, a rms spectrum can be created using  individually modelled spectra (see Section \ref{sec:uncertain}). However, if the SNR in each spectrum is not sufficient to accurately decompose the individual spectra, the mean spectrum is modelled instead and the width is measured accordingly. This technique was performed by \citet{DeRosa2015} for NGC5548.

On average, $\sigma_{R}/R\sim$20\% (0.08 dex in $M_{\rm BH}$) and no significant dependence on redshift is apparent. Thus, for the baseline OzDES setup we expect a median formal uncertainty in $M_{\rm BH}$ of 0.09 dex (0.16-0.21 dex) in a random sample of OzDES targets using the mean (rms) spectrum, ignoring the uncertainty in $f$.

\subsection{Recovery of the $R-L$ relationship}
 One of the major scientific goals of this survey is to derive the $R-L$ relationship for all three emission lines. In this section, we investigate how to optimally select our target quasar sample to recover the most accurate and precise $R-L$ relationship. To recover the $R-L$ relationship for all three lines we require a substantial calibration sample with lag measurements for two emission lines so that the relationships can be put on the same relative luminosity scale. This is crucial for both black hole mass estimates and any attempts to use quasars as standard candles \citep{King2014}. Therefore we decided to monitor all 69 of the quasars that fall into a redshift range where two lines can be observed simultaneously (i.e. $0.54<z<0.62$ and $1.78<z<1.96$). 
 
The optimal strategy for constraining the $R-L$ relation is to observe objects over a broad luminosity range. To do so we need to observe both faint low-$z$ objects and bright high-$z$ objects. However, we also want to optimise the number of recovered lags. For the rest of this section, we determine whether selecting our targets randomly to cover a broad range of properties, or based on their expected acceptance fraction or accuracy, leads to a better estimate of the $R-L$ relationship.
We separated our ten AGN mock catalogues into two groups, a training sample and a observed sample. The training sample was used to calculate which redshift-magnitude bins had the highest recovery rates and accuracy.  AGN were then selected from the observed sample according to their magnitude and redshift and used to calculate the $R-L$ relationship for each emission line. The separation was made in an attempt to avoid any biases in the resulting constraints from particularly favourable or unfavourable bins that do not follow the general magnitude and redshift trends. 

The training sample was broken into separate redshift and magnitude bins and the recovered fraction and accuracy, $\sigma_{\Delta}$, was calculated for each bin.  The resulting distributions were quite noisy, so to remove the influence of spurious bins that may skew our results we fit a low order polynomial surface to both the recovered fraction and $\sigma_{\Delta}$ training distributions, using the IDL\footnote{Interactive Data Language (Exelis Visual Information Solutions, Boulder, Colorado)} function SFIT. Due to the significant differences between  \mgii\ and \civ\ lags, we split the redshift range into a low- ($z<1.78$) and a high-redshift ($z>1.96$) group and fit each group separately. The recovered fraction distribution does not appear to have any major structure besides a decline towards bright high-redshift objects, so we simply fit a linear distribution. The distribution of  $\sigma_{\Delta}$ does exhibit several significant features due to the lag dependences found in the previous section, so we fit a 3rd order polynomial surface to the low redshift group and a 4th order polynomial surface to account for the apparent structure in the high redshift group. The residual distributions of these fits did not show any significant underlying structure. 

The observed sample of 500 sources was then chosen from specific redshift and magnitude bins according to either recovered fraction or accuracy restrictions.   The criteria tested were: (a) the recovered-percentage/acceptance was greater than 50\%; (b) The recovered-percentage/acceptance was greater than 40\%; (c) $\Delta$ was less that 0.05; or (d) $\Delta$ was less than 0.10.

\begin{figure*}
\centering 
\subfloat[H$\beta$: $\alpha$]{\label{fig:alphahb}\includegraphics[width=0.4\textwidth]{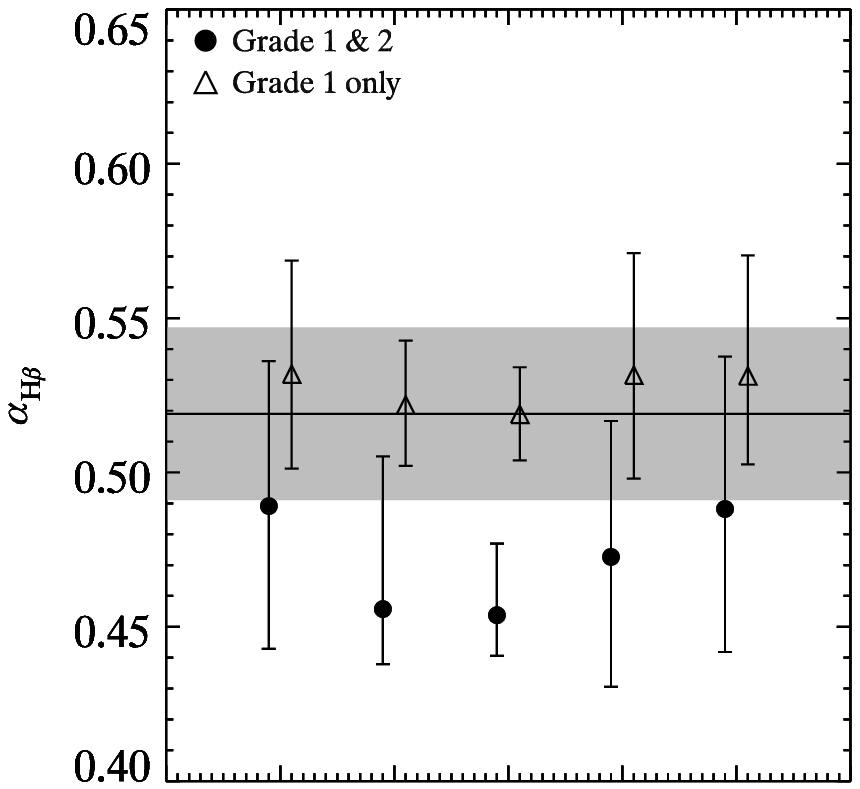}}
\subfloat[H$\beta$: $K$]{\label{fig:alphahb}\includegraphics[width=0.4\textwidth]{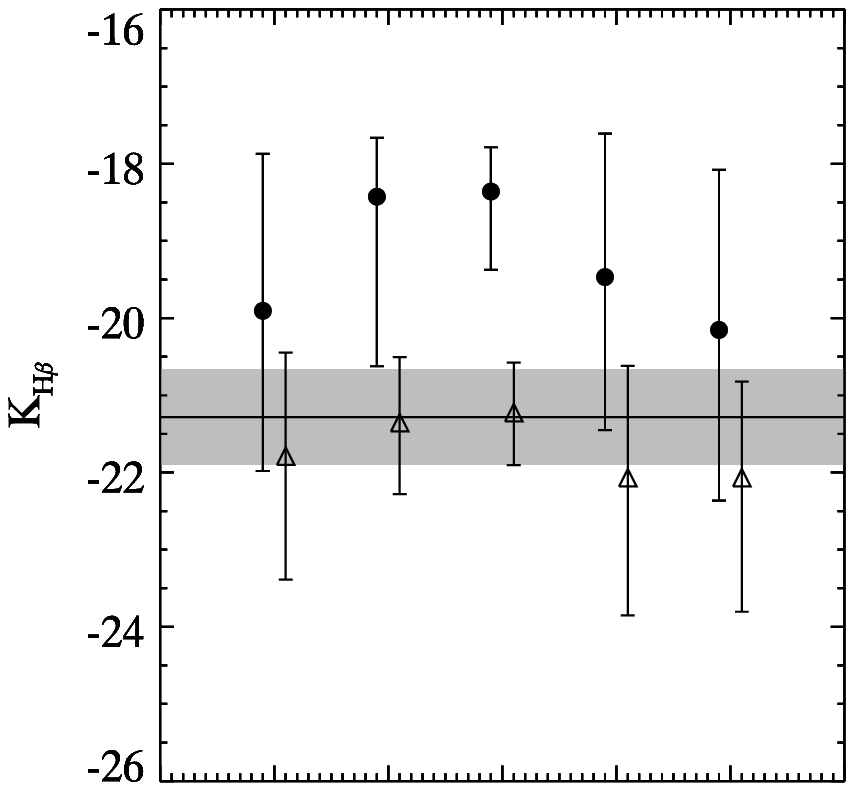}}\\
\subfloat[\mgii: $\alpha$]{\label{fig:alphamgii}\includegraphics[width=0.4\textwidth]{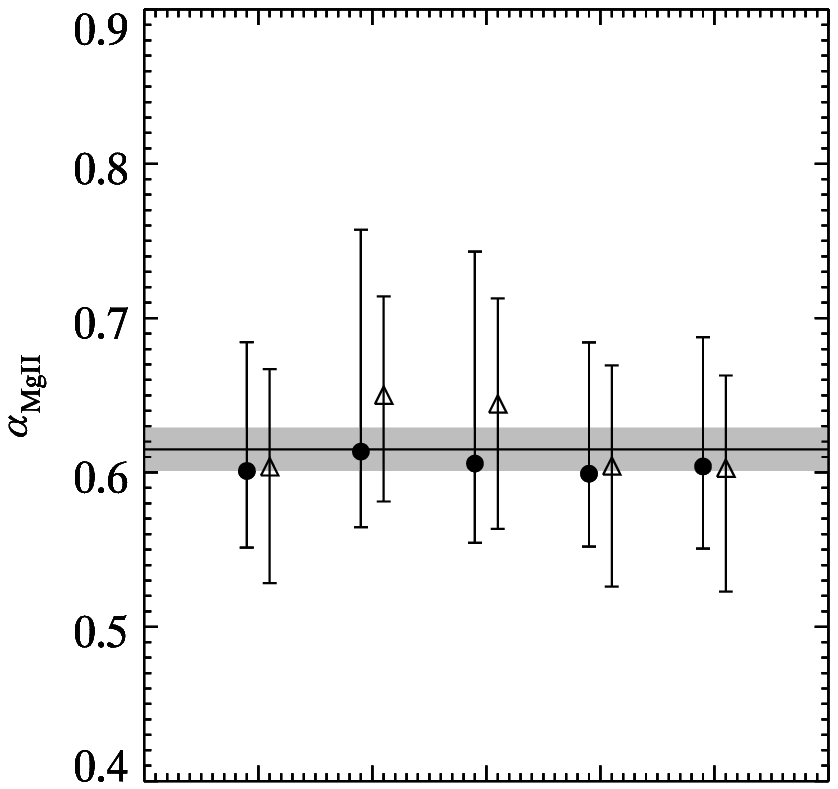}}
\subfloat[\mgii: $K$]{\label{fig:alphamgii}\includegraphics[width=0.4\textwidth]{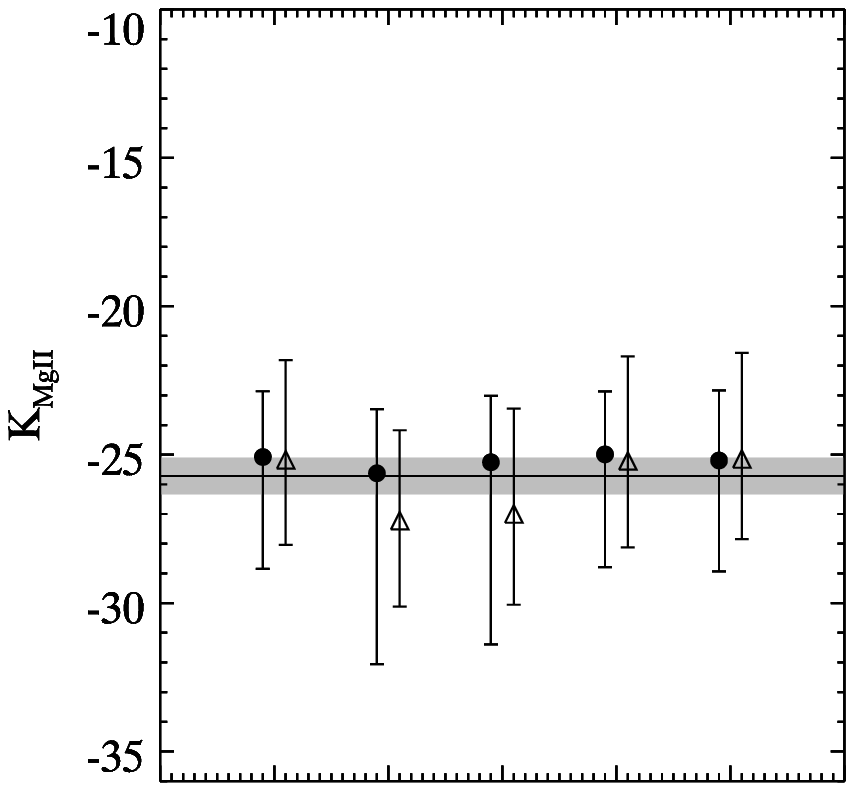}}\\
\subfloat[\civ: $\alpha$]{\label{fig:alphaciv}\includegraphics[width=0.4\textwidth]{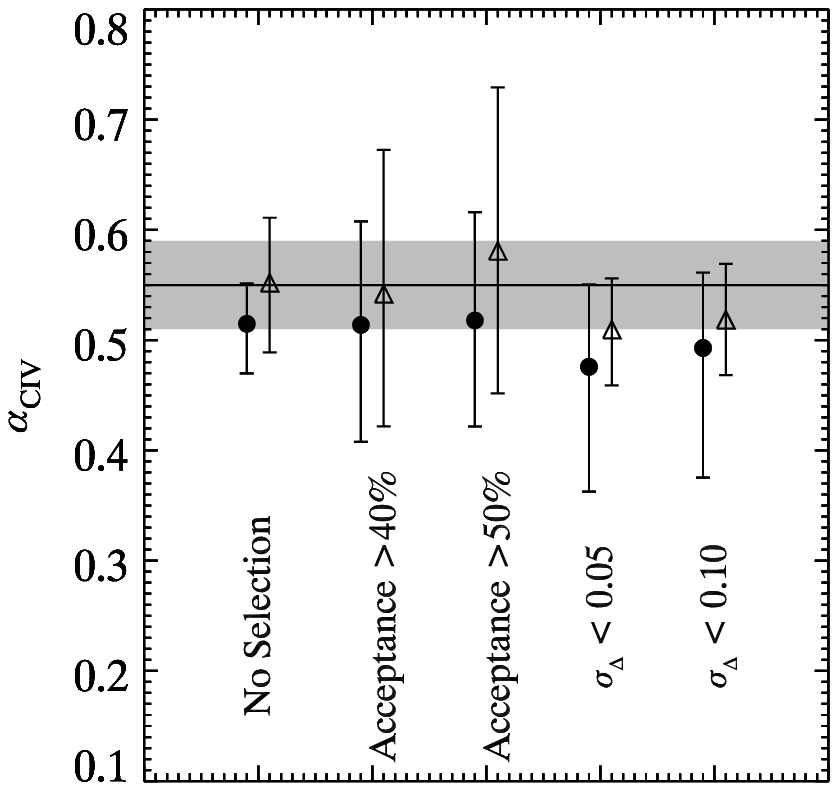}}
\subfloat[\civ: $K$]{\label{fig:alphaciv}\includegraphics[width=0.4\textwidth]{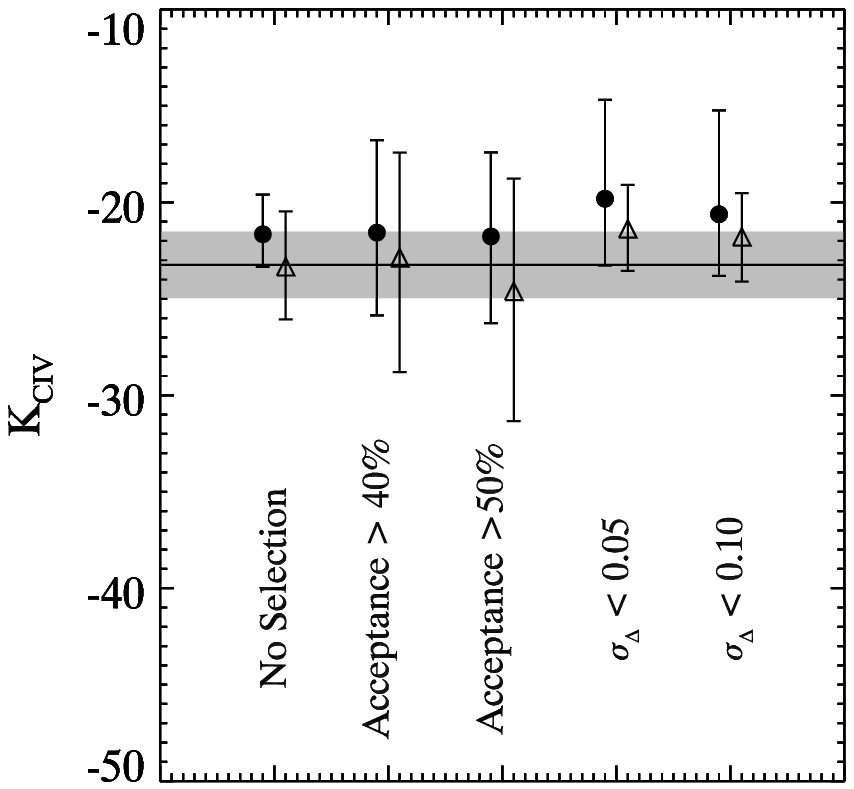}}
\caption{The median recovered gradient ($\alpha$: left) and intercept ($K$; right) of the $R-L$ relation for H$\beta$ (top), \mgii\  (middle) and \civ\ (bottom) for both the full accepted sample (circles) and the accepted - grade 1 sample (triangles) for each sample selection method. The horizontal line and grey shaded regions show the input values for $\alpha$ and $K$ and their current uncertainties.}
\label{fig:alpha}
\end{figure*}

For each sample we fit a $R-L$ relation of the form 
\begin{equation}
~~~~\log (R/ 1 ~{\rm lt~year}) = K + \alpha\log (\lambda L_{\lambda}/1~{\rm erg~s}^{-1}),
\end{equation}
 assuming the luminosity of each object is known exactly. 
The $\alpha$ and $K$ parameters were then calculated for 1000 different possible samples of observed quasars.
 The resulting median estimates of $\alpha$ and $K$ and their uncertainties are shown in Fig. \ref{fig:alpha} and Table 
\ref{tab:RL}. Following \citet{Bentz2013}, we also define the scatter in the sample as the standard deviation of the residuals around the best fit relation, and its value is given in Table \ref{tab:RL}. Table \ref{tab:RL} also includes the median observed and expected recovered number of H$\beta$, \mgii, and \civ\ lags for each selection criterion.

For the H$\beta$ case, there appears to be a systematic bias towards a shallower slope in the recovered $R-L$ relationship than expected. This is due to the poor lag accuracy for very short lags, associated with the lowest luminosity, low-redshift objects. We attempted to minimise this effect by restricting our sample to accepted grade 1 lags only. The resulting $\alpha$ and $K$ values are shown by the triangles in Fig. \ref{fig:alpha}. After this quality cut, the bias is no longer apparent and the recovered values are more tightly constrained. Therefore we should only include the objects with clear RM lag signals. We can also note, that the previous RM campaigns constrain the low luminosity, low redshift objects well \citep{Bentz2013} and their contribution to the $R-L$ relation fit has not been considered here.

  The \mgii\ $R-L$ relation constraints for both the full accepted sample and accepted grade 1 only sample were found to be reasonably consistent, although the scatter around the grade 1 best fit was $\sim$30\% smaller than for the full sample. Therefore the \mgii\ sample is not systematically biased by spurious lag values, but the higher constraining power gained by including more objects is balanced by the increase in the scatter around the relationship. 
  
  Like H$\beta$, the recovered \civ\ $R-L$ parameters also suffered from a bias towards a shallower $R-L$ slope with the whole accepted sample, which is resolved by restricting the sample to grade 1.  The resulting constraints accurately recover the input $R-L$ relation parameters in all cases except for samples chosen for acceptance. This is due to the reduced number of \civ\ lags used for the parameter constraints. The original bias towards a shallower $R-L$ slope is likely due to the short luminosity baseline for  \civ\ $R-L$ relationship, which is more sensitive to misidentified lags than the corresponding \mgii\ $R-L$ relation fit. Despite both lines sharing a similar lag distribution and therefore overall lag accuracy.
  
   The accuracy and precision of the parameter estimates appear to be relatively independent of  whether the sample was selected randomly or with an accuracy constraint. However, a sample chosen for high recovered fraction created tighter constraints for H$\beta$ and poor and possibly misleading constraints in \mgii\ and \civ. This is primarily driven by the number of lags recovered, although the luminosity baseline of the $R-L$ relation fit is also restricted by this choice, which will affect the recovery of the $R-L$ constraints. Therefore in order to accurately and precisely recover the $R-L$ relationship for all three lines it is preferable to get an even distribution of targets over the total redshift and magnitude distribution rather than maximising the total number of recovered lags.

\begin{figure*}
\centering 
\subfloat[H$\beta$: $\alpha$]{\label{fig:alphahb}\includegraphics[width=0.4\textwidth]{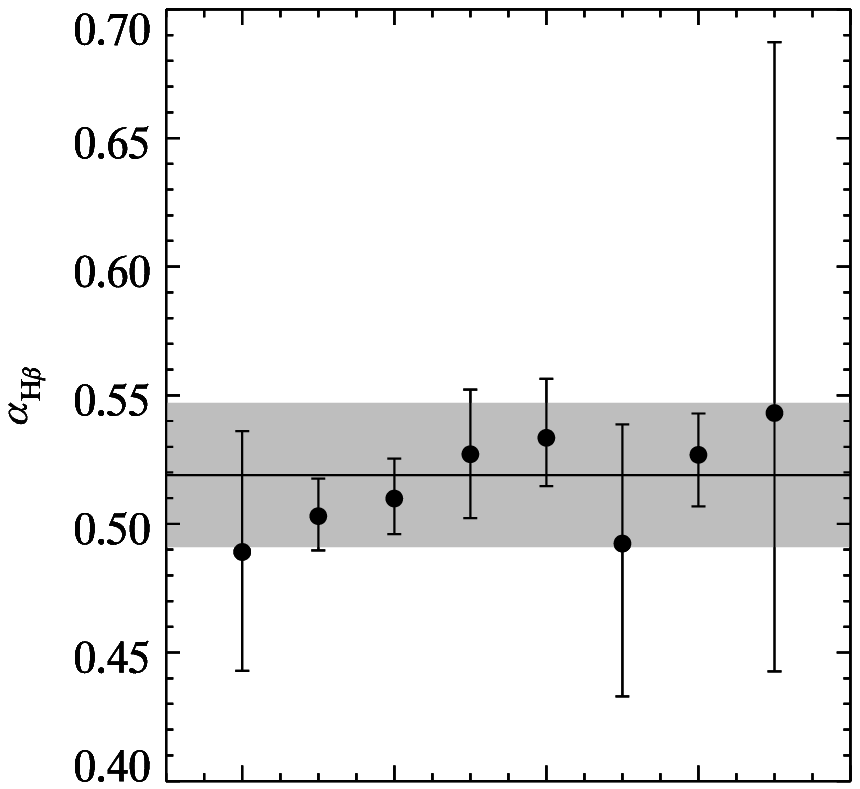}}
\subfloat[H$\beta$: $K$]{\label{fig:alphahb}\includegraphics[width=0.4\textwidth]{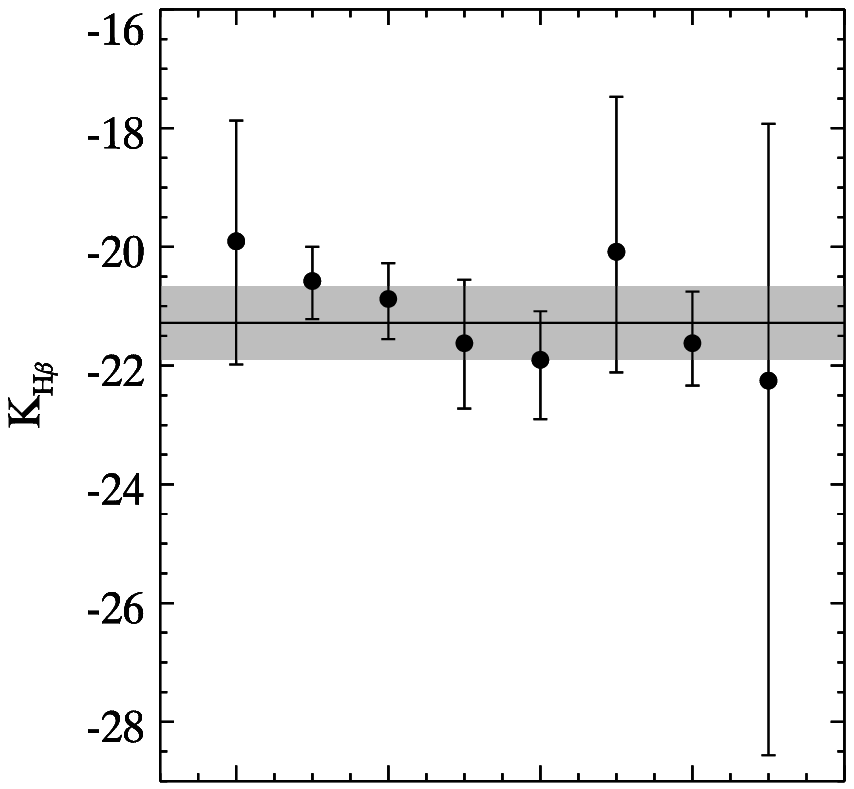}}\\
\subfloat[\mgii: $\alpha$]{\label{fig:alphamgii}\includegraphics[width=0.4\textwidth]{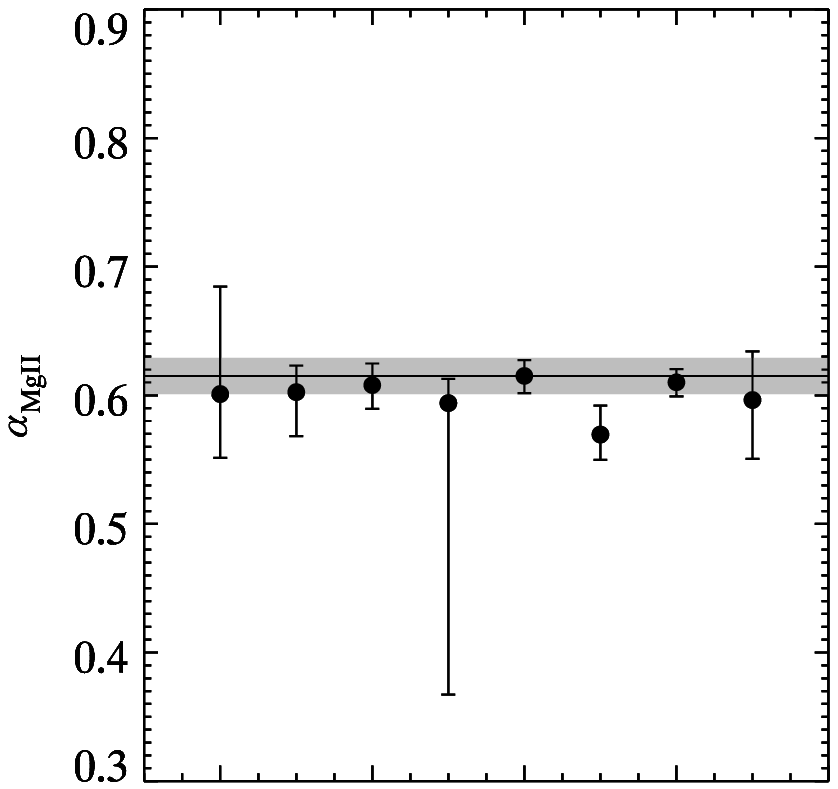}}
\subfloat[\mgii: $K$]{\label{fig:alphamgii}\includegraphics[width=0.4\textwidth]{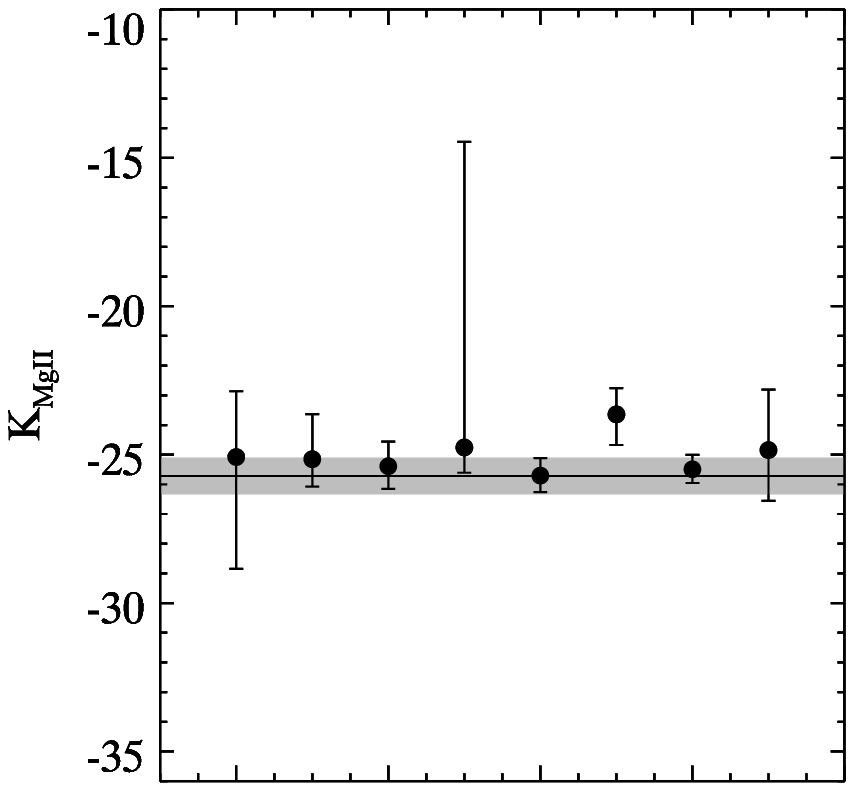}}\\
\subfloat[\civ: $\alpha$]{\label{fig:alphaciv}\includegraphics[width=0.4\textwidth]{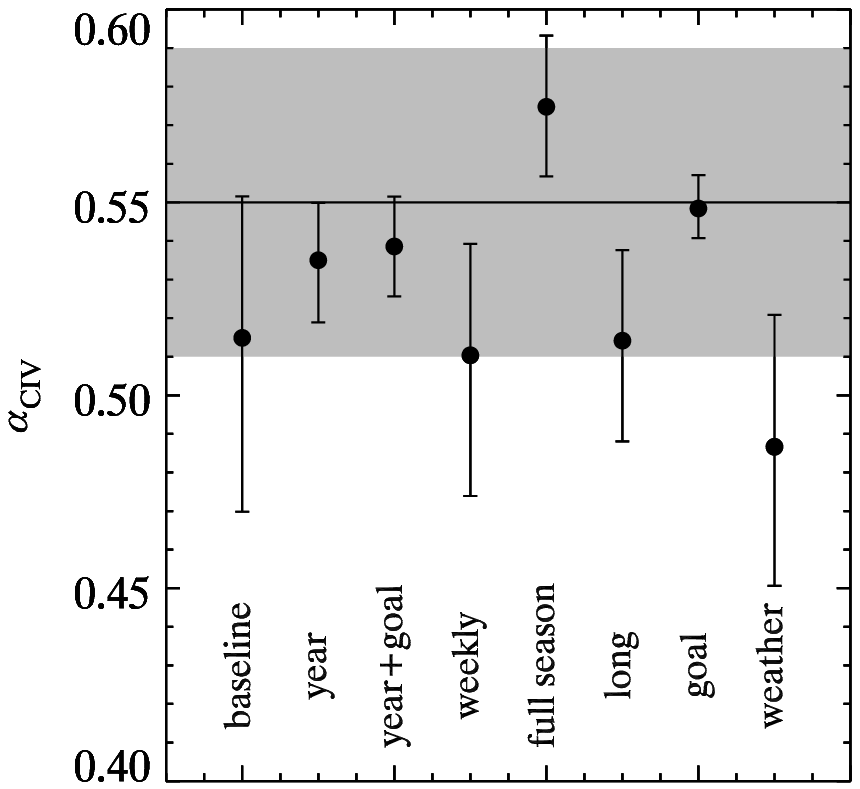}}
\subfloat[\civ: $K$]{\label{fig:alphaciv}\includegraphics[width=0.4\textwidth]{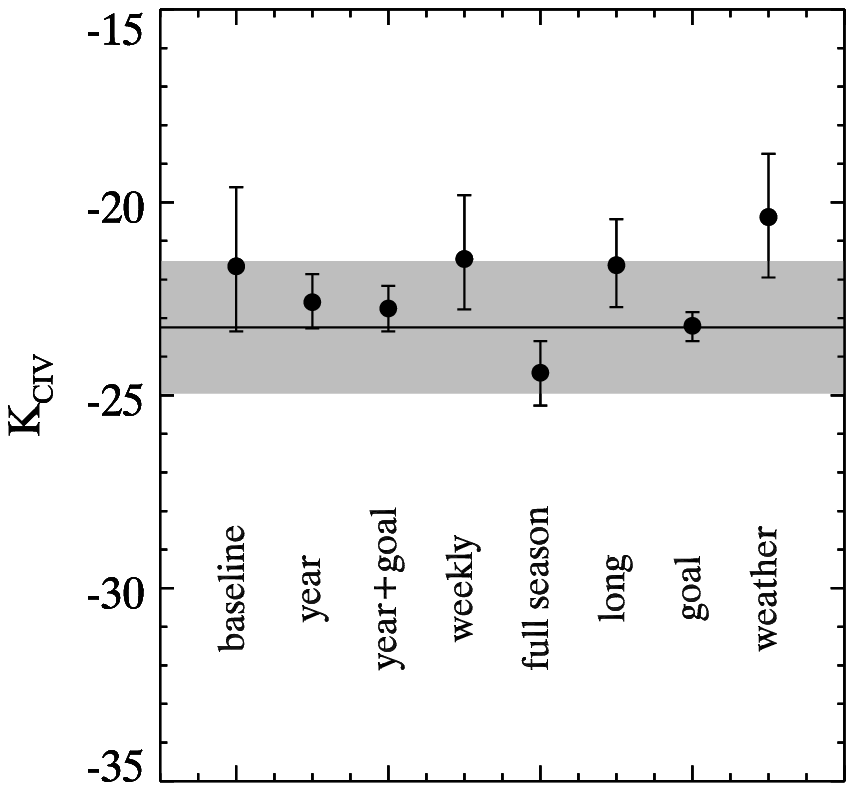}}
\caption{The median recovered gradients ($\alpha$: left) and intercepts ($K$; right) of the $R-L$ relation for the various survey extensions. Results are shown for grade-1 H$\beta$ (top), \mgii\ (middle) and \civ\ (bottom) lines. The horizontal line and grey shaded regions show the input values for $\alpha$ and $K$ and their current uncertainty.}
\label{fig:alpha_ext}
\end{figure*}

\begin{table*}
\footnotesize
\caption{Recovered $R-L$ parameters for the five selection criteria.}
\label{tab:RL}
\begin{center}
\begin{tabular}{lllllr@{}lr@{}l}
\hline
Selection Criteria &Line &$K$ & $\alpha$ & Scatter & Total~  &lags &Accepted~ & lags \\
\hline
None & H$\beta$&$-19.6_{-2.1}^{+1.9}$ &$0.48_{-0.04}^{+0.05}$ &$0.121_{-0.016}^{+0.020}$ &$80$&$\pm8$ &$51\pm6$\\
None & H$\beta$-grade 1&$-21.5_{-1.7}^{+1.3}$ &$0.52_{-0.03}^{+0.04}$ &$0.107_{-0.023}^{+0.036}$ &$80$&$\pm8$ &$28\pm5$\\
None & \mgii&$-25.6_{-3.6}^{+2.1}$ &$0.61_{-0.05}^{+0.08}$ &$0.122_{-0.016}^{+0.019}$ &$212$&$\pm10$ &$87\pm8$\\
None & \mgii-grade 1&$-25.5_{-2.7}^{+2.6}$ &$0.61_{-0.06}^{+0.06}$ &$0.090_{-0.013}^{+0.017}$ &$212$&$\pm10$ &$34\pm5$\\
None & \civ &$-21.1_{-1.9}^{+1.9}$ &$0.50_{-0.04}^{+0.04}$ &$0.129_{-0.014}^{+0.013}$ &$277$&$\pm10$ &$96\pm9$\\
None & \civ-grade 1&$-23.1_{-2.7}^{+2.8}$ &$0.55_{-0.06}^{+0.06}$ &$0.098_{-0.016}^{+0.015}$ &$277$&$\pm10$ &$33\pm5$\\
Acceptance $>$ 40\% &H$\beta$&$-18.3_{-1.2}^{+0.6}$ &$0.45_{-0.01}^{+0.03}$ &$0.127_{-0.010}^{+0.011}$ &$266$&$\pm10$ &$169\pm10$\\
Acceptance $>$ 40\% &H$\beta$- grade 1&$-21.1_{-0.8}^{+0.8}$ &$0.52_{-0.02}^{+0.02}$ &$0.114_{-0.015}^{+0.018}$ &$266$&$\pm10$ &
$95\pm8$\\
Acceptance $>$ 40\% &\mgii&$-26.6_{-7.2}^{+3.3}$ &$0.63_{-0.07}^{+0.16}$ &$0.125_{-0.017}^{+0.020}$ &$145$&$\pm9$ &$64\pm7$\\
Acceptance $>$ 40\% &\mgii-grade 1&$-26.9_{-3.7}^{+4.0}$ &$0.64_{-0.09}^{+0.08}$ &$0.091_{-0.015}^{+0.021}$ &$145$&$\pm9$ &$27\pm5$\\
Acceptance $>$ 40\% &\civ&$-20.9_{-5.5}^{+5.2}$ &$0.50_{-0.11}^{+0.12}$ &$0.131_{-0.015}^{+0.014}$ &$158$&$\pm8$ &$76\pm8$\\
Acceptance $>$ 40\% &\civ-grade 1&$-29.1_{-6.5}^{+5.4}$ &$0.68_{-0.12}^{+0.14}$ &$0.096_{-0.013}^{+0.013}$ &$158$&$\pm8$ &$31\pm5$\\
Acceptance $>$ 50\% &H$\beta$&$-18.2_{-0.7}^{+0.5}$ &$0.45_{-0.01}^{+0.02}$ &$0.128_{-0.008}^{+0.008}$ &$417$&$\pm4$ &$264\pm11$\\
Acceptance $>$ 50\% &H$\beta$- grade 1&$-21.1_{-0.6}^{+0.5}$ &$0.52_{-0.01}^{+0.01}$ &$0.117_{-0.014}^{+0.012}$ &$417$&$\pm4$ &
$148\pm10$\\
Acceptance $>$ 50\% &\mgii&$-26.1_{-6.9}^{+3.2}$ &$0.62_{-0.07}^{+0.15}$ &$0.108_{-0.019}^{+0.028}$ &$91$&$\pm5$ &$37\pm6$\\
Acceptance $>$ 50\% &\mgii-grade 1&$-26.4_{-3.7}^{+4.5}$ &$0.63_{-0.10}^{+0.08}$ &$0.081_{-0.014}^{+0.017}$ &$91$&$\pm5$ &$17\pm5$\\
Acceptance $>$ 50\% &\civ&$-22.0_{-5.7}^{+6.5}$ &$0.52_{-0.15}^{+0.13}$ &$0.155_{-0.027}^{+0.027}$ &$62$&$\pm0$ &$26\pm3$\\
Acceptance $>$ 50\% &\civ-grade 1&$-33.1_{-11.9}^{+8.0}$ &$0.77_{-0.18}^{+0.26}$ &$0.107_{-0.042}^{+0.030}$ &$62$&$\pm0$ &$11\pm2$\\
$\Delta<0.05$ & H$\beta$ &$-18.4_{-2.2}^{+1.6}$ &$0.45_{-0.04}^{+0.05}$ &$0.116_{-0.017}^{+0.016}$ &$75$&$\pm8$ &$49\pm6$\\
$\Delta<0.05$ & H$\beta$ - grade 1 &$-21.3_{-1.6}^{+2.0}$ &$0.52_{-0.04}^{+0.04}$ &$0.113_{-0.022}^{+0.022}$ &$75$&$\pm8$ &$27\pm5$\\
$\Delta<0.05$ & \mgii &$-26.0_{-3.5}^{+1.9}$ &$0.62_{-0.04}^{+0.08}$ &$0.111_{-0.013}^{+0.014}$ &$241$&$\pm11$ &$100\pm10$\\
$\Delta<0.05$ & \mgii-grade 1 &$-26.3_{-2.0}^{+2.1}$ &$0.63_{-0.05}^{+0.04}$ &$0.086_{-0.011}^{+0.011}$ &$241$&$\pm11$ &$41\pm6$\\
$\Delta<0.05$ & \civ &$-20.6_{-2.1}^{+2.2}$ &$0.49_{-0.05}^{+0.05}$ &$0.122_{-0.016}^{+0.020}$ &$253$&$\pm10$ &$78\pm9$\\
$\Delta<0.05$ & \civ-grade 1 &$-22.2_{-2.8}^{+2.7}$ &$0.53_{-0.06}^{+0.06}$ &$0.098_{-0.018}^{+0.017}$ &$253$&$\pm10$ &$28\pm5$\\
$\Delta<0.10$ & H$\beta$ &$-18.8_{-2.0}^{+1.7}$ &$0.46_{-0.04}^{+0.04}$ &$0.114_{-0.018}^{+0.024}$ &$69$&$\pm7$ &$44\pm6$\\
$\Delta<0.10$ & H$\beta$ - grade 1 &$-21.3_{-1.6}^{+1.3}$ &$0.52_{-0.03}^{+0.04}$ &$0.101_{-0.022}^{+0.040}$ &$69$&$\pm7$ &$25\pm5$\\
$\Delta<0.10$ & \mgii &$-25.4_{-4.3}^{+2.0}$ &$0.61_{-0.05}^{+0.10}$ &$0.119_{-0.013}^{+0.016}$ &$212$&$\pm11$ &$87\pm8$\\
$\Delta<0.10$ & \mgii-grade 1 &$-25.8_{-2.5}^{+2.9}$ &$0.62_{-0.07}^{+0.06}$ &$0.092_{-0.014}^{+0.016}$ &$212$&$\pm11$ &$35\pm5$\\
$\Delta<0.10$ & \civ &$-21.2_{-1.8}^{+1.7}$ &$0.50_{-0.04}^{+0.04}$ &$0.130_{-0.014}^{+0.015}$ &$287$&$\pm11$ &$98\pm9$\\
$\Delta<0.10$ & \civ-grade 1 &$-22.6_{-2.9}^{+3.0}$ &$0.54_{-0.06}^{+0.06}$ &$0.098_{-0.015}^{+0.015}$ &$287$&$\pm11$ &$34\pm5$\\
\hline
\end{tabular}
\end{center}
\end{table*}%

\subsection{Extensions}\label{sec:prospectsext}
We also investigated the improvement in both $\sigma_\tau/\tau$ and the recovered $R-L$ parameters associated with the different survey extensions using a random target selection process (equivalent to the `None' selection process in previous section). The results are summarised in Table \ref{tab:RL_ext}. The loss of epochs due to weather had a universally detrimental effect on the recovery of $M_{\rm BH}$ and the $R-L$ parameters. The precision in the black hole mass measurements was most significantly improved by a reduction in the spectral measurement uncertainty (goal), however great improvement was also observed in the case where no seasonal gaps are present (year). The `goal' extension is expected to have a median $M_{\rm BH}$ uncertainty of 0.05 dex, assuming no improvement in the $\Delta {V}$ measurement.  Interestingly, the precision in mass from the `goal' case was found to be superior to the `year+goal' case, though the difference is not highly significant. The difference is likely due to the slightly longer photometric baseline of the baseline OzDES survey from the DES science verification monitoring in 2012. Note that in the previous section we found the accuracy of the lag measurement to be consistent between the `goal' and `year+goal' extensions.

 The $R-L$ relationship parameter constraints for all lines were most improved from a reduction in the measurement uncertainty (goal; Fig. \ref{fig:alpha_ext}). The precision and accuracy in the $H\beta$ and \civ\ $R-L$ parameter constraints were   also significantly  improved in the cases where the seasonal gaps were reduced or removed completely (i.e. year and full season). In the \mgii\ case, the distribution of constraints for the `weekly' extension exhibited a large tail towards a shallower $R-L$ relationship. This is a consequence of the enhanced lag uncertainty created by the seasonal gap. If only grade 1 lags are used in the construction of the $R-L$ relationship, this tail disappears.
 Additionally, the $R-L$ parameter constraints were only marginal improved by extending the survey by an extra two years (long).

 \begin{table*}
 \footnotesize
\caption{Recovered $R-L$ parameters for the survey extensions.  }
\label{tab:RL_ext}
\begin{tabular}{@{}m{1.45cm}m{1.1cm}m{1.1cm}m{1.3cm}m{1.1cm}m{1.1cm}m{1.3cm}m{1.1cm}m{1.1cm}m{1.3cm}m{1.1cm}m{0cm}}
\hline
\footnotesize
Extension&$K_{\rm H\beta}$ & $\alpha_{\rm H\beta}$ & Scatter$_{\rm H\beta}$&$K_{\rm \mgii}$ & $\alpha_{\rm \mgii}$ & Scatter$_{\rm \mgii}$&$K_{\rm \civ}$ & $\alpha_{\rm \civ}$ & Scatter$_{\rm \civ}$& $\sigma_{\rm R}/{\rm R}$& \\[0.5cm]
\hline
Default &$-19.6_{-2.0}^{+1.9}$ &$0.48_{-0.04}^{+0.05}$ &$0.122_{-0.017}^{+0.020}$ &$-25.6_{-3.6}^{+2.2}$ &$0.61_{-0.05}^{+0.08}$ &$0.123_{-0.016}^{+0.018}$ &
 $-21.1_{-1.9}^{+1.9}$ &$0.50_{-0.04}^{+0.04}$ &$0.129_{-0.014}^{+0.013}$  &$0.213_{-0.104}^{+0.127}$&\\[0.5cm]
Weather &$-22.3_{-6.5}^{+4.3}$ &$0.54_{-0.10}^{+0.15}$ &$0.153_{-0.039}^{+0.050}$ & $-25.1_{-1.7}^{+2.2}$ &$0.60_{-0.05}^{+0.04}$ &
$0.108_{-0.013}^{+0.014}$ & $-20.6_{-1.5}^{+1.7}$ &$0.49_{-0.04}^{+0.03}$ &$0.107_{-0.009}^{+0.009}$ &$0.211_{-0.111}^{+0.124}$&\\[0.5cm]
Goal &$-21.5_{-0.8}^{+0.8}$ &$0.53_{-0.02}^{+0.02}$ &$0.065_{-0.025}^{+0.023}$& $-25.5_{-0.5}^{+0.5}$ &$0.61_{-0.01}^{+0.01}$ &
$0.046_{-0.005}^{+0.005}$ & $-23.3_{-0.4}^{+0.4}$ &$0.55_{-0.01}^{+0.01}$ &$0.046_{-0.005}^{+0.005}$ &$0.075_{-0.030}^{+0.063}$&\\[0.5cm]
Long &$-19.7_{-2.2}^{+2.4}$ &$0.48_{-0.05}^{+0.05}$ &$0.167_{-0.039}^{+0.047}$ & $-23.7_{-1.0}^{+1.0}$ &$0.57_{-0.02}^{+0.02}$ &
$0.108_{-0.013}^{+0.015}$ & $-21.8_{-1.1}^{+1.2}$ &$0.52_{-0.03}^{+0.02}$ &$0.097_{-0.008}^{+0.008}$ &$0.165_{-0.073}^{+0.084}$&\\[0.5cm]
Weekly &$-21.6_{-1.1}^{+1.0}$ &$0.53_{-0.02}^{+0.03}$ &$0.092_{-0.021}^{+0.026}$& $-24.5_{-1.0}^{+9.7}$ &$0.59_{-0.21}^{+0.02}$ &
$0.091_{-0.016}^{+0.027}$ & $-20.8_{-1.5}^{+1.7}$ &$0.50_{-0.04}^{+0.03}$ &$0.106_{-0.013}^{+0.015}$&$0.119_{-0.045}^{+0.089}$&\\[0.5cm]
Full Season &$-21.9_{-1.2}^{+0.9}$ &$0.53_{-0.02}^{+0.03}$ &$0.125_{-0.027}^{+0.032}$ & $-25.2_{-0.6}^{+0.7}$ &$0.60_{-0.01}^{+0.01}$ &
$0.115_{-0.049}^{+0.044}$ & $-24.1_{-0.8}^{+0.8}$ &$0.57_{-0.02}^{+0.02}$ &$0.082_{-0.013}^{+0.012}$  &$0.123_{-0.055}^{+0.086}$&\\[0.5cm]
Year &$-20.6_{-0.6}^{+0.6}$ &$0.50_{-0.01}^{+0.01}$ &$0.058_{-0.007}^{+0.008}$ & $-24.9_{-0.9}^{+1.5}$ &$0.60_{-0.03}^{+0.02}$ &
$0.049_{-0.005}^{+0.007}$ & $-22.7_{-0.7}^{+0.7}$ &$0.54_{-0.02}^{+0.02}$ &$0.060_{-0.008}^{+0.008}$ &$0.077_{-0.035}^{+0.057}$&\\[0.5cm]
Year+goal &$-20.9_{-0.7}^{+0.6}$ &$0.51_{-0.01}^{+0.02}$ &$0.073_{-0.012}^{+0.013}$ & $-25.9_{-0.8}^{+0.9}$ &$0.62_{-0.02}^{+0.02}$ &
$0.063_{-0.013}^{+0.046}$ & $-22.9_{-0.5}^{+0.6}$ &$0.54_{-0.01}^{+0.01}$ &$0.046_{-0.005}^{+0.006}$ &$0.076_{-0.034}^{+0.060}$&\\[0.5cm]
\hline
\end{tabular}
Note: In this case, the $H\beta$ sample corresponds to the $H\beta$ - grade 1 sample.
\end{table*}%

\section{Discussion and Conclusions}\label{sec:discussion}
The baseline OzDES survey is expected to observe $\sim500$ AGN for the full five year period, spanning a redshift range of $0<z\lesssim4$ and luminosity range of $10^{39}<\lambda L_{\lambda}<10^{47}$ erg s$^{-1}$. If the final 500 AGN are randomly selected from the target quasar catalogue we expect a lag recovery rate of $\sim35-45\%$ (Table \ref{tab:RL}). This would represent a four-fold increase in the number of measured lags compared to the current sample \citep{Bentz2013}, and a more than ten-fold increase in redshift range. A higher acceptance rate ($\sim60\%$) can be achieved if the final targets are selected based on their expected lag length and their current light curves. 

This sample will enable direct $M_{\rm BH}$ measurements over a broad range of quasar properties, environments and black hole masses. It will also help constrain the $R-L$ relationship for multiple emission lines and test the robustness of this relationship over a broad AGN population, including an investigation into the recently observed Eddington ratio dependence of the $R-L$ relationship discovered by \citep{Du2015}. The current reverberation mapping sample is biased towards local, low luminosity objects, which are not representative of typical quasars \citep{Richards2011}. This had lead to concerns that the existing RM sample may be biased compared to the broader quasar population \citep[e.g.,][]{Shen2008,Richards2011,Denney2012}.  The OzDES sample will probe a similar redshift and luminosity range to the quasar samples in which the secondary mass estimate techniques are applied and thus minimise any potential biases. 

The baseline OzDES reverberation mapping campaign is predicted to accurately recover the $R-L$ relationship for all three lines when only the accepted grade 1 sample is used and a wide distribution of targets are selected (Figure \ref{fig:alpha}). In general, the spread in the recovered $R-L$ parameters is larger than the uncertainties associated with the input relationships. However, the current \mgii\ $R-L$ relationship is not based on direct \mgii\ lag measurements \citep[see][]{Trakhtenbrot2012}, while OzDES will constrain it directly, and the \civ\ $R-L$ relationship is based only on a small number of objects. 

The large spread in the \civ\ $R-L$ relationship parameters is due to the combination of the low acceptance rate, low lag accuracy and relatively short luminosity baseline of the \civ\ sample. However, any of the survey extensions significantly improve the $R-L$ relationship parameter constraints for all lines (Figure \ref{fig:alpha_ext} and Table \ref{tab:RL_ext}). Accurate estimates of all three $R-L$ relationships are crucial for single epoch mass estimates and measuring distances to the highest redshifts. Consequently, there are significant gains from pursuing one or more of the survey extensions we have simulated.

One of the greatest concerns with using AGN as standard candles is whether the $R-L$ relationship evolves in redshift. The OzDES sample may enable an investigation into any trends in the $R-L$ due to redshift, metallicity, Eddington ratio and many other properties.  If the $R-L$ relationship appears to be consistent over the observed sample of quasars, then the OzDES reverberation sample will provide the first physically motivated distance measurement based on a single method from the present day back to redshift four. Unfortunately, the statistical power of the OzDES sample is not expected to be competitive with existing cosmological probes as the predicted uncertainty of the $R-L$ will still be too large to rival the precision in current SNe and BAO measurements. However, it has the potential to uncover unexpected expansion behaviour if large deviations to $\Lambda$CDM are present at high redshifts and will provide a strong base for future surveys. 

The Sloan Digital Sky Survey (SDSS) is currently running a comparable campaign \citep{Shen2014} on a much shorter time scale (6 months in 2014). They are observing 849 quasars in a 7 deg$^2$ field of view using the SDSS-III Baryon Oscillation Spectroscopic Survey (BOSS) spectrograph. Their sample is flux-limited to  $i_{psf}=21.7$ mag (best-fit point-spread function (PSF) magnitudes from SDSS), has $\sim$30 epochs of spectroscopic data over the duration of the survey (an $\sim$4 day cadence), photometric monitoring approximately every 2 days, and includes quasars up to redshift 4.5. They expect to recover lags for 10\% of their sample out to a redshift of 2, with a possible extension to z$\sim$4 with the inclusion of 3 years of photometry obtained over 2011-2013 from the Panoramic Survey Telescope and Rapid Response System 1 Survey \citep[PanSTARRS,][]{Kaiser2010}. 
The OzDES and SDSS reverberation mapping campaigns will produce quite complementary measurements. SDSS has finer temporal sampling and a shorter timeline, enabling the recovery of a broad range of faint AGN with shorter lags, while the DES/OzDES sample will be able to more efficiently recover the brighter and higher redshift AGN with longer lags.

\subsection{Target selection criteria}
Selecting our sample based on the expected accuracy  did not significantly change the predicted $R-L$ parameter constraints, although, when the sample was chosen for a higher recovered fraction we observe a significant tightening in the H$\beta$ constraints and degradation of the \mgii\ and \civ\ constraints (Figure \ref{fig:alpha}). This is primarily due to the relative number of lags used to constrain the $R-L$ relationships.
We find it very advantageous to prioritise targets with multiple lines present in their spectra to calibrate between the H$\beta$ $R-L$ relationship, which is quite tightly constrained by the current RM sample \citep{Bentz2013}, with the $R-L$ relationship of the higher ionisation lines. It is  also important to cover a diverse range in redshift and magnitude. 

\subsection{Extensions}
Of all the survey extensions, improving the measurement uncertainty, through better spectroscopic calibrations, was the most efficient means of improving the overall results. This is because the constraints on each emission line data point are much stronger, leading to less ambiguity in the lags. The goal measurement uncertainty of 3\% is quite optimistic but may be possible to achieve. During the span of the survey so far, major upgrades have been made in both the AAOmega instrument \footnote{http://www.aao.gov.au/science/instruments/current/status} and its pipeline \footnote{http://www.aao.gov.au/science/software/2dfdr}.   Another method of reducing these uncertainties is to increase the number of calibrating F stars monitored in each field. 

In terms of improved scientific results, reducing the measurement uncertainty was closely followed by closing the seasonal gaps, even by 3 months. By extending the observing season beyond the 6-month season of the standard survey we break the `half year degeneracy' and allow common variability features to be probed by both continuum and emission line light curves. However, the improvement in the lag recovery was generally limited to the shorter lags.  This result is consistent with previous findings \citep{Horne2004}, namely that the recovery of longer lags requires a longer baseline of observation rather than finer sampling.

Extending the observation timeline by 2 years (long), only marginally improves the predicted scientific results, despite expectations to the contrary. A longer program does enable longer lags to be recovered (as seen in Figure \ref{fig:wabs_ext}), which allows a broader luminosity baseline from which to constrain both the \mgii\ and \civ\ $R-L$ relationships. However, due to the random selection process we employed and the lower number density of brighter objects, this extension had a trivial influence on the $R-L$ parameter constraints. It is likely that brighter objects will have relatively higher priority in our target selection (Fig. \ref{fig:ozdesdist}), so in the real survey this improvement may be more substantial. The longest lag, highest redshift objects are also best studied with a long program like OzDES, and thus there are clear advantages to monitoring as many of those objects as possible in the hope that an extension of the survey will prove possible. The overall ranking of the extensions is shown in Table \ref{tab:extrank}. The rank is determined from the median value of each criterion and the final ranking is calculated from the overall sum of the other ranking values. The criteria tested were, the number of additional hours of observation required (referred to as resources), recovered fraction, accuracy ($\sigma_{\Delta}$), precision in lag measurement ($\sigma_\tau/\tau$), and the precision and accuracy of $R-L$ parameter recovery. The extensions from best to worst ranking are: Goal, Year, Full Season, Long and Weekly. The final ranking was based on the overall sum of the other ranking values.
\begin{table*}
\footnotesize
\caption{Survey extension rankings.}
\label{tab:extrank}
\begin{tabular}{cccccccc}
\hline
Extension&	Resources&	Recovered &	$\sigma_{\Delta}$&	$\sigma_\tau/\tau$&	$R-L$ Accuracy&	$R-L$ Residual Scatter&	Overall Rank\\
&	(Add. Hours)&	Fraction &	&	&	&	&	\\
\hline
Goal	&1&	1&	1&	1&	1&	1&	1\\
Full season&	2&	3&	2&	4&	3&	3&	3\\
Long	&3&	4&	5&	5&	4&	4&	4\\
Weekly&	5&	5&	4&	3&	5&	5&	5\\
Year	&4&	2&	3&	2&	2&	2&	2\\
\hline
\end{tabular}
\\
The rankings (1-best to 5-worst) are given for the required number of additional hours of observation (resources), recovered fraction, accuracy, $\sigma_\tau/\tau$, and $R-L$ parameter recovery. The rank is determined from the median value of each criterion and the final ranking is calculated from the overall sum of the other ranking values. \\
\end{table*}

Losing an additional 3-5 scheduled epochs of spectroscopic data did not significantly affect the recovered fraction and accuracy of the overall mock catalogue sample, but did severely affect both the precision of black hole mass measurements and $R-L$ relationship parameters. Therefore it is important to minimise the number of epochs lost over the observation period. One way OzDES is working to minimise potential losses is by working in close collaboration with the 2dFLenS survey, also using the 2dF instrument,  to make supplementary observations of the SNe fields when weather restricts OzDES observations. In exchange, OzDES will observe 2dFLenS fields when the DES SNe fields are at high airmass. 
\subsection{Alternative analysis techniques}
On completion, OzDES will be one of the longest running reverberation mapping campaign to date and in number of AGN monitored, it is second only to SDSS. Nevertheless, the expected number of OzDES spectroscopic epochs is small compared to traditional reverberation mapping campaigns \citep[e.g.][]{Peterson2002,Bentz2009,Denney2009,Barth2011}. This leads to the relatively low recovered fraction and accuracy predicted. However, one way to maximise the output of the OzDES data is to stack the lag signals for multiple objects of similar redshifts and magnitudes \citep[`composite reverberation mapping',][]{Fine2012, Brewer2014}. \citet{Fine2012} found that by stacking the continuum and emission-line light curve cross correlation signals of objects with similar redshifts and magnitudes, and therefore similar expected lags, a mean signal is recovered even if no lag signal is present in the individual cross correlations.  The large number of OzDES targets makes it a good sample to perform this type of analysis. 

We will also perform reverberation mapping on other lines in the spectrum beyond H$\beta$ $\lambda$4861, \mgii\ $\lambda$2798, and \civ\ $\lambda$1549. Only these three lines were mentioned in this work as they are the three lines traditionally used for single epoch mass determination, and calibration of the R-L relationship for single epoch masses is one of the main science drivers of this survey.

\subsection{Limitations of survey simulations}
There are several limitations to our survey simulation setup. The first is our continuum luminosity determination for the individual objects. We have simply used the existing SDSS template, and have not taken into account host galaxy contamination or extinction. Our choice of quasar template was based on the similar redshift and magnitude range covered by the SDSS sample and the OzDES target sample. 
The slope of the spectral energy distribution (SED) has been found to vary considerably between individual objects and different samples \citep[e.g][]{Richards2006} so using a single template is a simplification. However, it is sufficient for our use as we are investigating the efficiency of the survey for the bulk of the quasar sample.

Not taking into account host galaxy contamination overestimates the bolometric luminosity of the AGN, resulting in reduced sensitivity in our observations, an overestimation of the lag length, and an underestimation of characteristic variability. However, this is only significant for AGN of similar luminosities to their host, and in general, the expected decrease in lag recovery due to lower sensitivity in light curve variation measurements is  counteracted by the increased variability and shortened lag length of lower luminosity quasars.
  \citet{Shen2014} took this into account in the SDSS reverberation mapping campaign by assigning a constant host galaxy contribution of $8 \times10^{43}$ergs$^{-1}$  at 5100\AA\ and  negligible contribution at $L_{3000\text\AA}$ and $L_{1350\text\AA}$. If we follow the approach of \citet{Richards2006} (with $L_{bol}/L_{Edd} = 0.25$ based on the results of \citet{Kollmeier2006}) to estimate the host galaxy contribution of our sample, we found that only low redshift objects are expected to have a significant host galaxy component and the majority of our targets will not be affected appreciably by neglecting host galaxy light due to their intrinsic luminosities. To confirm this, we tested how the recovered fraction and accuracy of a mock sample were affected by host galaxy contamination and found the results to be consistent with the baseline simulation. However, the inclusion of host galaxy contamination does result in a systematic drop in the expected lag length and again the effect is dominant at low-redshifts.

   Ignoring internal extinction also underestimates the bolometric luminosity. However, this is only expected to be a 20\% luminosity correction based on extinction estimates for the SDSS DR9 quasar sample \citep{Paris2012}, which will not cause significant changes to the simulation results.

\begin{figure}
\centering 
\includegraphics[width=0.45\textwidth]{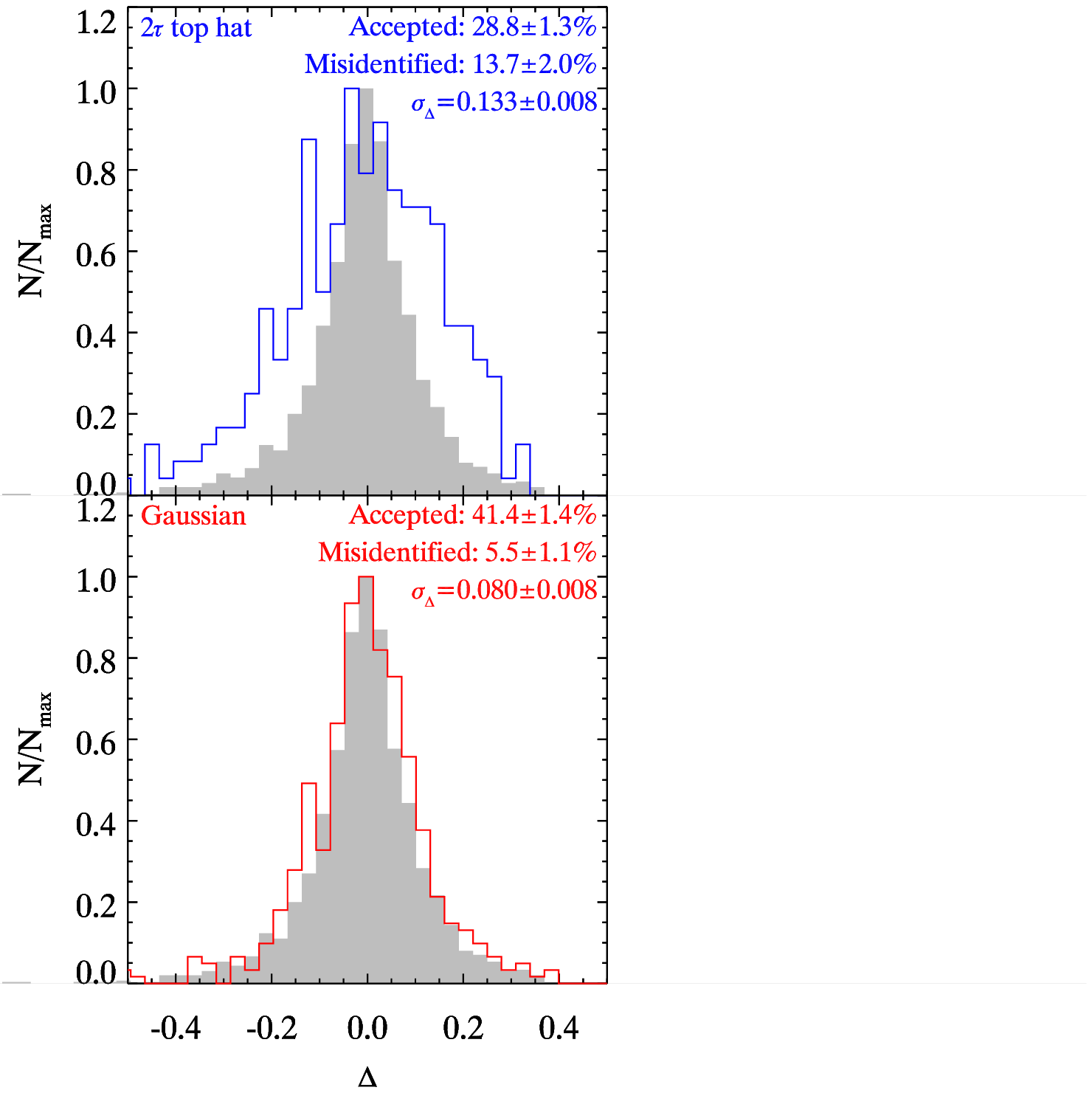}
\caption{The effects of using a very broad top hat ($w=\tau$) or a Gaussian ($\sigma=0.1\tau$) transfer function on the simulated results. The results for the default $0.1\tau$ top hat transfer function distribution is shown in grey. For comparison, the baseline scenario statistics are: Accepted: $39.7\pm0.7\%$, Misidentified : $6.4\pm0.6\%$, and $\sigma_{\Delta}=0.083\pm0.004$.}
\label{fig:transfer}
\end{figure}

Another potential limitation of our simulation is our choice of transfer function. A top hat transfer function is a good conservative choice in this type of investigation, as it spreads the lag response more dramatically than more Gaussian transfer functions. Nonetheless, it could be argued that by choosing a top hat transfer function, when JAVELIN is based on a top hat transfer function, we are biasing our results. Our choice of top hat width may also affect our results. To test both of these issues we consider two alternative transfer functions. First, we considered a Gaussian transfer function, with a width of 0.1$\tau$, motivated by existing reverberation mapping data \citep{Grier2013} and the SDSS survey simulations \citep{Shen2014}.  The recovered fraction and accuracy were consistent or slightly better than the top hat predictions (Fig. \ref{fig:transfer}). We also investigated a worst-case top hat transfer function scenario, where the top hat width is $2\tau$ instead of $0.2\tau$. In this case we do see a significant drop in the recovered fraction and accuracy due to the extreme smoothing of the emission line signal. This case can be considered as the worst-case result for the baseline OzDES survey.  We expect negligible effects from changing the width of the transfer function by a factor of 2-3, based on the findings of \citet{Shen2014}, who found that lag recovery is not significantly affected by changes in this range.

Additionally, the choice of top-hat scaling factor will have a strong effect on the lag recovery. Our choice of unity was based on \citet{Zu2011} findings for NGC5548, which were on the order of one. However, they also found a possible correlation between luminosity and amplitude that we have neglected in our analysis. The choice of amplitude directly relates to the responsivity of the line. As mentioned previously, we theoretically expect different emission lines to respond by different degrees to variations in the ionising continuum \citep{Goad1993,Korista2000,Korista2004} and this behaviour is witnessed observationally with sometimes contradictory trends. \citet{Cackett2015} recently found that the responsivity of \mgii\ was low in NGC5548, and displayed virtually no response to the continuum variability over an period of 170 days, despite a near-UV continuum variability amplitude of  $F_{VAR}=0.33$. This finding agrees with the observations of \citet{Clavel1991} who found a smaller variability amplitude in \mgii\ flux compared to the other UV lines observed. On the other-hand, \citet{Woo2008}  found low \mgii\ responsivity in only one of their objects and reasonable \mgii\ responsivity in the other four objects. Therefore, defining the responsivity for any line or object appears to be complex.  Additionally, some objects have also displayed very non-linear responses to continuum variations (e.g., NGC7469 \citealp{Peterson2014}; NGC5548 \citealp{DeRosa2015} during the second half of the campaign; and J080131 \citealp{Du2015}). Until the mechanisms that drive changes in responsivity are well understood we can only simulate the transfer function in a reasonable fashion, as we have done in this work. Based on our knowledge from previous RM programs and the diversity of the OzDES survey, it is hard to determine whether the OzDES will perform better or worse than  predicted in these simulations.

 Although host galaxy contamination is not expected to significantly affect the recovered fraction or lag accuracy, not accounting for host galaxy contamination will affect the determination of the $R-L$ relation parameters. Host galaxy contamination can be estimated, for the observed OzDES data, using a combination of spectral decomposition \citep{VandenBerk2006}, and high resolution imaging as performed by \citet{Bentz2009,Bentz2013}.

\subsection{Other considerations}
We expect certain sections of the extracted spectra to be suboptimal for flux measurements, in particular the region near the dichroic split of the  AAOmega spectrograph at 570nm and the Fraunhofer A+B absorption bands at 759nm and 687nm. 
We recommend prioritising targets whose emission lines avoid these wavelengths by 2$\times$ the FWHM of the line. This corresponds to approximate redshifts of $z\sim0.17$ and $z\sim0.56$ for H$\beta$, $z\sim1.0$ and $ z\sim 1.7$ for \mgii\, and $z\sim2.7$ and $z\sim 3.9$ for \civ.  Avoiding the Fraunhofer bands is of less concern than the dichroic split.

\subsection{Summary}
We have generated mock catalogues of AGN and corresponding light curves according to the expected OzDES sampling and quasar properties. We attempted to recover the input lag from the simulated light curves to quantify the efficiency and accuracy of the lag recovery. These results were used to predict the expected performance and scientific output of the OzDES reverberation mapping project and several proposed survey extensions.
We expect OzDES to yield lags for  $\sim35-45\%$ of the monitored quasars.  The resulting direct $M_{\rm BH}$ measurements are expected to have formal uncertainties of 0.16-0.21 dex and the baseline OzDES reverberation mapping campaign will accurately recover the $R-L$ relationship parameters for H$\beta$, \mgii, and \civ. However, substantial improvements can be gained if we either increase the spectroscopic cadence, extend the survey season, or improve the spectroscopic measurement accuracy of the survey.  

\section{Acknowledgements}
We thanks Roberto Assef for useful discussions about the design and the robustness of the survey simulations.
AK would like to
acknowledge the support provided by the University of Queensland and Australian Commonwealth Government via the Australian Postgraduate Award.
The Dark Cosmology Centre is funded by the Danish National Research Foundation. 
Parts of this research were conducted by the Australian Research Council Centre of Excellence for All-sky Astrophysics (CAASTRO), through project number CE110001020. TMD acknowledges the support of the Australian Research Council through Future Fellowship, FT100100595. CSK is supported by NSF grant AST-1009756. MV gratefully acknowledges support from the Danish Council for Independent Research via grant no. DFF 4002-00275. Funding for the DES Projects has been provided by the U.S. Department of Energy, the U.S. National Science Foundation, the Ministry of Science and Education of Spain, 
the Science and Technology Facilities Council of the United Kingdom, the Higher Education Funding Council for England, the National Center for Supercomputing 
Applications at the University of Illinois at Urbana-Champaign, the Kavli Institute of Cosmological Physics at the University of Chicago, 
the Center for Cosmology and Astro-Particle Physics at the Ohio State University,
the Mitchell Institute for Fundamental Physics and Astronomy at Texas A\&M University, Financiadora de Estudos e Projetos, 
Funda{\c c}{\~a}o Carlos Chagas Filho de Amparo {\`a} Pesquisa do Estado do Rio de Janeiro, Conselho Nacional de Desenvolvimento Cient{\'i}fico e Tecnol{\'o}gico and 
the Minist{\'e}rio da Ci{\^e}ncia e Tecnologia, the Deutsche Forschungsgemeinschaft and the Collaborating Institutions in the Dark Energy Survey. 

The DES data management system is supported by the National Science Foundation under Grant Number AST-1138766.
The DES participants from Spanish institutions are partially supported by MINECO under grants AYA2012-39559, ESP2013-48274, FPA2013-47986, and Centro de Excelencia Severo Ochoa SEV-2012-0234, 
some of which include ERDF funds from the European Union.

The Collaborating Institutions are Argonne National Laboratory, the University of California at Santa Cruz, the University of Cambridge, Centro de Investigaciones Energeticas, 
Medioambientales y Tecnologicas-Madrid, the University of Chicago, University College London, the DES-Brazil Consortium, the Eidgen{\"o}ssische Technische Hochschule (ETH) Z{\"u}rich, 
Fermi National Accelerator Laboratory, the University of Edinburgh, the University of Illinois at Urbana-Champaign, the Institut de Ciencies de l'Espai (IEEC/CSIC), 
the Institut de Fisica d'Altes Energies, Lawrence Berkeley National Laboratory, the Ludwig-Maximilians Universit{\"a}t and the associated Excellence Cluster Universe, 
the University of Michigan, the National Optical Astronomy Observatory, the University of Nottingham, The Ohio State University, the University of Pennsylvania, the University of Portsmouth, 
SLAC National Accelerator Laboratory, Stanford University, the University of Sussex, and Texas A\&M University.

This paper has gone through internal review by the DES collaboration.

\bibliographystyle{mn2e}
\bibliography{agnbib}

\begin{thebibliography}{129}
\expandafter\ifx\csname natexlab\endcsname\relax\def\natexlab#1{#1}\fi

\bibitem[{{Anderson} {et~al}\mbox{.}(2012){Anderson}, {Aubourg}, {Bailey},
  {Bizyaev}, {Blanton}, {Bolton}, {Brinkmann}, {Brownstein}, {Burden},
  {Cuesta}, {da Costa}, {Dawson}, {de Putter}, {Eisenstein}, {Gunn}, {Guo},
  {Hamilton}, {Harding}, {Ho}, {Honscheid}, {Kazin}, {Kirkby}, {Kneib},
  {Labatie}, {Loomis}, {Lupton}, {Malanushenko}, {Malanushenko}, {Mandelbaum},
  {Manera}, {Maraston}, {McBride}, {Mehta}, {Mena}, {Montesano}, {Muna},
  {Nichol}, {Nuza}, {Olmstead}, {Oravetz}, {Padmanabhan},
  {Palanque-Delabrouille}, {Pan}, {Parejko}, {P{\^a}ris}, {Percival},
  {Petitjean}, {Prada}, {Reid}, {Roe}, {Ross}, {Ross}, {Samushia},
  {S{\'a}nchez}, {Schlegel}, {Schneider}, {Sc{\'o}ccola}, {Seo}, {Sheldon},
  {Simmons}, {Skibba}, {Strauss}, {Swanson}, {Thomas}, {Tinker}, {Tojeiro},
  {Maga{\~n}a}, {Verde}, {Wagner}, {Wake}, {Weaver}, {Weinberg}, {White}, {Xu},
  {Y{\`e}che}, {Zehavi}, \& {Zhao}}]{Anderson2012}
{Anderson} L. {et~al.}, 2012, \mnras, 427, 3435

\bibitem[{{Banerji} {et~al}\mbox{.}(2015){Banerji}, {Jouvel}, {Lin}, {McMahon},
  {Lahav}, {Castander}, {Abdalla}, {Bertin}, {Bosman}, {Carnero}, {Kind}, {da
  Costa}, {Gerdes}, {Gschwend}, {Lima}, {Maia}, {Merson}, {Miller}, {Ogando},
  {Pellegrini}, {Reed}, {Saglia}, {S{\'a}nchez}, {Allam}, {Annis}, {Bernstein},
  {Bernstein}, {Bernstein}, {Capozzi}, {Childress}, {Cunha}, {Davis}, {DePoy},
  {Desai}, {Diehl}, {Doel}, {Findlay}, {Finley}, {Flaugher}, {Frieman},
  {Gaztanaga}, {Glazebrook}, {Gonz{\'a}lez-Fern{\'a}ndez}, {Gonzalez-Solares},
  {Honscheid}, {Irwin}, {Jarvis}, {Kim}, {Koposov}, {Kuehn}, {Kupcu-Yoldas},
  {Lagattuta}, {Lewis}, {Lidman}, {Makler}, {Marriner}, {Marshall}, {Miquel},
  {Mohr}, {Neilsen}, {Peoples}, {Sako}, {Sanchez}, {Scarpine}, {Schindler},
  {Schubnell}, {Sevilla}, {Sharp}, {Soares-Santos}, {Swanson}, {Tarle},
  {Thaler}, {Tucker}, {Uddin}, {Wechsler}, {Wester}, {Yuan}, \&
  {Zuntz}}]{Banerji2015}
{Banerji} M. {et~al.}, 2015, \mnras, 446, 2523

\bibitem[{{Barth} {et~al}\mbox{.}(2011){Barth}, {Pancoast}, {Thorman},
  {Bennert}, {Sand}, {Li}, {Canalizo}, {Filippenko}, {Gates}, {Greene},
  {Malkan}, {Stern}, {Treu}, {Woo}, {Assef}, {Bae}, {Brewer}, {Buehler},
  {Cenko}, {Clubb}, {Cooper}, {Diamond-Stanic}, {Hiner}, {H{\"o}nig}, {Joner},
  {Kandrashoff}, {Laney}, {Lazarova}, {Nierenberg}, {Park}, {Silverman}, {Son},
  {Sonnenfeld}, {Tollerud}, {Walsh}, {Walters}, {da Silva}, {Fumagalli},
  {Gregg}, {Harris}, {Hsiao}, {Lee}, {Lopez}, {Rex}, {Suzuki}, {Trump},
  {Tytler}, {Worseck}, \& {Yesuf}}]{Barth2011}
{Barth} A.~J. {et~al.}, 2011, \apjl, 743, L4

\bibitem[{{Bentz} {et~al}\mbox{.}(2006){Bentz}, {Denney}, {Cackett},
  {Dietrich}, {Fogel}, {Ghosh}, {Horne}, {Kuehn}, {Minezaki}, {Onken},
  {Peterson}, {Pogge}, {Pronik}, {Richstone}, {Sergeev}, {Vestergaard},
  {Walker}, \& {Yoshii}}]{Bentz2006NGC4151}
{Bentz} M.~C. {et~al.}, 2006, \apj, 651, 775

\bibitem[{{Bentz} {et~al}\mbox{.}(2013){Bentz}, {Denney}, {Grier}, {Barth},
  {Peterson}, {Vestergaard}, {Bennert}, {Canalizo}, {De Rosa}, {Filippenko},
  {Gates}, {Greene}, {Li}, {Malkan}, {Pogge}, {Stern}, {Treu}, \&
  {Woo}}]{Bentz2013}
{Bentz} M.~C. {et~al.}, 2013, \apj, 767, 149

\bibitem[{{Bentz} {et~al}\mbox{.}(2009{\natexlab{a}}){Bentz}, {Peterson},
  {Netzer}, {Pogge}, \& {Vestergaard}}]{Bentz2009}
{Bentz} M.~C., {Peterson} B.~M., {Netzer} H., {Pogge} R.~W., {Vestergaard} M.,
  2009{\natexlab{a}}, \apj, 697, 160

\bibitem[{{Bentz} {et~al}\mbox{.}(2008){Bentz}, {Walsh}, {Barth}, {Baliber},
  {Bennert}, {Canalizo}, {Filippenko}, {Ganeshalingam}, {Gates}, {Greene},
  {Hidas}, {Hiner}, {Lee}, {Li}, {Malkan}, {Minezaki}, {Serduke}, {Shiode},
  {Silverman}, {Steele}, {Stern}, {Street}, {Thornton}, {Treu}, {Wang}, {Woo},
  \& {Yoshii}}]{Bentz2008}
{Bentz} M.~C. {et~al.}, 2008, \apjl, 689, L21

\bibitem[{{Bentz} {et~al}\mbox{.}(2009{\natexlab{b}}){Bentz}, {Walsh}, {Barth},
  {Baliber}, {Bennert}, {Canalizo}, {Filippenko}, {Ganeshalingam}, {Gates},
  {Greene}, {Hidas}, {Hiner}, {Lee}, {Li}, {Malkan}, {Minezaki}, {Sakata},
  {Serduke}, {Silverman}, {Steele}, {Stern}, {Street}, {Thornton}, {Treu},
  {Wang}, {Woo}, \& {Yoshii}}]{Bentz2009LAMP}
{Bentz} M.~C. {et~al.}, 2009{\natexlab{b}}, \apj, 705, 199

\bibitem[{{Blake} {et~al}\mbox{.}(2011{\natexlab{a}}){Blake}, {Davis}, {Poole},
  {Parkinson}, {Brough}, {Colless}, {Contreras}, {Couch}, {Croom},
  {Drinkwater}, {Forster}, {Gilbank}, {Gladders}, {Glazebrook}, {Jelliffe},
  {Jurek}, {Li}, {Madore}, {Martin}, {Pimbblet}, {Pracy}, {Sharp}, {Wisnioski},
  {Woods}, {Wyder}, \& {Yee}}]{Blake2011cos}
{Blake} C. {et~al.}, 2011{\natexlab{a}}, \mnras, 415, 2892

\bibitem[{{Blake} {et~al}\mbox{.}(2011{\natexlab{b}}){Blake}, {Kazin},
  {Beutler}, {Davis}, {Parkinson}, {Brough}, {Colless}, {Contreras}, {Couch},
  {Croom}, {Croton}, {Drinkwater}, {Forster}, {Gilbank}, {Gladders},
  {Glazebrook}, {Jelliffe}, {Jurek}, {Li}, {Madore}, {Martin}, {Pimbblet},
  {Poole}, {Pracy}, {Sharp}, {Wisnioski}, {Woods}, {Wyder}, \&
  {Yee}}]{Blake2011distz}
{Blake} C. {et~al.}, 2011{\natexlab{b}}, \mnras, 418, 1707

\bibitem[{{Blandford} \& {McKee}(1982)}]{Blandford1982}
{Blandford} R.~D., {McKee} C.~F., 1982, \apj, 255, 419

\bibitem[{{Brewer} \& {Elliott}(2014)}]{Brewer2014}
{Brewer} B.~J., {Elliott} T.~M., 2014, \mnras, 439, L31

\bibitem[{{Cackett} {et~al}\mbox{.}(2015){Cackett}, {Gultekin}, {Bentz},
  {Fausnaugh}, {Peterson}, \& {Troyer}}]{Cackett2015}
{Cackett} E.~M., {Gultekin} K., {Bentz} M.~C., {Fausnaugh} M.~M., {Peterson}
  B.~M., {Troyer} J., 2015, preprint (arXiv:1503.02029)

\bibitem[{{Clavel} {et~al}\mbox{.}(1991){Clavel}, {Reichert}, {Alloin},
  {Crenshaw}, {Kriss}, {Krolik}, {Malkan}, {Netzer}, {Peterson}, {Wamsteker},
  {Altamore}, {Baribaud}, {Barr}, {Beck}, {Binette}, {Bromage}, {Brosch},
  {Diaz}, {Filippenko}, {Fricke}, {Gaskell}, {Giommi}, {Glass}, {Gondhalekar},
  {Hackney}, {Halpern}, {Hutter}, {Joersaeter}, {Kinney}, {Kollatschny},
  {Koratkar}, {Korista}, {Laor}, {Lasota}, {Leibowitz}, {Maoz}, {Martin},
  {Mazeh}, {Meurs}, {Nair}, {O'Brien}, {Pelat}, {Perez}, {Perola}, {Ptak},
  {Rodriguez-Pascual}, {Rosenblatt}, {Sadun}, {Santos-Lleo}, {Shaw}, {Smith},
  {Stirpe}, {Stoner}, {Sun}, {Ulrich}, {van Groningen}, \&
  {Zheng}}]{Clavel1991}
{Clavel} J. {et~al.}, 1991, \apj, 366, 64

\bibitem[{{Collier} {et~al}\mbox{.}(1998){Collier}, {Horne}, {Kaspi}, {Netzer},
  {Peterson}, {Wanders}, {Alexander}, {Bertram}, {Comastri}, {Gaskell},
  {Malkov}, {Maoz}, {Mignoli}, {Pogge}, {Pronik}, {Sergeev}, {Snedden},
  {Stirpe}, {Bochkarev}, {Burenkov}, {Shapovalova}, \& {Wagner}}]{Collier1998}
{Collier} S.~J. {et~al.}, 1998, \apj, 500, 162

\bibitem[{{Conley} {et~al}\mbox{.}(2011){Conley}, {Guy}, {Sullivan},
  {Regnault}, {Astier}, {Balland}, {Basa}, {Carlberg}, {Fouchez}, {Hardin},
  {Hook}, {Howell}, {Pain}, {Palanque-Delabrouille}, {Perrett}, {Pritchet},
  {Rich}, {Ruhlmann-Kleider}, {Balam}, {Baumont}, {Ellis}, {Fabbro},
  {Fakhouri}, {Fourmanoit}, {Gonz{\'a}lez-Gait{\'a}n}, {Graham}, {Hudson},
  {Hsiao}, {Kronborg}, {Lidman}, {Mourao}, {Neill}, {Perlmutter}, {Ripoche},
  {Suzuki}, \& {Walker}}]{Conley2011}
{Conley} A. {et~al.}, 2011, \apjs, 192, 1

\bibitem[{{Czerny} {et~al}\mbox{.}(2013){Czerny}, {Hryniewicz}, {Maity},
  {Schwarzenberg-Czerny}, {{\.Z}ycki}, \& {Bilicki}}]{Czerny2013}
{Czerny} B., {Hryniewicz} K., {Maity} I., {Schwarzenberg-Czerny} A.,
  {{\.Z}ycki} P.~T., {Bilicki} M., 2013, \aap, 556, A97

\bibitem[{{Davidson}(1972)}]{Davidson1972}
{Davidson} K., 1972, \apj, 171, 213

\bibitem[{{De Rosa} {et~al}\mbox{.}(2015){De Rosa}, {Peterson}, {Ely}, {Kriss},
  {Crenshaw}, {Horne}, {Korista}, {Netzer}, {Pogge}, {Ar{\'e}valo}, {Barth},
  {Bentz}, {Brandt}, {Breeveld}, {Brewer}, {Dalla Bont{\`a}}, {De
  Lorenzo-C{\'a}ceres}, {Denney}, {Dietrich}, {Edelson}, {Evans}, {Fausnaugh},
  {Gehrels}, {Gelbord}, {Goad}, {Grier}, {Grupe}, {Hall}, {Kaastra}, {Kelly},
  {Kennea}, {Kochanek}, {Lira}, {Mathur}, {McHardy}, {Nousek}, {Pancoast},
  {Papadakis}, {Pei}, {Schimoia}, {Siegel}, {Starkey}, {Treu}, {Uttley},
  {Vaughan}, {Vestergaard}, {Villforth}, {Yan}, {Young}, \& {Zu}}]{DeRosa2015}
{De Rosa} G. {et~al.}, 2015, \apj, 806, 128

\bibitem[{{Denney}(2012)}]{Denney2012}
{Denney} K.~D., 2012, \apj, 759, 44

\bibitem[{{Denney} {et~al}\mbox{.}(2009){Denney}, {Peterson}, {Dietrich},
  {Vestergaard}, \& {Bentz}}]{Denney2009}
{Denney} K.~D., {Peterson} B.~M., {Dietrich} M., {Vestergaard} M., {Bentz}
  M.~C., 2009, \apj, 692, 246

\bibitem[{{Denney} {et~al}\mbox{.}(2010){Denney}, {Peterson}, {Pogge}, {Adair},
  {Atlee}, {Au-Yong}, {Bentz}, {Bird}, {Brokofsky}, {Chisholm}, {Comins},
  {Dietrich}, {Doroshenko}, {Eastman}, {Efimov}, {Ewald}, {Ferbey}, {Gaskell},
  {Hedrick}, {Jackson}, {Klimanov}, {Klimek}, {Kruse}, {Lad{\'e}route}, {Lamb},
  {Leighly}, {Minezaki}, {Nazarov}, {Onken}, {Petersen}, {Peterson},
  {Poindexter}, {Sakata}, {Schlesinger}, {Sergeev}, {Skolski}, {Stieglitz},
  {Tobin}, {Unterborn}, {Vestergaard}, {Watkins}, {Watson}, \&
  {Yoshii}}]{Denney2010}
{Denney} K.~D. {et~al.}, 2010, \apj, 721, 715

\bibitem[{{Di Matteo} {et~al}\mbox{.}(2008){Di Matteo}, {Colberg}, {Springel},
  {Hernquist}, \& {Sijacki}}]{DiMatteo2008}
{Di Matteo} T., {Colberg} J., {Springel} V., {Hernquist} L., {Sijacki} D.,
  2008, \apj, 676, 33

\bibitem[{{Di Matteo}, {Springel} \& {Hernquist}(2005){Di Matteo}, {Springel},
  \& {Hernquist}}]{DiMatteo2005}
{Di Matteo} T., {Springel} V., {Hernquist} L., 2005, \nat, 433, 604

\bibitem[{{Dietrich} \& {Kollatschny}(1995)}]{Dietrich1995}
{Dietrich} M., {Kollatschny} W., 1995, \aap, 303, 405

\bibitem[{{Du} {et~al}\mbox{.}(2015){Du}, {Hu}, {Lu}, {Huang}, {Cheng}, {Qiu},
  {Li}, {Zhang}, {Fan}, {Bai}, {Bian}, {Yuan}, {Kaspi}, {Ho}, {Netzer}, {Wang},
  \& {SEAMBH Collaboration}}]{Du2015}
{Du} P. {et~al.}, 2015, \apj, 806, 22

\bibitem[{{Du} {et~al}\mbox{.}(2014){Du}, {Hu}, {Lu}, {Wang}, {Qiu}, {Li},
  {Bai}, {Kaspi}, {Netzer}, {Wang}, \& {SEAMBH Collaboration}}]{Du2014}
{Du} P. {et~al.}, 2014, \apj, 782, 45

\bibitem[{{Edelson} \& {Krolik}(1988)}]{Edelson1988}
{Edelson} R.~A., {Krolik} J.~H., 1988, \apj, 333, 646

\bibitem[{{Ferrarese} \& {Ford}(2005)}]{Ferrarese2005}
{Ferrarese} L., {Ford} H., 2005, \ssr, 116, 523

\bibitem[{{Ferrarese} \& {Merritt}(2000)}]{Ferrarese2000}
{Ferrarese} L., {Merritt} D., 2000, \apjl, 539, L9

\bibitem[{Fine {et~al}\mbox{.}(2010)Fine, Croom, Bland-Hawthorn, Pimbblet,
  Ross, Schneider, \& Shanks}]{Fine2010}
Fine S., Croom S.~M., Bland-Hawthorn J., Pimbblet K.~A., Ross N.~P., Schneider
  D.~P., Shanks T., 2010, Monthly Notices of the Royal Astronomical Society,
  409, 591

\bibitem[{{Fine} {et~al}\mbox{.}(2012){Fine}, {Shanks}, {Croom}, {Green},
  {Kelly}, {Berger}, {Chornock}, {Burgett}, {Magnier}, \& {Price}}]{Fine2012}
{Fine} S. {et~al.}, 2012, \mnras, 427, 2701

\bibitem[{{Flaugher} {et~al}\mbox{.}(2010){Flaugher}, {Abbott}, {Annis},
  {Antonik}, {Bailey}, {Ballester}, {Bernstein}, {Bernstein}, {Bonati},
  {Bremer}, {Briones}, {Brooks}, {Buckley-Geer}, {Campa}, {Cardiel-Sas},
  {Castander}, {Castilla}, {Cease}, {Chappa}, {Chi}, {da Costa}, {DePoy},
  {Derylo}, {De Vicente}, {Diehl}, {Doel}, {Estrada}, {Eiting}, {Elliott},
  {Finley}, {Frieman}, {Gaztanaga}, {Gerdes}, {Gladders}, {Guarino},
  {Gutierrez}, {Grudzinski}, {Hanlon}, {Hao}, {Holland}, {Honscheid},
  {Huffman}, {Jackson}, {Karliner}, {Kau}, {Kent}, {Krempetz}, {Krider},
  {Kozlovsky}, {Kubik}, {Kuehn}, {Kuhlmann}, {Kuk}, {Lahav}, {Lewis}, {Lin},
  {Lorenzon}, {Marshall}, {Mart{\'{\i}}nez}, {McKay}, {Merritt}, {Meyer},
  {Miquel}, {Morgan}, {Moore}, {Moore}, {Nord}, {Ogando}, {Olsen}, {Peoples},
  {Plazas}, {Roe}, {Roodman}, {Rossetto}, {Sanchez}, {Scarpine}, {Schalk},
  {Schindler}, {Schmidt}, {Schmitt}, {Schubnell}, {Schultz}, {Selen},
  {Serrano}, {Shaw}, {Simaitis}, {Slaughter}, {Smith}, {Spinka}, {Stefanik},
  {Stuermer}, {Sypniewski}, {Talaga}, {Tarle}, {Thaler}, {Tucker}, {Walker},
  {Weaverdyck}, {Wester}, {Woods}, {Worswick}, \& {Zhao}}]{Flaugher2010}
{Flaugher} B.~L. {et~al.}, 2010, in Society of Photo-Optical Instrumentation
  Engineers (SPIE) Conference Series, Vol. 7735, Society of Photo-Optical
  Instrumentation Engineers (SPIE) Conference Series, p.~0

\bibitem[{{Gaskell} \& {Sparke}(1986)}]{Gaskell1986}
{Gaskell} C.~M., {Sparke} L.~S., 1986, \apj, 305, 175

\bibitem[{{Gebhardt} {et~al}\mbox{.}(2000){Gebhardt}, {Bender}, {Bower},
  {Dressler}, {Faber}, {Filippenko}, {Green}, {Grillmair}, {Ho}, {Kormendy},
  {Lauer}, {Magorrian}, {Pinkney}, {Richstone}, \& {Tremaine}}]{Gebhardt2000}
{Gebhardt} K. {et~al.}, 2000, \apjl, 539, L13

\bibitem[{{Goad}, {O'Brien} \& {Gondhalekar}(1993){Goad}, {O'Brien}, \&
  {Gondhalekar}}]{Goad1993}
{Goad} M.~R., {O'Brien} P.~T., {Gondhalekar} P.~M., 1993, \mnras, 263, 149

\bibitem[{{Graham} {et~al}\mbox{.}(2001){Graham}, {Erwin}, {Caon}, \&
  {Trujillo}}]{Graham2001}
{Graham} A.~W., {Erwin} P., {Caon} N., {Trujillo} I., 2001, \apjl, 563, L11

\bibitem[{{Grier} {et~al}\mbox{.}(2013){Grier}, {Peterson}, {Horne}, {Bentz},
  {Pogge}, {Denney}, {De Rosa}, {Martini}, {Kochanek}, {Zu}, {Shappee},
  {Siverd}, {Beatty}, {Sergeev}, {Kaspi}, {Araya Salvo}, {Bird}, {Bord},
  {Borman}, {Che}, {Chen}, {Cohen}, {Dietrich}, {Doroshenko}, {Efimov}, {Free},
  {Ginsburg}, {Henderson}, {King}, {Mogren}, {Molina}, {Mosquera}, {Nazarov},
  {Okhmat}, {Pejcha}, {Rafter}, {Shields}, {Skowron}, {Szczygiel}, {Valluri},
  \& {van Saders}}]{Grier2013}
{Grier} C.~J. {et~al.}, 2013, \apj, 764, 47

\bibitem[{{Grier} {et~al}\mbox{.}(2012){Grier}, {Peterson}, {Pogge}, {Denney},
  {Bentz}, {Martini}, {Sergeev}, {Kaspi}, {Minezaki}, {Zu}, {Kochanek},
  {Siverd}, {Shappee}, {Stanek}, {Araya Salvo}, {Beatty}, {Bird}, {Bord},
  {Borman}, {Che}, {Chen}, {Cohen}, {Dietrich}, {Doroshenko}, {Drake},
  {Efimov}, {Free}, {Ginsburg}, {Henderson}, {King}, {Koshida}, {Mogren},
  {Molina}, {Mosquera}, {Nazarov}, {Okhmat}, {Pejcha}, {Rafter}, {Shields},
  {Skowron}, {Szczygiel}, {Valluri}, \& {van Saders}}]{Grier2012}
{Grier} C.~J. {et~al.}, 2012, \apj, 755, 60

\bibitem[{{Hinshaw} {et~al}\mbox{.}(2013){Hinshaw}, {Larson}, {Komatsu},
  {Spergel}, {Bennett}, {Dunkley}, {Nolta}, {Halpern}, {Hill}, {Odegard},
  {Page}, {Smith}, {Weiland}, {Gold}, {Jarosik}, {Kogut}, {Limon}, {Meyer},
  {Tucker}, {Wollack}, \& {Wright}}]{Hinshaw2012}
{Hinshaw} G. {et~al.}, 2013, \apjs, 208, 19

\bibitem[{{Honscheid}, {DePoy} \& {for the DES Collaboration}(2008){Honscheid},
  {DePoy}, \& {for the DES Collaboration}}]{Honscheid2008}
{Honscheid} K., {DePoy} D.~L., {for the DES Collaboration}, 2008, preprint
  (arXiv:0810.3600)

\bibitem[{{Hopkins} {et~al}\mbox{.}(2013){Hopkins}, {Driver}, {Brough},
  {Owers}, {Bauer}, {Gunawardhana}, {Cluver}, {Colless}, {Foster},
  {Lara-L{\'o}pez}, {Roseboom}, {Sharp}, {Steele}, {Thomas}, {Baldry}, {Brown},
  {Liske}, {Norberg}, {Robotham}, {Bamford}, {Bland-Hawthorn}, {Drinkwater},
  {Loveday}, {Meyer}, {Peacock}, {Tuffs}, {Agius}, {Alpaslan}, {Andrae},
  {Cameron}, {Cole}, {Ching}, {Christodoulou}, {Conselice}, {Croom}, {Cross},
  {De Propris}, {Delhaize}, {Dunne}, {Eales}, {Ellis}, {Frenk}, {Graham},
  {Grootes}, {H{\"a}u{\ss}ler}, {Heymans}, {Hill}, {Hoyle}, {Hudson}, {Jarvis},
  {Johansson}, {Jones}, {van Kampen}, {Kelvin}, {Kuijken},
  {L{\'o}pez-S{\'a}nchez}, {Maddox}, {Madore}, {Maraston}, {McNaught-Roberts},
  {Nichol}, {Oliver}, {Parkinson}, {Penny}, {Phillipps}, {Pimbblet}, {Ponman},
  {Popescu}, {Prescott}, {Proctor}, {Sadler}, {Sansom}, {Seibert},
  {Staveley-Smith}, {Sutherland}, {Taylor}, {Van Waerbeke}, {V{\'a}zquez-Mata},
  {Warren}, {Wijesinghe}, {Wild}, \& {Wilkins}}]{Hopkins2013}
{Hopkins} A.~M. {et~al.}, 2013, \mnras, 430, 2047

\bibitem[{{Horne} {et~al}\mbox{.}(2004){Horne}, {Peterson}, {Collier}, \&
  {Netzer}}]{Horne2004}
{Horne} K., {Peterson} B.~M., {Collier} S.~J., {Netzer} H., 2004, \pasp, 116,
  465

\bibitem[{{Jensen, Jens J.}(2012)}]{Jensen2012}
{Jensen, Jens J.}, 2012, Master's thesis, Dark Cosmology Centre, University of
  Copenhagen

\bibitem[{{Kaiser} {et~al}\mbox{.}(2010){Kaiser}, {Burgett}, {Chambers},
  {Denneau}, {Heasley}, {Jedicke}, {Magnier}, {Morgan}, {Onaka}, \&
  {Tonry}}]{Kaiser2010}
{Kaiser} N. {et~al.}, 2010, in Society of Photo-Optical Instrumentation
  Engineers (SPIE) Conference Series, Vol. 7733, Society of Photo-Optical
  Instrumentation Engineers (SPIE) Conference Series, p.~0

\bibitem[{{Kaspi} {et~al}\mbox{.}(2007){Kaspi}, {Brandt}, {Maoz}, {Netzer},
  {Schneider}, \& {Shemmer}}]{Kaspi2007}
{Kaspi} S., {Brandt} W.~N., {Maoz} D., {Netzer} H., {Schneider} D.~P.,
  {Shemmer} O., 2007, \apj, 659, 997

\bibitem[{{Kaspi} {et~al}\mbox{.}(2000){Kaspi}, {Smith}, {Netzer}, {Maoz},
  {Jannuzi}, \& {Giveon}}]{Kaspi2000}
{Kaspi} S., {Smith} P.~S., {Netzer} H., {Maoz} D., {Jannuzi} B.~T., {Giveon}
  U., 2000, \apj, 533, 631

\bibitem[{{Kelly}, {Bechtold} \& {Siemiginowska}(2009){Kelly}, {Bechtold}, \&
  {Siemiginowska}}]{Kelly2009}
{Kelly} B.~C., {Bechtold} J., {Siemiginowska} A., 2009, \apj, 698, 895

\bibitem[{{Kelly} \& {Shen}(2013)}]{Kelly2013}
{Kelly} B.~C., {Shen} Y., 2013, \apj, 764, 45

\bibitem[{{Kilerci Eser} {et~al}\mbox{.}(2015){Kilerci Eser}, {Vestergaard},
  {Peterson}, {Denney}, \& {Bentz}}]{KilerciEser2014}
{Kilerci Eser} E., {Vestergaard} M., {Peterson} B.~M., {Denney} K.~D., {Bentz}
  M.~C., 2015, \apj, 801, 8

\bibitem[{{King}(2003)}]{King2003}
{King} A., 2003, \apjl, 596, L27

\bibitem[{{King}(2005)}]{King2005}
{King} A., 2005, \apjl, 635, L121

\bibitem[{{King} {et~al}\mbox{.}(2014){King}, {Davis}, {Denney}, {Vestergaard},
  \& {Watson}}]{King2014}
{King} A.~L., {Davis} T.~M., {Denney} K.~D., {Vestergaard} M., {Watson} D.,
  2014, \mnras, 441, 3454

\bibitem[{{Kollmeier} {et~al}\mbox{.}(2006){Kollmeier}, {Onken}, {Kochanek},
  {Gould}, {Weinberg}, {Dietrich}, {Cool}, {Dey}, {Eisenstein}, {Jannuzi}, {Le
  Floc'h}, \& {Stern}}]{Kollmeier2006}
{Kollmeier} J.~A. {et~al.}, 2006, \apj, 648, 128

\bibitem[{{Koratkar} \& {Gaskell}(1991)}]{Koratkar1991}
{Koratkar} A.~P., {Gaskell} C.~M., 1991, \apjl, 370, L61

\bibitem[{{Korista} \& {Goad}(2000)}]{Korista2000}
{Korista} K.~T., {Goad} M.~R., 2000, \apj, 536, 284

\bibitem[{{Korista} \& {Goad}(2004)}]{Korista2004}
{Korista} K.~T., {Goad} M.~R., 2004, \apj, 606, 749

\bibitem[{{Kormendy} \& {Gebhardt}(2001)}]{Kormendy2001}
{Kormendy} J., {Gebhardt} K., 2001, in American Institute of Physics Conference
  Series, Vol. 586, 20th Texas Symposium on relativistic astrophysics,
  {Wheeler} J.~C., {Martel} H., eds., pp. 363--381

\bibitem[{{Kormendy} \& {Richstone}(1995)}]{Kormendy1995}
{Kormendy} J., {Richstone} D., 1995, \araa, 33, 581

\bibitem[{{Koz{\l}owski} {et~al}\mbox{.}(2010){Koz{\l}owski}, {Kochanek},
  {Udalski}, {Wyrzykowski}, {Soszy{\'n}ski}, {Szyma{\'n}ski}, {Kubiak},
  {Pietrzy{\'n}ski}, {Szewczyk}, {Ulaczyk}, {Poleski}, \& {OGLE
  Collaboration}}]{Kozlowski2010}
{Koz{\l}owski} S. {et~al.}, 2010, \apj, 708, 927

\bibitem[{{Krawczyk} {et~al}\mbox{.}(2013){Krawczyk}, {Richards}, {Mehta},
  {Vogeley}, {Gallagher}, {Leighly}, {Ross}, \& {Schneider}}]{Krawczyk2013}
{Krawczyk} C.~M., {Richards} G.~T., {Mehta} S.~S., {Vogeley} M.~S., {Gallagher}
  S.~C., {Leighly} K.~M., {Ross} N.~P., {Schneider} D.~P., 2013, \apjs, 206, 4

\bibitem[{{Krolik} \& {McKee}(1978)}]{Krolik1978}
{Krolik} J.~H., {McKee} C.~F., 1978, \apjs, 37, 459

\bibitem[{{Laor}(1998)}]{Laor1998}
{Laor} A., 1998, \apjl, 505, L83

\bibitem[{{MacLeod} {et~al}\mbox{.}(2010){MacLeod}, {Ivezi{\'c}}, {Kochanek},
  {Koz{\l}owski}, {Kelly}, {Bullock}, {Kimball}, {Sesar}, {Westman}, {Brooks},
  {Gibson}, {Becker}, \& {de Vries}}]{Macleod2010}
{MacLeod} C.~L. {et~al.}, 2010, \apj, 721, 1014

\bibitem[{{Marconi} \& {Hunt}(2003)}]{Marconi2003}
{Marconi} A., {Hunt} L.~K., 2003, \apjl, 589, L21

\bibitem[{{McConnell} \& {Ma}(2013)}]{McConnell2013}
{McConnell} N.~J., {Ma} C.-P., 2013, \apj, 764, 184

\bibitem[{{McLure} \& {Jarvis}(2002)}]{McLure2002}
{McLure} R.~J., {Jarvis} M.~J., 2002, \mnras, 337, 109

\bibitem[{{McMahon} {et~al}\mbox{.}(2013){McMahon}, {Banerji}, {Gonzalez},
  {Koposov}, {Bejar}, {Lodieu}, {Rebolo}, \& {VHS Collaboration}}]{McMahon2013}
{McMahon} R.~G., {Banerji} M., {Gonzalez} E., {Koposov} S.~E., {Bejar} V.~J.,
  {Lodieu} N., {Rebolo} R., {VHS Collaboration}, 2013, The Messenger, 154, 35

\bibitem[{{Metzroth}, {Onken} \& {Peterson}(2006){Metzroth}, {Onken}, \&
  {Peterson}}]{Metzroth2006}
{Metzroth} K.~G., {Onken} C.~A., {Peterson} B.~M., 2006, \apj, 647, 901

\bibitem[{{Mortlock} {et~al}\mbox{.}(2011){Mortlock}, {Warren}, {Venemans},
  {Patel}, {Hewett}, {McMahon}, {Simpson}, {Theuns}, {Gonz{\'a}les-Solares},
  {Adamson}, {Dye}, {Hambly}, {Hirst}, {Irwin}, {Kuiper}, {Lawrence}, \&
  {R{\"o}ttgering}}]{Mortlock2011}
{Mortlock} D.~J. {et~al.}, 2011, \nat, 474, 616

\bibitem[{{Murray}, {Quataert} \& {Thompson}(2005){Murray}, {Quataert}, \&
  {Thompson}}]{Murray2005}
{Murray} N., {Quataert} E., {Thompson} T.~A., 2005, \apj, 618, 569

\bibitem[{{Padmanabhan} {et~al}\mbox{.}(2012){Padmanabhan}, {Xu}, {Eisenstein},
  {Scalzo}, {Cuesta}, {Mehta}, \& {Kazin}}]{Padmanabhan2012}
{Padmanabhan} N., {Xu} X., {Eisenstein} D.~J., {Scalzo} R., {Cuesta} A.~J.,
  {Mehta} K.~T., {Kazin} E., 2012, \mnras, 427, 2132

\bibitem[{{Pancoast}, {Brewer} \& {Treu}(2011){Pancoast}, {Brewer}, \&
  {Treu}}]{Pancoast2011}
{Pancoast} A., {Brewer} B.~J., {Treu} T., 2011, \apj, 730, 139

\bibitem[{{Pancoast} {et~al}\mbox{.}(2012){Pancoast}, {Brewer}, {Treu},
  {Barth}, {Bennert}, {Canalizo}, {Filippenko}, {Gates}, {Greene}, {Li},
  {Malkan}, {Sand}, {Stern}, {Woo}, {Assef}, {Bae}, {Buehler}, {Cenko},
  {Clubb}, {Cooper}, {Diamond-Stanic}, {Hiner}, {H{\"o}nig}, {Joner},
  {Kandrashoff}, {Laney}, {Lazarova}, {Nierenberg}, {Park}, {Silverman}, {Son},
  {Sonnenfeld}, {Thorman}, {Tollerud}, {Walsh}, \& {Walters}}]{Pancoast2012}
{Pancoast} A. {et~al.}, 2012, \apj, 754, 49

\bibitem[{{P{\^a}ris} {et~al}\mbox{.}(2012){P{\^a}ris}, {Petitjean}, {Aubourg},
  {Bailey}, {Ross}, {Myers}, {Strauss}, {Anderson}, {Arnau}, {Bautista},
  {Bizyaev}, {Bolton}, {Bovy}, {Brandt}, {Brewington}, {Browstein}, {Busca},
  {Capellupo}, {Carithers}, {Croft}, {Dawson}, {Delubac}, {Ebelke},
  {Eisenstein}, {Engelke}, {Fan}, {Filiz Ak}, {Finley}, {Font-Ribera}, {Ge},
  {Gibson}, {Hall}, {Hamann}, {Hennawi}, {Ho}, {Hogg}, {Ivezi{\'c}}, {Jiang},
  {Kimball}, {Kirkby}, {Kirkpatrick}, {Lee}, {Le Goff}, {Lundgren}, {MacLeod},
  {Malanushenko}, {Malanushenko}, {Maraston}, {McGreer}, {McMahon},
  {Miralda-Escud{\'e}}, {Muna}, {Noterdaeme}, {Oravetz},
  {Palanque-Delabrouille}, {Pan}, {Perez-Fournon}, {Pieri}, {Richards},
  {Rollinde}, {Sheldon}, {Schlegel}, {Schneider}, {Slosar}, {Shelden}, {Shen},
  {Simmons}, {Snedden}, {Suzuki}, {Tinker}, {Viel}, {Weaver}, {Weinberg},
  {White}, {Wood-Vasey}, \& {Y{\`e}che}}]{Paris2012}
{P{\^a}ris} I. {et~al.}, 2012, \aap, 548, A66

\bibitem[{{Park} {et~al}\mbox{.}(2015){Park}, {Woo}, {Bennert}, {Treu},
  {Auger}, \& {Malkan}}]{Park2014}
{Park} D., {Woo} J.-H., {Bennert} V.~N., {Treu} T., {Auger} M.~W., {Malkan}
  M.~A., 2015, \apj, 799, 164

\bibitem[{{Park} {et~al}\mbox{.}(2013){Park}, {Woo}, {Denney}, \&
  {Shin}}]{Park2013}
{Park} D., {Woo} J.-H., {Denney} K.~D., {Shin} J., 2013, \apj, 770, 87

\bibitem[{{Peterson}(1993)}]{Peterson1993}
{Peterson} B.~M., 1993, \pasp, 105, 247

\bibitem[{{Peterson}(1999)}]{Peterson1999}
{Peterson} B.~M., 1999, in Astronomical Society of the Pacific Conference
  Series, Vol. 175, Structure and Kinematics of Quasar Broad Line Regions,
  {Gaskell} C.~M., {Brandt} W.~N., {Dietrich} M., {Dultzin-Hacyan} D.,
  {Eracleous} M., eds., p.~49

\bibitem[{{Peterson} {et~al}\mbox{.}(1999){Peterson}, {Barth}, {Berlind},
  {Bertram}, {Bischoff}, {Bochkarev}, {Burenkov}, {Cheng}, {Dietrich},
  {Filippenko}, {Giannuzzo}, {Ho}, {Huchra}, {Hunley}, {Kaspi}, {Kollatschny},
  {Leonard}, {Malkov}, {Matheson}, {Mignoli}, {Nelson}, {Papaderos}, {Peters},
  {Pogge}, {Pronik}, {Sergeev}, {Sergeeva}, {Shapovalova}, {Stirpe}, {Tokarz},
  {Wagner}, {Wanders}, {Wei}, {Wilkes}, {Wu}, {Xue}, \&
  {Zou}}]{Peterson1999_NGC5548}
{Peterson} B.~M. {et~al.}, 1999, \apj, 510, 659

\bibitem[{{Peterson} {et~al}\mbox{.}(2002){Peterson}, {Berlind}, {Bertram},
  {Bischoff}, {Bochkarev}, {Borisov}, {Burenkov}, {Calkins}, {Carrasco},
  {Chavushyan}, {Chornock}, {Dietrich}, {Doroshenko}, {Ezhkova}, {Filippenko},
  {Gilbert}, {Huchra}, {Kollatschny}, {Leonard}, {Li}, {Lyuty}, {Malkov},
  {Matheson}, {Merkulova}, {Mikhailov}, {Modjaz}, {Onken}, {Pogge}, {Pronik},
  {Qian}, {Romano}, {Sergeev}, {Sergeeva}, {Shapovalova}, {Spiridonova}, {Tao},
  {Tokarz}, {Valdes}, {Vlasiuk}, {Wagner}, \& {Wilkes}}]{Peterson2002}
{Peterson} B.~M. {et~al.}, 2002, \apj, 581, 197

\bibitem[{{Peterson} {et~al}\mbox{.}(2004){Peterson}, {Ferrarese}, {Gilbert},
  {Kaspi}, {Malkan}, {Maoz}, {Merritt}, {Netzer}, {Onken}, {Pogge},
  {Vestergaard}, \& {Wandel}}]{Peterson2004RM}
{Peterson} B.~M. {et~al.}, 2004, \apj, 613, 682

\bibitem[{{Peterson} {et~al}\mbox{.}(2014){Peterson}, {Grier}, {Horne},
  {Pogge}, {Bentz}, {De Rosa}, {Denney}, {Martini}, {Sergeev}, {Kaspi},
  {Minezaki}, {Zu}, {Kochanek}, {Siverd}, {Shappee}, {Araya Salvo}, {Beatty},
  {Bird}, {Bord}, {Borman}, {Che}, {Chen}, {Cohen}, {Dietrich}, {Doroshenko},
  {Drake}, {Efimov}, {Free}, {Ginsburg}, {Henderson}, {King}, {Koshida},
  {Mogren}, {Molina}, {Mosquera}, {Motohara}, {Nazarov}, {Okhmat}, {Pejcha},
  {Rafter}, {Shields}, {Skowron}, {Skowron}, {Valluri}, {van Saders}, \&
  {Yoshii}}]{Peterson2014}
{Peterson} B.~M. {et~al.}, 2014, \apj, 795, 149

\bibitem[{{Peterson} \& {Horne}(2004)}]{Peterson2004}
{Peterson} B.~M., {Horne} K., 2004, Astronomische Nachrichten, 325, 248

\bibitem[{{Peterson} {et~al}\mbox{.}(1998){Peterson}, {Wanders}, {Bertram},
  {Hunley}, {Pogge}, \& {Wagner}}]{Peterson1998}
{Peterson} B.~M., {Wanders} I., {Bertram} R., {Hunley} J.~F., {Pogge} R.~W.,
  {Wagner} R.~M., 1998, \apj, 501, 82

\bibitem[{{Planck Collaboration} {et~al}\mbox{.}(2014){Planck Collaboration},
  {Ade}, {Aghanim}, {Armitage-Caplan}, {Arnaud}, {Ashdown}, {Atrio-Barandela},
  {Aumont}, {Baccigalupi}, {Banday}, \& et~al.}]{Planck2013CosmoParams}
{Planck Collaboration} {et~al.}, 2014, \aap, 571, A16

\bibitem[{{Press} {et~al}\mbox{.}(1992){Press}, {Teukolsky}, {Vetterling}, \&
  {Flannery}}]{Press1992}
{Press} W.~H., {Teukolsky} S.~A., {Vetterling} W.~T., {Flannery} B.~P., 1992,
  {Numerical recipes in FORTRAN. The art of scientific computing}. Cambridge:
  University Press, |c1992, 2nd ed.

\bibitem[{{Rafiee} \& {Hall}(2011)}]{Rafiee2011}
{Rafiee} A., {Hall} P.~B., 2011, \apjs, 194, 42

\bibitem[{{Rafter} {et~al}\mbox{.}(2011){Rafter}, {Kaspi}, {Behar},
  {Kollatschny}, \& {Zetzl}}]{Rafter2011}
{Rafter} S.~E., {Kaspi} S., {Behar} E., {Kollatschny} W., {Zetzl} M., 2011,
  \apj, 741, 66

\bibitem[{{Rafter} {et~al}\mbox{.}(2013){Rafter}, {Kaspi}, {Chelouche},
  {Sabach}, {Karl}, \& {Behar}}]{Rafter2013}
{Rafter} S.~E., {Kaspi} S., {Chelouche} D., {Sabach} E., {Karl} D., {Behar} E.,
  2013, \apj, 773, 24

\bibitem[{{Reichert} {et~al}\mbox{.}(1994){Reichert}, {Rodriguez-Pascual},
  {Alloin}, {Clavel}, {Crenshaw}, {Kriss}, {Krolik}, {Malkan}, {Netzer},
  {Peterson}, {Wamsteker}, {Altamore}, {Altieri}, {Anderson}, {Blackwell},
  {Boisson}, {Brosch}, {Carone}, {Dietrich}, {England}, {Evans}, {Filippenko},
  {Gaskell}, {Goad}, {Gondhalekar}, {Horne}, {Kazanas}, {Kollatschny},
  {Koratkar}, {Korista}, {MacAlpine}, {Maoz}, {Mazeh}, {McCollum}, {Miller},
  {Mendes de Oliveira}, {O'Brien}, {Pastoriza}, {Pelat}, {Perez}, {Perola},
  {Pogge}, {Ptak}, {Recondo-Gonzalez}, {Rodriguez-Espinosa}, {Rosenblatt},
  {Sadun}, {Santos-Lleo}, {Shields}, {Shrader}, {Shull}, {Simkin}, {Sitko},
  {Snijders}, {Sparke}, {Stirpe}, {Stoner}, {Storchi-Bergmann}, {Sun}, {Wang},
  {Welsh}, {White}, {Winge}, \& {Zheng}}]{Reichert1994}
{Reichert} G.~A. {et~al.}, 1994, \apj, 425, 582

\bibitem[{{Richards} {et~al}\mbox{.}(2011){Richards}, {Kruczek}, {Gallagher},
  {Hall}, {Hewett}, {Leighly}, {Deo}, {Kratzer}, \& {Shen}}]{Richards2011}
{Richards} G.~T. {et~al.}, 2011, \aj, 141, 167

\bibitem[{{Richards} {et~al}\mbox{.}(2006){Richards}, {Strauss}, {Fan}, {Hall},
  {Jester}, {Schneider}, {Vanden Berk}, {Stoughton}, {Anderson}, {Brunner},
  {Gray}, {Gunn}, {Ivezi{\'c}}, {Kirkland}, {Knapp}, {Loveday}, {Meiksin},
  {Pope}, {Szalay}, {Thakar}, {Yanny}, {York}, {Barentine}, {Brewington},
  {Brinkmann}, {Fukugita}, {Harvanek}, {Kent}, {Kleinman}, {Krzesi{\'n}ski},
  {Long}, {Lupton}, {Nash}, {Neilsen}, {Nitta}, {Schlegel}, \&
  {Snedden}}]{Richards2006}
{Richards} G.~T. {et~al.}, 2006, \aj, 131, 2766

\bibitem[{{Richstone}(1998)}]{Richstone1998}
{Richstone} D., 1998, in IAU Symposium, Vol. 184, The Central Regions of the
  Galaxy and Galaxies, {Sofue} Y., ed., p. 451

\bibitem[{{Robinson}(1994)}]{Robinson1994}
{Robinson} A., 1994, in Astronomical Society of the Pacific Conference Series,
  Vol.~69, Reverberation Mapping of the Broad-Line Region in Active Galactic
  Nuclei, {Gondhalekar} P.~M., {Horne} K., {Peterson} B.~M., eds., p. 147

\bibitem[{{Rodr{\'{\i}}guez-Pascual}
  {et~al}\mbox{.}(1997){Rodr{\'{\i}}guez-Pascual}, {Alloin}, {Clavel},
  {Crenshaw}, {Horne}, {Kriss}, {Krolik}, {Malkan}, {Netzer}, {O'Brien},
  {Peterson}, {Reichert}, {Wamsteker}, {Alexander}, {Barr}, {Blandford},
  {Bregman}, {Carone}, {Clements}, {Courvoisier}, {Robertis}, {Dietrich},
  {Dottori}, {Edelson}, {Filippenko}, {Gaskell}, {Huchra}, {Hutchings},
  {Kollatschny}, {Koratkar}, {Korista}, {Laor}, {MacAlpine}, {Martin}, {Maoz},
  {McCollum}, {Morris}, {Perola}, {Pogge}, {Ptak}, {Recondo-Gonz{\'a}lez},
  {J.~M.~Rodr{\'{\i}}guez-Espinoza}, {Rokaki}, {Santos-Lle{\'o}}, {Sekiguchi},
  {Shull}, {Snijders}, {Sparke}, {Stirpe}, {Stoner}, {Sun}, {Wagner},
  {Wanders}, {Wilkes}, {Winge}, \& {Zheng}}]{RodriguezPascual1997}
{Rodr{\'{\i}}guez-Pascual} P.~M. {et~al.}, 1997, \apjs, 110, 9

\bibitem[{{Saunders} {et~al}\mbox{.}(2004){Saunders}, {Bridges}, {Gillingham},
  {Haynes}, {Smith}, {Whittard}, {Churilov}, {Lankshear}, {Croom}, {Jones}, \&
  {Boshuizen}}]{Saunders2004}
{Saunders} W. {et~al.}, 2004, in Society of Photo-Optical Instrumentation
  Engineers (SPIE) Conference Series, Vol. 5492, Ground-based Instrumentation
  for Astronomy, {Moorwood} A.~F.~M., {Iye} M., eds., pp. 389--400

\bibitem[{{Schulze} \& {Wisotzki}(2010)}]{Schulze2010}
{Schulze} A., {Wisotzki} L., 2010, \aap, 516, A87

\bibitem[{{Shen} {et~al}\mbox{.}(2015){Shen}, {Brandt}, {Dawson}, {Hall},
  {McGreer}, {Anderson}, {Chen}, {Denney}, {Eftekharzadeh}, {Fan}, {Gao},
  {Green}, {Greene}, {Ho}, {Horne}, {Jiang}, {Kelly}, {Kinemuchi}, {Kochanek},
  {P{\^a}ris}, {Peters}, {Peterson}, {Petitjean}, {Ponder}, {Richards},
  {Schneider}, {Seth}, {Smith}, {Strauss}, {Tao}, {Trump}, {Wood-Vasey}, {Zu},
  {Eisenstein}, {Pan}, {Bizyaev}, {Malanushenko}, {Malanushenko}, \&
  {Oravetz}}]{Shen2014}
{Shen} Y. {et~al.}, 2015, \apjs, 216, 4

\bibitem[{{Shen} {et~al}\mbox{.}(2008){Shen}, {Greene}, {Strauss}, {Richards},
  \& {Schneider}}]{Shen2008}
{Shen} Y., {Greene} J.~E., {Strauss} M.~A., {Richards} G.~T., {Schneider}
  D.~P., 2008, \apj, 680, 169

\bibitem[{{Shen} \& {Liu}(2012)}]{Shen2012}
{Shen} Y., {Liu} X., 2012, \apj, 753, 125

\bibitem[{{Shen} {et~al}\mbox{.}(2011){Shen}, {Richards}, {Strauss}, {Hall},
  {Schneider}, {Snedden}, {Bizyaev}, {Brewington}, {Malanushenko},
  {Malanushenko}, {Oravetz}, {Pan}, \& {Simmons}}]{Shen2011}
{Shen} Y. {et~al.}, 2011, \apjs, 194, 45

\bibitem[{{Stirpe} {et~al}\mbox{.}(1994){Stirpe}, {Winge}, {Altieri}, {Alloin},
  {Aguero}, {Anupama}, {Ashley}, {Bertram}, {Calderon}, {Catchpole}, {Corradi},
  {Covino}, {Dottori}, {Feast}, {Ghosh}, {Hutton}, {Glass}, {Grebel}, {Jorda},
  {Koen}, {Laney}, {Maia}, {Marang}, {Mayya}, {Morrell}, {Nakada}, {Pastoriza},
  {Pati}, {Pelat}, {Peterson}, {Prabhu}, {Roberts}, {Sagar}, {Salamanca},
  {Sekiguchi}, {Storchi-Bergmann}, {Subramaniam}, {Van Winckel}, {van Wyk},
  {Villada}, {Wagner}, {Whitelock}, {Winkler}, {Clavel}, {Dietrich},
  {Kollatschny}, {O'Brien}, {Perola}, {Recondo-Gonzalez}, {Rodriguez-Pascual},
  \& {Santos-Lleo}}]{Stirpe1994}
{Stirpe} G.~M. {et~al.}, 1994, \apj, 425, 609

\bibitem[{{Sutherland} {et~al}\mbox{.}(2015){Sutherland}, {Emerson}, {Dalton},
  {Atad-Ettedgui}, {Beard}, {Bennett}, {Bezawada}, {Born}, {Caldwell}, {Clark},
  {Craig}, {Henry}, {Jeffers}, {Little}, {McPherson}, {Murray}, {Stewart},
  {Stobie}, {Terrett}, {Ward}, {Whalley}, \& {Woodhouse}}]{Sutherland2014}
{Sutherland} W. {et~al.}, 2015, \aap, 575, A25

\bibitem[{{Trakhtenbrot} \& {Netzer}(2012)}]{Trakhtenbrot2012}
{Trakhtenbrot} B., {Netzer} H., 2012, \mnras, 427, 3081

\bibitem[{{Trevese} {et~al}\mbox{.}(2014){Trevese}, {Perna}, {Vagnetti},
  {Saturni}, \& {Dadina}}]{Trevese2014}
{Trevese} D., {Perna} M., {Vagnetti} F., {Saturni} F.~G., {Dadina} M., 2014,
  \apj, 795, 164

\bibitem[{{Trump} {et~al}\mbox{.}(2011){Trump}, {Impey}, {Kelly}, {Civano},
  {Gabor}, {Diamond-Stanic}, {Merloni}, {Urry}, {Hao}, {Jahnke}, {Nagao},
  {Taniguchi}, {Koekemoer}, {Lanzuisi}, {Liu}, {Mainieri}, {Salvato}, \&
  {Scoville}}]{Trump2011}
{Trump} J.~R. {et~al.}, 2011, \apj, 733, 60

\bibitem[{{Vanden Berk} {et~al}\mbox{.}(2004){Vanden Berk}, {Yip}, {Connolly},
  {Jester}, \& {Stoughton}}]{VandenBerk2004}
{Vanden Berk} D., {Yip} C., {Connolly} A., {Jester} S., {Stoughton} C., 2004,
  in Astronomical Society of the Pacific Conference Series, Vol. 311, AGN
  Physics with the Sloan Digital Sky Survey, {Richards} G.~T., {Hall} P.~B.,
  eds., p.~21

\bibitem[{{Vanden Berk} {et~al}\mbox{.}(2001){Vanden Berk}, {Richards},
  {Bauer}, {Strauss}, {Schneider}, {Heckman}, {York}, {Hall}, {Fan}, {Knapp},
  {Anderson}, {Annis}, {Bahcall}, {Bernardi}, {Briggs}, {Brinkmann}, {Brunner},
  {Burles}, {Carey}, {Castander}, {Connolly}, {Crocker}, {Csabai}, {Doi},
  {Finkbeiner}, {Friedman}, {Frieman}, {Fukugita}, {Gunn}, {Hennessy},
  {Ivezi{\'c}}, {Kent}, {Kunszt}, {Lamb}, {Leger}, {Long}, {Loveday}, {Lupton},
  {Meiksin}, {Merelli}, {Munn}, {Newberg}, {Newcomb}, {Nichol}, {Owen}, {Pier},
  {Pope}, {Rockosi}, {Schlegel}, {Siegmund}, {Smee}, {Snir}, {Stoughton},
  {Stubbs}, {SubbaRao}, {Szalay}, {Szokoly}, {Tremonti}, {Uomoto}, {Waddell},
  {Yanny}, \& {Zheng}}]{VandenBerk2001}
{Vanden Berk} D.~E. {et~al.}, 2001, \aj, 122, 549

\bibitem[{{Vanden Berk} {et~al}\mbox{.}(2006){Vanden Berk}, {Shen}, {Yip},
  {Schneider}, {Connolly}, {Burton}, {Jester}, {Hall}, {Szalay}, \&
  {Brinkmann}}]{VandenBerk2006}
{Vanden Berk} D.~E. {et~al.}, 2006, \aj, 131, 84

\bibitem[{{Vestergaard}(2002)}]{Vestergaard2002}
{Vestergaard} M., 2002, \apj, 571, 733

\bibitem[{{Vestergaard}(2004)}]{Vestergaard2004}
{Vestergaard} M., 2004, \apj, 601, 676

\bibitem[{{Vestergaard} {et~al}\mbox{.}(2008){Vestergaard}, {Fan}, {Tremonti},
  {Osmer}, \& {Richards}}]{Vestergaard2008}
{Vestergaard} M., {Fan} X., {Tremonti} C.~A., {Osmer} P.~S., {Richards} G.~T.,
  2008, \apjl, 674, L1

\bibitem[{{Vestergaard} \& {Osmer}(2009)}]{Vestergaard2009}
{Vestergaard} M., {Osmer} P.~S., 2009, \apj, 699, 800

\bibitem[{{Vestergaard} \& {Peterson}(2006)}]{Vestergaard2006}
{Vestergaard} M., {Peterson} B.~M., 2006, \apj, 641, 689

\bibitem[{{Vestergaard} \& {Wilkes}(2001)}]{Vestergaard2001}
{Vestergaard} M., {Wilkes} B.~J., 2001, \apjs, 134, 1

\bibitem[{{Wandel}, {Peterson} \& {Malkan}(1999){Wandel}, {Peterson}, \&
  {Malkan}}]{Wandel1999}
{Wandel} A., {Peterson} B.~M., {Malkan} M.~A., 1999, \apj, 526, 579

\bibitem[{{Wanders} {et~al}\mbox{.}(1997){Wanders}, {Peterson}, {Alloin},
  {Ayres}, {Clavel}, {Crenshaw}, {Horne}, {Kriss}, {Krolik}, {Malkan},
  {Netzer}, {O'Brien}, {Reichert}, {Rodr{\'{\i}}guez-Pascual}, {Wamsteker},
  {Alexander}, {Anderson}, {Benitez}, {Bochkarev}, {Burenkov}, {Cheng},
  {Collier}, {Comastri}, {Dietrich}, {Dultzin-Hacyan}, {Espey}, {Filippenko},
  {Gaskell}, {George}, {Goad}, {Ho}, {Kaspi}, {Kollatschny}, {Korista}, {Laor},
  {MacAlpine}, {Mignoli}, {Morris}, {Nandra}, {Penton}, {Pogge}, {Ptak},
  {Rodr{\'{\i}}guez-Espinoza}, {Santos-Lle{\'o}}, {Shapovalova}, {Shull},
  {Snedden}, {Sparke}, {Stirpe}, {Sun}, {Turner}, {Ulrich}, {Wang}, {Wei},
  {Welsh}, {Xue}, \& {Zou}}]{Wanders1997}
{Wanders} I. {et~al.}, 1997, \apjs, 113, 69

\bibitem[{{Warren}, {Hewett} \& {Foltz}(2000){Warren}, {Hewett}, \&
  {Foltz}}]{Warren2000}
{Warren} S.~J., {Hewett} P.~C., {Foltz} C.~B., 2000, \mnras, 312, 827

\bibitem[{{Watson} {et~al}\mbox{.}(2011){Watson}, {Denney}, {Vestergaard}, \&
  {Davis}}]{Watson2011}
{Watson} D., {Denney} K.~D., {Vestergaard} M., {Davis} T.~M., 2011, \apjl, 740,
  L49

\bibitem[{{White} \& {Peterson}(1994)}]{White1994}
{White} R.~J., {Peterson} B.~M., 1994, \pasp, 106, 879

\bibitem[{{Whittle}(1985)}]{Whittle1985}
{Whittle} M., 1985, \mnras, 213, 1

\bibitem[{{Willott} {et~al}\mbox{.}(2010){Willott}, {Albert}, {Arzoumanian},
  {Bergeron}, {Crampton}, {Delorme}, {Hutchings}, {Omont}, {Reyl{\'e}}, \&
  {Schade}}]{Willott2010}
{Willott} C.~J. {et~al.}, 2010, \aj, 140, 546

\bibitem[{{Woo}(2008)}]{Woo2008}
{Woo} J.-H., 2008, \aj, 135, 1849

\bibitem[{{Woo} {et~al}\mbox{.}(2010){Woo}, {Treu}, {Barth}, {Wright}, {Walsh},
  {Bentz}, {Martini}, {Bennert}, {Canalizo}, {Filippenko}, {Gates}, {Greene},
  {Li}, {Malkan}, {Stern}, \& {Minezaki}}]{Woo2010}
{Woo} J.-H. {et~al.}, 2010, \apj, 716, 269

\bibitem[{{Wright} {et~al}\mbox{.}(2010){Wright}, {Eisenhardt}, {Mainzer},
  {Ressler}, {Cutri}, {Jarrett}, {Kirkpatrick}, {Padgett}, {McMillan},
  {Skrutskie}, {Stanford}, {Cohen}, {Walker}, {Mather}, {Leisawitz}, {Gautier},
  {McLean}, {Benford}, {Lonsdale}, {Blain}, {Mendez}, {Irace}, {Duval}, {Liu},
  {Royer}, {Heinrichsen}, {Howard}, {Shannon}, {Kendall}, {Walsh}, {Larsen},
  {Cardon}, {Schick}, {Schwalm}, {Abid}, {Fabinsky}, {Naes}, \&
  {Tsai}}]{Wright2010}
{Wright} E.~L. {et~al.}, 2010, \aj, 140, 1868

\bibitem[{{Yuan} {et~al}\mbox{.}(2015){Yuan}, {Lidman}, {Davis}, {Childress},
  {Abdalla}, {Banerji}, {Buckley-Geer}, {Carnero Rosell}, {Carollo},
  {Castander}, {D'Andrea}, {Diehl}, {Cunha}, {Foley}, {Frieman}, {Glazebrook},
  {Gschwend}, {Hinton}, {Jouvel}, {Kessler}, {Kim}, {King}, {Kuehn},
  {Kuhlmann}, {Lewis}, {Lin}, {Martini}, {McMahon}, {Mould}, {Nichol},
  {Norris}, {O'Neill}, {Ostrovski}, {Papadopoulos}, {Parkinson}, {Reed},
  {Romer}, {Rooney}, {Rozo}, {Rykoff}, {Sako}, {Scalzo}, {Schmidt}, {Scolnic},
  {Seymour}, {Sharp}, {Sobreira}, {Sullivan}, {Thomas}, {Tucker}, {Uddin},
  {Wechsler}, {Wester}, {Wilcox}, {Zhang}, {Abbott}, {Allam}, {Bauer},
  {Benoit-Levy}, {Bertin}, {Brooks}, {Burke}, {Carrasco Kind}, {Covarrubias},
  {Crocce}, {da Costa}, {DePoy}, {Desai}, {Doel}, {Eifler}, {Evrard}, {Fausti
  Neto}, {Flaugher}, {Fosalba}, {Gaztanaga}, {Gerdes}, {Gruen}, {Gruendl},
  {Honscheid}, {James}, {Kuropatkin}, {Lahav}, {Li}, {Maia}, {Makler},
  {Marshall}, {Miller}, {Miquel}, {Ogando}, {Plazas}, {Roodman}, {Sanchez},
  {Scarpine}, {Schubnell}, {Sevilla-Noarbe}, {Smith}, {Soares-Santos},
  {Suchyta}, {Swanson}, {Tarle}, {Thaler}, \& {Walker}}]{Yuan2015}
{Yuan} F. {et~al.}, 2015, preprint (arXiv:1504.03039)

\bibitem[{{Zu} {et~al}\mbox{.}(2013){Zu}, {Kochanek}, {Koz{\l}owski}, \&
  {Udalski}}]{Zu2013}
{Zu} Y., {Kochanek} C.~S., {Koz{\l}owski} S., {Udalski} A., 2013, \apj, 765,
  106

\bibitem[{{Zu}, {Kochanek} \& {Peterson}(2011){Zu}, {Kochanek}, \&
  {Peterson}}]{Zu2011}
{Zu} Y., {Kochanek} C.~S., {Peterson} B.~M., 2011, \apj, 735, 80

\end{thebibliography}

\appendix

\section{Quality of OzDES spectra}\label{sec:Spectra}

To quantify the quality of the OzDES spectra we estimated the SNR of the emission-line flux in the rest-frame spectrum, as follows: 
\begin{enumerate}
\item For each line,  the continuum flux density ($S^C$) is defined by a linear function of wavelength anchored by the median pixel value in the upper and lower continuum regions, defined in Table~\ref{tab:wavelengthlimits}. 
\item The line flux, $F_L $, is measured as the integrated flux above the linear continuum flux within the specified line integration wavelength limits such that
\begin{equation}
F_L = \sum_{i_1}^{i_2} [S_i-S^C_i],
\label{eq:flux}
\end{equation}
where $S_i$ is the uncalibrated flux density value at pixel $i$, and $i_1$ and $i_2$ are the pixel values that correspond to the defined minimum and maximum emission-line wavelength range.
\item The associated variance in the line flux, $\sigma^2(F_L)$, is measured as the integrated variance within the emission-line wavelength region. The SNR is then calculated as,
\begin{equation}
{\rm SNR} = \frac{F_L}{\sqrt{\sigma^2(F_L)}}.
\end{equation}
\end{enumerate}

\begin{figure}
\begin{center}
\includegraphics[width=0.45\textwidth]{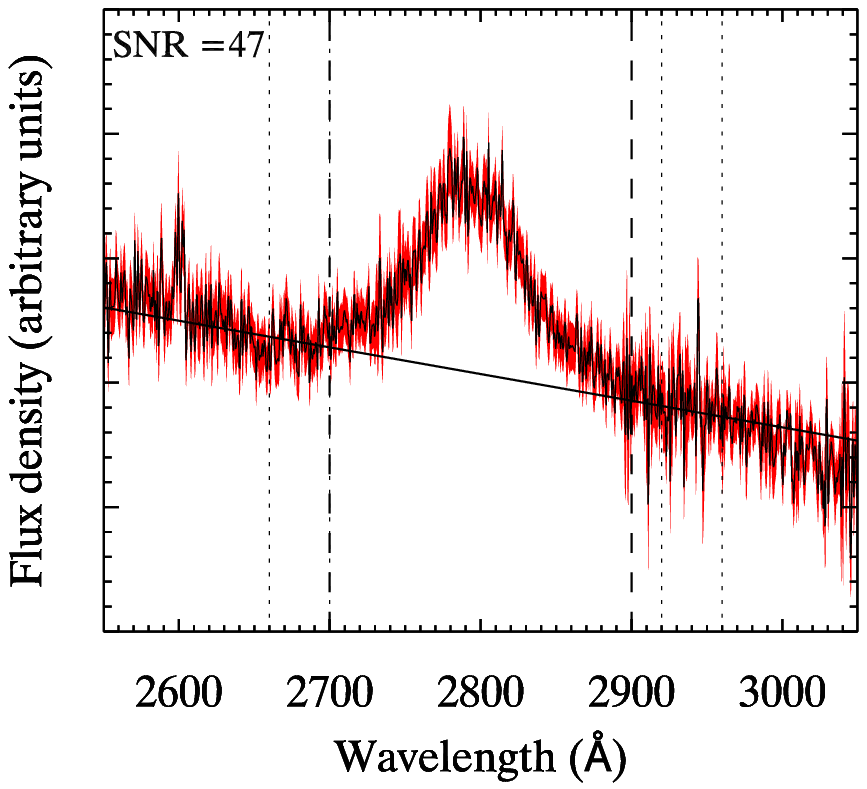}
\caption{An example of the line and continuum decomposition implemented in the SNR measurement process. The diagrams shows only the \mgii\ $\lambda$2798 emission line. The spectrum is shown by the black solid curve. The associated flux density uncertainty, $\sqrt{{\rm Var}(F_L)}$,  is shown by the surrounding red shaded region. The line summation limits and associated continuum regions are shown by the dashed and dotted vertical lines, respectively. The linear continuum fit is shown by the solid line and the resulting SNR in the emission-line flux measurement for this example is 47.}
\label{fig:hbetaexample}
\end{center}
\end{figure}

\begin{figure*}
\centering 
\includegraphics[width=0.8\textwidth]{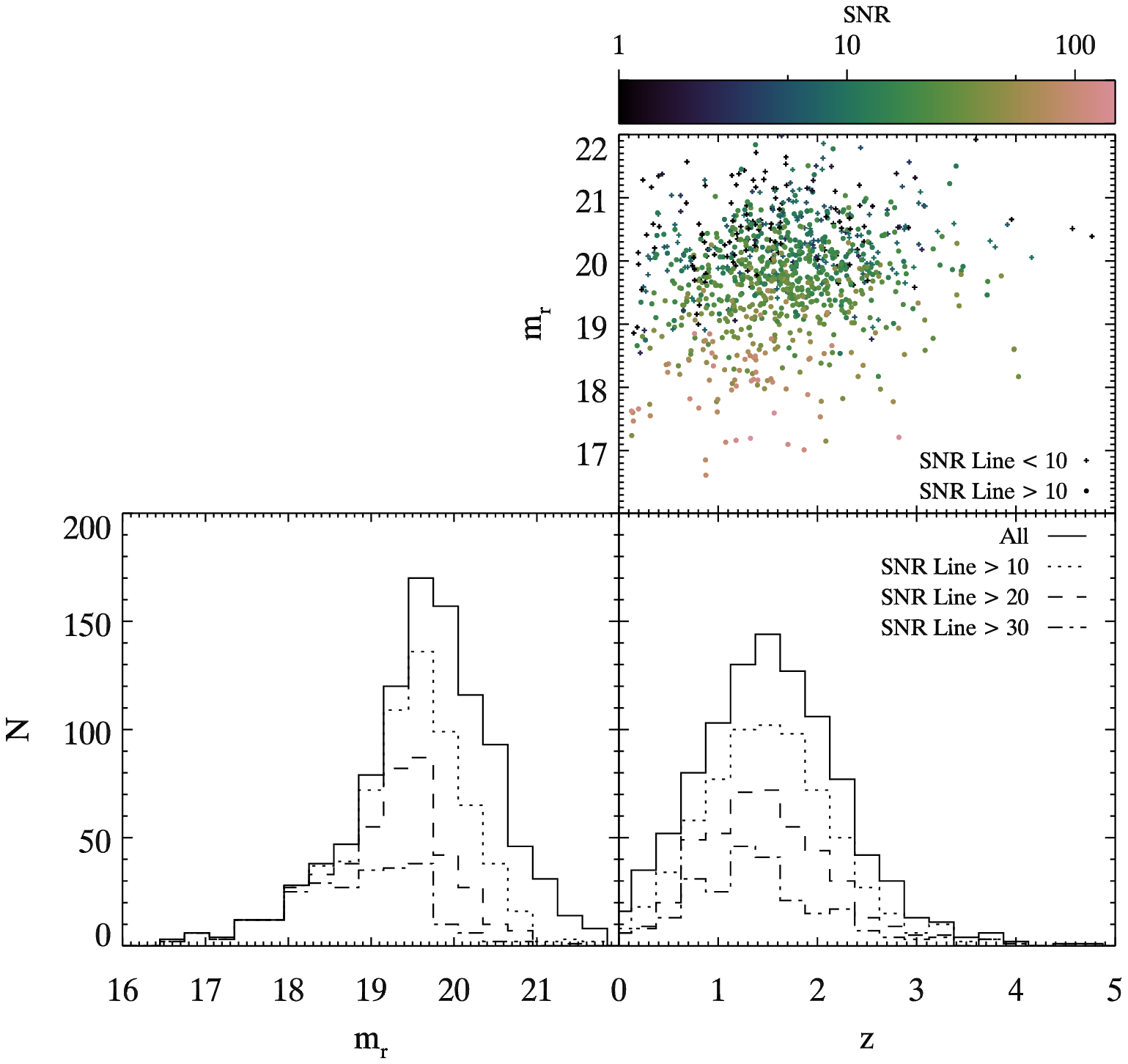}
\caption{The top panel shows the distribution in characteristic emission-line flux measurement signal-to-noise (SNR) values for each object, the dots represent SNR$>$10 and crosses represent SNR$<$10. To achieve the baseline emission-line uncertainty of 0.1 mag a bare minimum of SNR$>$10 is required. The histograms show the magnitude distribution (left) and redshift distribution (right) of the whole sample (solid) compared to objects with SNR values greater that 10 (dotted), 20 (dashed), or 30 (dot--dashed).}
\label{fig:snrdist30_10}
\end{figure*}

An example of a spectrum and the associated line summation limits and continuum fits is given in Fig. \ref{fig:hbetaexample}. The SNR for a  single object can vary markedly between lines and epochs  due to changes in observation conditions and individual line strength. Therefore, we allocate an SNR value for each QSO based on the median SNR of the best measured line in the spectra. Fig. \ref{fig:snrdist30_10} shows the magnitude and redshift distribution of quasars with different SNR value cut-offs. There appears to be no marked difference between the redshift distributions of high and low SNR objects; however, a significant shift in magnitude is observed such that higher  SNR  objects tend to be brighter, as expected. 

\begin{table}
\caption{Line Summation Limits}
\label{tab:wavelengthlimits}
\begin{center}
\begin{tabular}{cccc}
\hline
Emission &Line  &	Continuum  &	Continuum \\
line&integration&window&window\\
&limits (\AA) & lower (\AA)& upper (\AA)\\
\hline
H$\beta$ $\lambda$4861& 4810--4940 &4770--4800&5100--5130\\
\mgii\ $\lambda$2798\footnotemark & 2700--2900&2660--2700&2920--2960\\
\civ\ $\lambda$1549 & 1470--1620&1440--1470&1700--1730\\
\hline
\end{tabular}
\end{center}
\end{table}%
\footnotetext{The regions around \mgii\ are severely affected by Fe {\sc ii} contamination so this choice of pseudo-continuum wavelength range should be used critically \citep{Vestergaard2001}.}

These signal-to-noise measurements are preliminary, and only serve to quickly quantify the expected distribution of emission line measurement quality within the OzDES sample.  Line strength measurements will improve with the more thorough methods we will utilize for the final RM analysis, including decomposition of the spectra to remove contamination.

\end{document}